\documentclass[a4paper,12pt]{article}

\usepackage{amsmath}
\usepackage{amssymb}
\usepackage{amsfonts}
\usepackage{graphicx}

\newcommand{\Slash}[1]{{\ooalign{\hfil/\hfil\crcr$#1$}}}

\newcommand{\n}{\notag \\}

\renewcommand{\theequation}{\thesection.\arabic{equation}}

\allowdisplaybreaks[3]

\begin{document}
\begin{titlepage}

\begin{flushright}
\begin{tabular}{l}
KEK-TH-1296 \\
December 2008
\end{tabular}
\end{flushright}

\begin{Large}
\begin{center}
 {\bf Vertex operators of Type IIB matrix model
    \\ via calculation of disk amplitudes}
\end{center}
\end{Large}
\vspace{1cm}
\begin{center}
 Yoshihisa K{\sc itazawa}$^{1),2)}$
 \footnote{E-mail address: kitazawa@post.kek.jp}
 and
 Toshikazu N{\sc egi}$^{2)}$
 \footnote{E-mail address: tnegi@post.kek.jp}\\
\vspace{0.5cm}
  $^{1)}$ {\it High Energy Accelerator Research Organization (KEK),} \\
          {\it Tsukuba, Ibaraki 305-0801, Japan} \\
  $^{2)}$ {\it Department of Particle and Nuclear Physics,} \\
          {\it The Graduate University for Advanced Studies (SOKENDAI),} \\
          {\it Tsukuba, Ibaraki 305-0801, Japan} \\
\end{center}
\vspace{2cm}
\begin{abstract}
 We investigate the vertex operators of the supergravity multiplet in
 IIB(IKKT) matrix model by
 calculating the disk amplitudes, exploiting the technique of conformal
 field theory.
 The vertex operators of IIB matrix model are given as the coupling
 of a closed string and the open strings which are introduced by the existence of
 D($-1$)-branes.
 We consider the most generic couplings which involve both the bosonic
 and fermionic open strings.
 Our results are consistent with the previous results based on
 supersymmetry.
 We thus confirm the structure of the IIB matrix model vertex operators
 from the first principle.

\end{abstract}
\end{titlepage}
\section{Introduction}
Type IIB (IKKT) matrix model \cite{IKKT} is one of the proposal to
define the superstring theory nonperturbatively. It is formulated as
the zero-dimensional reduced model of the maximally supersymmetric
Yang-Mills theory. In the case of finite matrix size $N$, it can be
thought as the effective theory of D-instantons.

Action of the type IIB matrix model is :
\begin{equation}
 S = - \frac{1}{g^2} Tr
     \left[   \frac{1}{4} [A_\mu, A_\nu] [A^\mu, A^\nu]
            + \frac{1}{2} \bar{\epsilon} \gamma^\mu [A_\mu, \epsilon]
    \right] ,
\label{action}
\end{equation}
where $A_\mu$ and $\epsilon$ are $N \times N$ Hermitian matrices.
$A_\mu$ is a ten-dimensional vector and $\epsilon$ is a ten-dimensional
Majorana-Weyl spinor field respectively.

To calculate the correlators in type IIB matrix model, it is
necessary to construct the vertex operators. Their bosonic terms are
determined first by one of the authors \cite{Kitazawa2002} by exploiting
supersymmetry. Subsequently a systematic procedure is developed
through the construction of the supersymmetric Wilson loop operator. 
In this way, the vertex operator is determined
completely up to the 4-th rank antisymmetric tensor
\cite{ITU,ISTU}. Recently there was a further progress in
determining the precise form of vertex operators for the matrix
model \cite{KMS}.

In this paper, we investigate the vertex operators by using
conformal field theoretical technique. In this method, the bosonic
part of the vertex operators is investigated in
\cite{Kitazawa2002,OkawaOoguri1,OkawaOoguri2} in the presence of $N$
D(-1)-branes. The investigation of the fermionic part is carried out
in \cite{GreenGutperle} for the case of a single D(-1)-brane. We extend these
investigations into the most generic case. Namely, we consider
both the fermionic and bosonic open strings in the presence
of $N$ D(-1)-branes.
In the case of a single D(-1)-brane \cite{GreenGutperle}, the Majorana-Weyl fermions were not
matrices. In the presence of $N$ D(-1)-branes, Majorana-Weyl
fermions become $N \times N$ matrices. The matrix model vertex operator becomes the
Wilson lines due to the multiple insertions of the bosonic open string vertex
operators. Type IIB superstring theory is a closed string theory.
The type IIB matrix model is multiple D(-1)-brane theory for finite
$N$. The existence of the D-branes introduces open strings. The disk
amplitude technique relies on the fact that the the
vertex operator couples the closed strings to open strings.
Disk amplitude technique was also used in the study of supersymmetric 
four-dimensional effective gauge theories with instantons from type IIB 
superstring theory such as in \cite{Billo:2002hm}.

In comparison to the previous investigations using supersymmetric
transformation in IIB matrix model \cite{ITU,ISTU,KMS}, we can check
that the resultant formulae give the consistent form of the vertex
operators for the IIB matrix model up to the 4-th rank antisymmetric tensor completely.

This paper is organized as follows. In section 2, we introduce the disk
amplitudes which couple closed strings in the bulk to open strings
on the boundary. In section 3, we define the conformal
field theory vertex operators that are necessary to calculate the
disk amplitude. From the section 4 to section 9, we investigate the
explicit structure of the vertex operators of the type IIB matrix model
via the calculation of the disk amplitudes. We conclude in section
10 with discussions.

\section{Disk amplitudes}
In this section, we introduce the disk amplitudes which couple closed strings 
in the bulk to open strings on the boundary. They will be
investigated in the subsequent sections.

\begin{figure}
 \centering
 \includegraphics[width=8cm,clip]{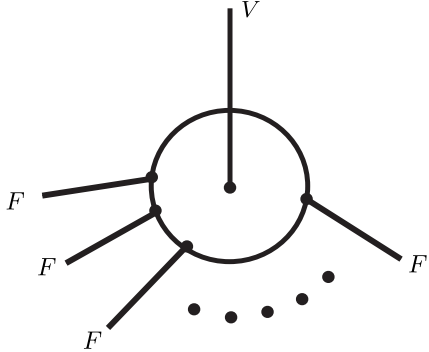}
 \caption{Disk amplitude in which $n$-th SUSY level closed
 string vertex operator couples to $n$ open strings.$V$ denotes the
 closed string (supergravity) vertex operator.
 $F$ denotes the vertex operator of the fermionic open string.}
 \label{DAmplitude1}
\end{figure}

When there exist D-branes, they introduce the boundary to the string
world-sheet. So the topology of the world-sheet becomes a disk and
it is conformaly equivalent to the upper half plane. We put open
string vertex operator on the boundary of the disk, namely the real
axis. We put a closed string vertex operator in the inner part of
the disk, i.e. at a location $z$ in the upper half plane. The closed
string vertex operator, bosonic and fermionic open string vertex
operator are denoted as $V$, $U$ and $F$ respectively. The concrete
form of the $V$, $U$ and $F$ are given in the next section. We
illustrate the disk amplitude in Figure.\ref{DAmplitude1}.

We focus on the supergravity
multiplet in type IIB superstring theory.
They are the BPS states which preserve the half SUSY (16
supercharges).
The IIB supergravity multiplet consists of a complex dilaton
$\mathbf{\Phi}$, a complex dilatino $\Lambda$, a complex
antisymmetric tensor $B_{\mu \nu}$, a complex gravitino
$\mathbf{\Psi}_\mu$, a real graviton $h_{\mu \nu}$ and a real 4-th
rank antisymmetric tensor $A_{\mu \nu \rho \sigma}$. 
They are generated from the complex dilaton by acting 16
broken SUSY generators. By regarding the complex dilaton as the highest
state in the supergravity multiplet, we can classify the type IIB
supergravity multiplet as Table.\ref{table:typeIIB}.

\begin{figure}
 \centering
 \includegraphics[width=8cm,clip]{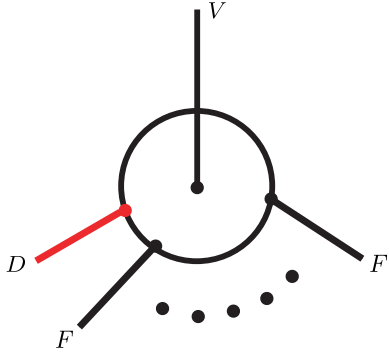}
 \caption{Alternative to $n$ fermionic open strings,
 the closed string vertex operator can couple to $n-2$ fermionic
 open strings and a single bosonic open string.}
 \label{DAmplitude2}
\end{figure}

For the vertex operator $V$ at the $n$-th SUSY level, we
may put $n$ fermionic open strings on the boundary of the
world-sheet. We may also put $m$ bosonic strings and $n-2m$ fermionic
open strings (where $ 2m \leq n $). For example we can insert a single bosonic open
string vertex operator on the boundary instead of $2$ fermionic
open strings. Such an amplitude is illustrated in
Figure.\ref{DAmplitude2}.

\begin{table}
 \caption{Massless multiplet of the type IIB superstring theory.}
\begin{center}
 \begin{tabular}{c|c}      \hline
  SUSY($n$-th level) & type IIB supergravity multiplet \\ \hline
   $n=0$          & complex scalar $\mathbf{\Phi}$ ((NS-NS dilaton) + (R-R axion))  \\
   $n=1$          & complex dilatino $\Lambda$    \\
   $n=2$          & complex antisymmetric tensor $B_{\mu \nu}$  \\
   $n=3$          & complex gravitino $\mathbf{\Psi_{\mu}}$  \\
   $n=4$          & real graviton $h_{\mu \nu}$
                    and real 4-th rank antisymmetric tensor $A_{\mu \nu \rho \sigma}$
 \end{tabular}
\end{center}
\label{table:typeIIB}
\end{table}

\newpage

\section{Vertex operators}
The vertex operator of the closed string that satisfies the NS-NS
boundary condition is
\begin{equation}
   V^{\mathrm{NN}}_{(-1,-1)} (z,\bar{z})
 = \zeta^{\mathrm{NN}}_{\mu \nu} \psi^\mu (z) e^{-\phi(z)}
  \bar{\psi}^\nu (\bar{z}) e^{- \bar{\phi}(\bar{z})} e^{i k X(z, \bar{z})},
\end{equation}
If $\zeta^{\mathrm{NN}}_{\mu \nu}$ is a traceless symmetric tensor,
it corresponds to the vertex operator for the graviton. In the case
of $(0)$-picture, instead of $e^{-\phi(z)} \psi(z)$, we use
$\partial X^\mu (z) + i k_{\rho} j^{\mu \rho}(z)$. For instance,
$V^{\text{NN}}_{(0,-1)}(z, \bar{z})$ becomes as follows,
\begin{equation}
   V^{\mathrm{NN}}_{(0,-1)} (z,\bar{z})
    = \zeta^{\mathrm{NN}}_{\mu \nu}
    (\partial X^\mu(z) + i k_{\rho} j^{\mu \rho}(z))
    \bar{\psi}(\bar{z}) e^{- \bar{\phi}(\bar{z})} e^{i k X(z,\bar{z})}.
\end{equation}
The R-R vertex operator involves the spin field $S$. For example,
\begin{equation}
 V^{\mathrm{RR}}_{(- \frac{1}{2}, - \frac{1}{2})} (z,\bar{z})
 = \zeta^{\mathrm{RR}}_{\mu \nu} S^a(z) e^{- \frac{1}{2} \phi(z)}
   (\gamma^{\mu \nu})_{a b}
   \bar{S}^b (\bar{z}) e^{- \frac{1}{2} \bar{\phi}(\bar{z})} e^{i k X(z,\bar{z})}.
\end{equation}
is the vertex operator for the second rank antisymmetric tensor.

 When D$(-1)$ -branes exist, we need to introduce open string
vertex operators in this theory. The bosonic one is
\begin{align}
   U_{0}(t)
 = & \Phi^{\mu}(X) i g_{\mu \nu} \partial_\perp X^{\nu} (t)
   - i g_{\mu \rho} g_{\nu \sigma} [\Phi^\rho(X), \Phi^\sigma(X)]
   \Psi^\mu \Psi^\nu(t) \n
 \equiv & W^{(0)}(t) + D^{(0)}(t) ,
\label{bosonicU}
\end{align}
where $\Phi^\mu$ is the bosonic field and it is related to $A^\mu$
(the bosonic degree of freedom in the type IIB matrix model) by
\begin{equation}
 2 \pi \alpha^\prime \Phi^\mu = A^\mu ,
\end{equation}
where we put
\begin{equation}
\alpha^\prime = 1
\end{equation}
in our convention.
We have decomposed $U^{(0)}$ into the two vertex operators $W^{(0)}$ and $D^{(0)}$.
\begin{align}
 &W^{(0)}(t) = \Phi^{\mu}(X) i g_{\mu \nu} \partial_\perp X^{\nu}(t) , \n
 &D^{(0)}(t) = - i g_{\mu \rho} g_{\nu \sigma}
     [\Phi^\rho(X), \Phi^\sigma(X)] \Psi^\mu \Psi^\nu(t) .
\end{align}
This vertex operator is introduced in
\cite{OkawaOoguri1,OkawaOoguri2}. Here we introduce $\Psi(t)$ as the fermionic field on the boundary.
The exponentiated term of $W^{(0)}$, namely  $ \exp (iW^{(0)})$ always appears in our calculation and introduces the
Wilson-line structure in the matrix model vertex operators. The
fermionic open string vertex operators are
\begin{equation}
  F_{- \frac{1}{2}}(x)
 = \epsilon_a S^a(x) e^{-\frac{1}{2} \phi (x)}
\end{equation}
and
\begin{equation}
  F_{\frac{1}{2}}(x)
 = \epsilon^a (\gamma_\mu)_{ab} S^b (x) \partial_\perp X^\mu (x) e^{\frac{1}{2} \phi (x)},
 \label{F(1/2)}
\end{equation}
where $\epsilon$'s are the Majorana-Weyl spinors. These vertex
operators are considered in \cite{GreenGutperle} for a single D
instanton. In this paper, we generalize them for $N$
D$(-1)$-branes. Therefore $A$ and $\epsilon$ become $N \times N$
matrices. 
In this process, we also introduce the vertex operator that contains the
commutator of $A$ and $\epsilon$. Because the conformal weight of the
operator $\Psi^\alpha \Psi^\beta(x)$ is the same with
$\partial_\mu$, we can add the following operator to (\ref{F(1/2)}):
\begin{equation}
   C_{+\frac{1}{2}}(x)
 = [\epsilon_a,\Phi_\alpha] (\gamma_\beta)^{a b}
   \Psi^\alpha \Psi^\beta (x)
   S^b(x) e^{+ \frac{1}{2} \phi (x)}.
\label{C(1/2)}
\end{equation}
This operator contains a bosonic string and a fermionic string. Thus we
can substitute this operator for $3$ fermionic strings.
Using these operators, we calculate the
disk amplitudes exploiting conformal field theory
technique\cite{Friedan1985}.

$W^{(0)}$, $F_{\frac{1}{2}}$, $D^{(0)}$ and $C_{\frac{1}{2}}$ contains the fields of type IIB matrix model as 
follows:
\begin{align}
 & W^{(0)}         \propto A, \n
 & F_{\frac{1}{2}} \propto \epsilon, \n
 & D^{(0)}         \propto [A,A], \n
 & C_{\frac{1}{2}} \propto [\epsilon,A] .
\end{align}
Just like the closed string vertex operators in Table.\ref{table:typeIIB} ,
we can classify the open string vertex operators by their SUSY level.
Namely, starting with $W^{(0)}$,
we can assign a SUSY level to each operator as in Table.\ref{table:WFDC}.
\begin{table}
 \caption{Open string vertex operators and fields.}
\begin{center}
 \begin{tabular}{c|c|c}      \hline
   Vertex operator & Field in type IIB matrix model &  SUSY($n$-th level) \\ \hline
   $W^{(0)}$                  & $A$        & $0$ \\
   $F_{\frac{1}{2}}$          & $\epsilon$     & $1$ \\
   $D^{(0)}$                  & $[A,A]$        & $2$ \\
   $C_{\frac{1}{2}}$          & $[A,\epsilon]$ & $3$ 
 \end{tabular}
\end{center}
\label{table:WFDC}
\end{table}

We calculate the disk amplitude in which 
$V$ is in the upper half plane and a group of  open string vertex operators
consisting of
$W^{(0)}$, $F_{\frac{1}{2}}$, $D^{(0)}$, $C_{\frac{1}{2}}$ are on the boundary.
In order to respect space-time SUSY in our calculation,
we impose the following condition for a group of the 
open string vertex operators;  i.e.
''SUSY level of a closed string vertex operator $V$ 
(shown in Table.\ref{table:typeIIB})
should be equal to the total SUSY level of a group of the open string vertex operators
 on the boundary 
(shown in \ref{table:WFDC})''.
Because we set the level of $W^{(0)}$ as $0$,
the closed string vertex operator $V$ always can couple to 
infinite number of $W^{(0)}$'s.
In fact $V$ always couples to
\begin{equation}
 \exp (iW^{(0)}) ,
\end{equation}
which corresponds to the Wilson-line operator in the matrix model.

In type IIB matrix model, the
highest state in the SUSY classification is a complex scalar.
In Table.\ref{table:WFDC}, we have assigned it null SUSY level.
Thus this field can couple only to $\exp(iW^{(0)})$.
The SUSY level for dilatino is $1$.
So the dilatino field can couple to $\exp(iW^{(0)})$ and one fermionic open string.
Since the SUSY level for B-field is $2$,
it can couple two $F_{\frac{1}{2}}$ or one $D^{(0)}$ in addition to $\exp(iW^{(0)})$.
We can calculate other disk amplitudes in a similar way.
With this rule, we will demonstrate that
the precise form of the vertex operators in type IIB matrix model
can be reproduced from conformal field theory.

\subsection{Propagators}
In order to calculate the disk amplitudes, we need the two point
functions for $X$'s and $\psi$'s. The two point function for $X$'s
is given as :
\begin{equation}
  \langle X^\mu (z) X^\nu (w) \rangle
  = - \left( g^{\mu \nu} \ln |z - w|
    - g^{\mu \nu} \ln |z - \bar{w}|
    + 2  G^{\mu \nu} \ln |z - \bar{w}| \right) ,
\end{equation}
where $g^{\mu \nu}$ is a closed string metric and $G^{\mu \nu}$ is
an open string metric. Here for the open string metric, we take
\begin{equation}
 G^{\mu \nu}=0.
\end{equation}
We consider $g_{\mu \nu}$ as Minkowski spacetime metric :
\begin{equation}
 g_{\mu \nu} = diag(-1,+1,+1,\dots,+1).
\end{equation}
So
\begin{equation}
 \langle X^\mu (z) X^\nu (w) \rangle = - g^{\mu \nu} \ln \left| \frac{z-t}{\bar{z}-t} \right|
.
\end{equation}
For later conveniences, we introduce a function $\tau(t,z)$ as
\begin{equation}
 \tau(t,z) = \frac{1}{2 \pi i}
             \ln \left( \frac{ t-z }{ t-\bar{z} } \right) .
\end{equation}
It is related to the Dirichlet propagator as
\begin{equation}
   i \partial_\perp \left[ \ln \left| \frac{z-t}{\bar{z}-t} \right| \right]
 = \frac{z-\bar{z}}{(z-t)(\bar{z}-t)}
 = 2 \pi i \frac{ \partial \tau(t,z) }{\partial t} .
\end{equation}
For a fixed value of $z$, $\tau (t,z)$ is a monotonically increasing
function of $t$ and
\begin{equation}
 \tau ( \infty, z ) - \tau ( - \infty, z ) = 1 .
\end{equation}

Next we write down the $2$-point functions of fermions,
\begin{align}
   & \langle \psi^\mu (z) \psi^\nu (w) \rangle
   = \frac{ 1 }{z-w} g^{\mu \nu} , \n
   & \langle \psi^\mu (z) \psi^\nu (\bar{w}) \rangle
   = \frac{1}{ z-\bar{w} } \left( -g^{\mu \nu} + 2 G^{\mu \nu} \right) , \n
   & \langle \bar{\psi}^\mu (\bar{z}) \psi^\nu (w) \rangle
   = \frac{1}{ \bar{z} - w } (- g^{\mu \nu} + 2 G^{\mu \nu} ) , \n
   & \langle \bar{\psi}^\mu (z) \bar{\psi}^\nu (\bar{w}) \rangle
   = \frac{1}{\bar{z} - \bar{w}}g^{\mu \nu} .
\end{align}
The boundary-boundary and bulk-boundary propagators for the
fermions are given by
\begin{align}
    & \langle \Psi^\mu (t) \Psi^\nu (t^\prime) \rangle =
     \frac{1}{t-t^\prime} g^{\mu \nu} , \n
    & \langle \psi^\mu (z) \Psi^\nu (t) \rangle =
     \frac{1}{z-t} g^{\mu \nu} , \n
    & \langle \bar{\psi}^\mu (\bar{z}) \Psi^\nu (t) \rangle =
     - \frac{1}{\bar{z}-t} g^{\mu \nu} .
 \end{align}

\section{Scalar field}
Having set up the disk amplitude calculation which involves 
closed string vertex operators in the bulk and open string vertex operators
on the boundary, we start explicit evaluations of these amplitudes
in conformal field theory. 
We start with the simplest case, namely the complex scalar which
is the highest state in the SUSY classification.
In type IIB supergravity multiplet, there are two kinds of Lorentz
scalars. They are dilaton and axion, which satisfy the NS-NS and R-R
boundary condition, respectively. Here we need the disk amplitude
for complex scalar. Therefore the amplitude that we want to consider
is the complex combination of the NS-NS part and R-R part and it
becomes as follows  :
\begin{equation}
 \langle \mathbf{\Phi} \rangle
 =     \langle \mathbf{\Phi}^{ \mathrm{NN} } \rangle
   + i \langle \mathbf{\Phi}^{ \mathrm{RR} } \rangle .
\end{equation}
Scalar field can couple only to bosonic open strings
that results in the Wilson line; i.e. the term given as follows :
\begin{equation}
   W^{(0)}(t_a)
 = \Phi^\rho i g_{\rho \sigma}
    \partial_\perp X^\sigma (t_a).
\end{equation}
The disk amplitude for the NS-NS dilation is given as
\begin{align}
  & \sum_{n=0}^{\infty}  2 \pi i Tr \mathbf{P} \int_{-\infty}^{+\infty}
     dt_2  dt_3  \dots
     dt_{n-1}  dt_n
    \langle c(z) \bar{c} (\bar{z}) c(t_1) \rangle
    \langle V^{ \mathrm{NN} }_{(-1, -1)}(z,\bar{z}) \frac{1}{n!} \prod_{a=1}^n 
    \left( i W^{(0)} (t_a) \right) \rangle \n
= & \sum_{n=0}^{\infty}  2 \pi i Tr \mathbf{P}
    \int_{-\infty}^{+\infty} dt_2  dt_3  \dots
    dt_{n-1} dt_n
    (z-\bar{z})(z-t_1)(\bar{z}-t_1)  \n
  & \times \langle \left[ \zeta^{ \mathrm{NN} }_{(0)} g_{\mu \nu} \psi^\mu (z) e^{-\phi}(z)
                 \bar{\psi}^\nu (\bar{z}) e^{-\bar{\phi}}(\bar{z}) e^{i k X(z,\bar{z})} \right]
    \prod_{a=1}^{n} \frac{1}{n!} \left( i \Phi^\rho i g_{\rho \sigma}
    \partial_{ \perp } X^\sigma (t_a) \right)
    \rangle \n
= & \sum_{n=0}^{\infty} 2 \pi i Tr \mathbf{P}
 \int_{-\infty}^{+\infty} dt_2 dt_3  \dots
 dt_{n-1} dt_n
 (z-\bar{z})(z-t_1)(\bar{z}-t_1) \n
 & \times \zeta^{ \mathrm{NN} }_{(0)}
   \frac{ - g_{\mu \nu} g^{\mu \nu} }{ (z-\bar{z})^2 }
   \prod_{a=1}^{n} \frac{1}{n!}
   \left[ - \Phi^\rho g_{\rho \sigma}
   [ - i k_\alpha g^{\alpha \sigma}
    \partial_\perp \ln \left| \frac{z-t_a}{\bar{z}-t_a} \right| ]  \right] \n
= & \sum_{n=0}^{\infty} 2 \pi i Tr \mathbf{P} \frac{1}{n!}
 \int_{-\infty}^{+\infty} dt_2 dt_3  \dots
 dt_{n-1} dt_n \n
 & \times \zeta^{ \mathrm{NN} }_{ (0) } \left[ ( z - \bar{z} )( z - t_1 )( \bar{z} - t_1 )
   \frac{-D}{ ( z-\bar{z} )^2 }
   \left( \Phi \cdot k \right)
   \frac{ z-\bar{z} }{ (z-t_1)(\bar{z}-t_1) } \right] \n
 & \times \prod_{a=2}^n \left[ \left( \Phi \cdot k \right)
   \frac{z-\bar{z}}{ (z-t_a)(\bar{z}-t_a) } \right] \n
= & \sum_{n=0}^{\infty} Tr \mathbf{P}
    \bigg[
    \frac{1}{n!} \int_{-\infty}^{+\infty} dt_2 dt_3  \dots
    dt_{n-1} dt_n \n
  & \times \bigg{ \{ } - \zeta^{ \mathrm{NN} }_{(0)} D \left( 2 \pi i \Phi \cdot k \right) \prod_{a=2}^n
      \left[ (\Phi \cdot k)
       2 \pi i \frac{\partial \tau(t_a,z)}{\partial t_a} \right]
      \bigg{ \} } \bigg] \n
 = & - \sum_{n=0}^{\infty} Tr \mathbf{P}
    D \zeta^{ \mathrm{NN} }_{(0)}
   \prod_{a=1}^{n}
   \left[ \frac{1}{n!} \int^{\infty}_{-\infty} d t_a
         \left[ i \left( 2 \pi \Phi \cdot k \right)
                \frac{\partial \tau(t_a,z)}{\partial t_a} \right]
   \right] \n
 = & -D \zeta^{ \mathrm{NN} }_{(0)} \sum_{n=0}^\infty Tr \mathbf{P}
     \prod_{a=1}^{n} \frac{1}{n!}
     \left[ \int^{1}_{0} d \tau \left[ i \left( A \cdot k \right)
             \right] \right] \n
 = & - D \zeta^{ \mathrm{NN} }_{(0)} STr \exp \left( i A \cdot k \right) ,
\end{align}
where $D$ is the number of spacetime dimensions and in the superstring
theory, we consider the case $ D = 10 $.
$Tr$ is the trace for $A$ and $\epsilon$ matrices.
We use $tr$ for the trace for $\gamma$ - matrices.
$STr$ denotes the symmetrized trace
defined in Appendix A.4.
$\mathbf{P}$ denotes the path-ordered operators.
Up to a normalization constant, we find
\begin{equation}
 \zeta^{ \mathrm{NN} }_{(0)} STr \exp \left( i A \cdot k \right) .
 \label{eq:DAforDL}
\end{equation}

Secondly, we calculate the disk amplitude for the axion.
The amplitude is
 \begin{align}
  & \sum_{n=0}^{\infty}  2 \pi i Tr \mathbf{P}
  \int_{-\infty}^{+\infty} dt_2 dt_3  \dots
  dt_{n-1} dt_n
  \langle c(z) \bar{c} (\bar{z}) c(t_1) \rangle
  \langle V^{ \mathrm{RR} }_{(-\frac{1}{2}, - \frac{3}{2})}(z,\bar{z})
  \frac{1}{n!} \prod_{a=1}^n \left( iW^{(0)}(t_a) \right) \rangle \n
  = & \sum_{n=0}^{\infty}  2 \pi i Tr \mathbf{P}
      \frac{1}{n!} \int_{-\infty}^{+\infty} dt_2 dt_3  \dots
      dt_{n-1} dt_n
      ( z-\bar{z} )( z-t_1 )(\bar{z}-t_1) \n
  & \times \langle \left[ \zeta^{ \mathrm{RR} }_{(0)} e^{ - \frac{1}{2} \phi } (z)
                          S^a (z) \delta_{a b} S^b (\bar{z})
                          e^{- \frac{3}{2} \bar{\phi}} (\bar{z}) e^{i k X(z,\bar{z})} \right]
    \prod_{a=1}^{n} [ i \Phi^\rho i g_{\rho \sigma}
    \partial_{ \perp } X^\sigma (t_a) ]
           \rangle .
 \end{align}
The OPE for 2-point function of the spin fields is given by (\ref{2spin}).
So the disk amplitude is
\begin{align}
 & \sum_{n=0}^{\infty} \frac{1}{n!} 2 \pi i Tr \mathbf{P}
 \int_{-\infty}^{+\infty} dt_2 dt_3  \dots
 dt_{n-1} dt_n
 (z-\bar{z})(z-t_1)(\bar{z}-t_1) \n
 & \times \zeta^{ \mathrm{RR} }_{(0)}
   \left[ \delta_{a b}
          (z-\bar{z})^{ - \frac{3}{4} } \right]
   \frac{ - \delta^{a b} }{ (z-\bar{z})^{ \frac{5}{4} } }
   \prod_{a=1}^{n}
   \left[ - \Phi^\rho g_{\rho \sigma}
   [ - i k_\alpha g^{\alpha \sigma}
    \partial_\perp \ln \left| \frac{z-t_a}{\bar{z}-t_a} \right| ]  \right] \n
= & \sum_{n=0}^{\infty} 2 \pi i Tr \mathbf{P}
 \frac{1}{n!} \int_{-\infty}^{+\infty} dt_2 dt_3 \dots
 dt_{n-1} dt_n \n
 & \times \zeta^{ \mathrm{RR} }_{ (0) } \left[ ( z - \bar{z} )( z - t_1 )( \bar{z} - t_1 )
   \frac{- 2^{\frac{D}{2}} }{ ( z-\bar{z} )^2 }
   \left( \Phi \cdot k \right)
   \frac{ z-\bar{z} }{ (z-t_1)(\bar{z}-t_1) } \right] \n
 & \times \prod_{a=2}^n \left[ \left( \Phi \cdot k \right)
   \frac{z-\bar{z}}{ (z-t_a)(\bar{z}-t_a) } \right] \n
= & \sum_{n=0}^{\infty} Tr \mathbf{P}
    \left[
    \frac{1}{n!} \int_{-\infty}^{+\infty} dt_2 dt_3  \dots
    dt_{n-1} dt_n
    \zeta^{ \mathrm{RR} }_{(0)} (- 2^{\frac{D}{2}}) \left( 2 \pi i \Phi \cdot k \right) \prod_{a=2}^n
    \left[ (\Phi \cdot k)
      2 \pi i \frac{\partial \tau(t_a,z)}{\partial t_a} \right]
    \right] \n
 = & - 2^{\frac{D}{2}} \zeta^{ \mathrm{RR} }_{(0)}
   \sum_{n=0}^{\infty} Tr \mathbf{P}
   \prod_{a=1}^{n} \frac{1}{n!}
   \left[ \int d t_a
         \left[ i \left( 2 \pi \Phi \cdot k \right)
                \frac{\partial \tau(t_a,z)}{\partial t_a} \right]
   \right] \n
 = & - 2^{\frac{D}{2}} \zeta^{ \mathrm{RR} }_{(0)} \sum_{n=0}^\infty Tr \mathbf{P}
     \prod_{a=1}^{n} \frac{1}{n!}
     \left[ \int^{1}_{0} d \tau \left[ i \left( A \cdot k \right)
             \right] \right] \n
  = & - 2^{\frac{D}{2}} \zeta^{ \mathrm{RR} }_{(0)} STr
      \exp \left( i \left( A \cdot k \right)
             \right) .
\end{align}
We thus find the identical result with (\ref{eq:DAforDL}),
\begin{equation}
 \zeta^{ \mathrm{RR} }_{(0)} STr \exp \left( i A \cdot k \right) .
 \label{eq:DAforAX}
\end{equation}
Therefore, by forming a linear combination (\ref{eq:DAforDL})$+ i$(\ref{eq:DAforAX}),
we can derive the vertex operator for the scalar field in typeIIB matrix
model as
\begin{equation}
 (\zeta^{ \mathrm{NN} }_{(0)} + i \zeta^{ \mathrm{RR} }_{(0)}) STr \exp \left( i k \cdot A \right)
 \equiv \mathbf{\Phi} (\lambda) \ V^{\mathbf{\Phi}} (A,\epsilon) .
\end{equation}
\section{Dilatino}
Next we calculate the disk amplitude for the dilatino.
The disk amplitude for the complex dilatino is given as
\begin{equation}
   \langle \Lambda \rangle
 =     \langle \Lambda^{ \mathrm{RN} } \rangle
   + i \langle \Lambda^{ \mathrm{NR} } \rangle .
\end{equation}
Dilatino couples to one fermionic open string.
Firstly, the disk amplitude for the NS-R dilatino field is
 \begin{align}
   & \langle \Lambda^{ \mathrm{NR} }  \rangle \n
 = & \langle c(z) \bar{c} (\bar{z})
             V^{ \mathrm{NR} }_{(-1,-\frac{1}{2})}(z,\bar{z}) c(x_1)
             Tr \mathbf{P}
             \exp \left( i \int d t W^{(0)} (t) \right)
             F_{- \frac{1}{2} } (x_1)
     \rangle \n
 = & \langle c(z) c(\bar{z})
      \zeta^{ \mathrm{NR} }_{\mu a}
      e^{-\phi}(z) \psi^\mu (z) \bar{S}^a (\bar{z}) e^{- \frac{1}{2} \bar{\phi}}(\bar{z})
      e^{i k X}(z,\bar{z}) \n
   & \times  Tr \mathbf{P} \exp \left( i \int d t \Phi^{\rho} i g_{\rho \sigma} \partial_\perp X^{\sigma} (t) \right)
      c(x_1) \epsilon_b S^b (x_1) e^{ - \frac{1}{2} \phi(x_1) }
     \rangle \n
 = & \zeta^{ \mathrm{NR} }_{\mu a} Tr \mathbf{P}
      \exp \left( - \int d t \Phi^{\alpha} (- ik_\alpha) \partial_\perp \ln \left| \frac{z-t}{\bar{z}-t} \right| \right) \n
   & \times
     \left[ (z-\bar{z})(z-x_1)(\bar{z}-x_1) \right]
     \left[ (z-\bar{z})^{ -\frac{1}{2} }
            (z-x_1)^{ -\frac{1}{2} }
            (\bar{z}-x_1)^{ -\frac{1}{4} } \right] \n
   & \times
     \left[
      (\gamma^\mu)^{a b}
      ( z-\bar{z} )^{- \frac{1}{2}}
      ( z-x_1 )^{- \frac{1}{2}}
      ( \bar{z}-x_1 )^{- \frac{3}{4}}
     \right]
     \epsilon_b \n
 = & \zeta^{ \mathrm{NR} }_{\mu a} (\gamma^\mu)^{a b} Tr \mathbf{P}
      \exp \left( \int d t \Phi^{\alpha} ik_\alpha \partial_\perp \ln \left| \frac{z-t}{\bar{z}-t} \right| \right)
     \epsilon_b \n
 = & \zeta^{ \mathrm{NR} }_{\mu a} (\gamma^\mu)^{a b} Tr \mathbf{P}
      \exp \left( i \int d t (2 \pi \Phi \cdot k)
                  \frac{\partial \tau}{\partial t}
           \right)
     \epsilon_b \n
 = & \zeta^{ \mathrm{NR} }_{\mu a} (\gamma^\mu)^{a b} STr
     \exp \left( i k \cdot A \right) \epsilon_b ,
 \label{NRdilatino}
 \end{align}
where we used the correlator
\begin{equation}
   \langle \psi^\mu (z_1) S^a (z_2) S^b (z_3) \rangle
 = ( z_1 - z_2 )^{ - \frac{1}{2} } ( z_1 - z_3 )^{ - \frac{1}{2} }
   (z_2 - z_3)^{ - \frac{3}{4} } (\gamma^\mu)^{a b} .
\end{equation}
The disk amplitude
for the R-NS dilatino field is similarly
calculated as
 \begin{align}
   & \langle \Lambda^{ \mathrm{RN} } \rangle \n
 = & \langle
      c(z) \bar{c} (\bar{z})
      V^{ \mathrm{RN} } (z,\bar{z})
      c(x_1) Tr \mathbf{P}
      \exp \left(
               i \int d t W^{(0)} (t)
           \right)
      F_{ -\frac{1}{2} } (x_1)
     \rangle \n
  = & - \zeta^{ \mathrm{RN} }_{a \mu}
     STr
     \left(
       \exp \left( i k \cdot A \right) (\gamma^\mu)^{a b} \epsilon_b
     \right) .
\label{RNdilatino}
\end{align}
From (\ref{RNdilatino})$ +i$(\ref{NRdilatino}),
we get
\begin{equation}
 (\zeta^{ \mathrm{RN} }_{\mu a} + i \zeta^{ \mathrm{NR} }_{a \mu}) (\gamma^\mu)^{a b}
  STr \left( e^{i k \cdot A}
  \epsilon_b \right) \equiv \Lambda(\lambda) V^{\Lambda}(A,\epsilon) ,
\end{equation}
where $\Lambda(\lambda)$ is the wave function of dilatino and it is defined as
\begin{equation}
    \Lambda^b (\lambda)
  \equiv (\zeta^{ \mathrm{RN} }_{\mu a}
         + i \zeta^{ \mathrm{NR} }_{a \mu}) (\gamma^\mu)^{a b} .
\end{equation}

\section{B-field}
Next, we consider the vertex operators for Kalb-Ramond B-field.
In type IIB theory, we have NS-NS B-field and R-R B-field.
The vertex operator of the matrix model is
given by the following disk amplitude:
\begin{equation}
 \langle B \rangle
 =     \langle B^{ \mathrm{NN} } \rangle
   + i \langle B^{ \mathrm{RR} } \rangle .
\end{equation}
Since B-field is the two times SUSY transformed field in type IIB supergravity
multiplet, it couples to two fermionic strings.
\subsection{B-field coupling to fermionic strings}
We calculate the disk amplitude in which the vertex operator of the
B-field couples to two fermionic open strings.
Firstly, we calculate NS-NS part.
It is given as follows,
\begin{align}
   & \langle B^{ \mathrm{NN} } \rangle \n
 = & \langle c (z) \bar{c} (\bar{z})
      V^{ \mathrm{NN} }_{(-1,0)} (z, \bar{z})
      c(x_1) Tr \mathbf{P}
     \exp
          \left(
           i W^{(0)}(t)
          \right)
     F_{-\frac{1}{2}} (x_1) \int dx_2 F_{-\frac{1}{2}} (x_2)
     \rangle \n
 \sim & \langle c (z) \bar{c}(\bar{z})
             \zeta^{ \mathrm{NN} }_{\mu \nu}
               e^{- \phi}(z) \psi(z)
              (\bar{\partial} X^\nu + i k_\lambda \bar{j}^{\lambda \nu} )(\bar{z})
              e^{i k X (z,\bar{z})} \n
     & \times Tr \mathbf{P}
     \exp
          \left(
           i \int dt \Phi^\rho i g_{\rho \sigma} \partial_\perp X^\sigma (t)
          \right)
               c(x_1) \epsilon_a S^a (x_1) e^{-\frac{1}{2} \phi}(x_1)
               \int dx_2 \epsilon_b S^b (x_2) e^{-\frac{1}{2} \phi}(x_2)
     \rangle \n
 = & \zeta^{ \mathrm{NN} }_{\mu \nu}
     Tr \mathbf{P}
        \exp
          \left(
           i \int d \tau k \cdot A
          \right)
     \epsilon_a \epsilon_b \int d x_2
     \langle c (z) \bar{c} (\bar{z}) c(x_1) \rangle
     \langle e^{- \phi}(z) e^{- \frac{1}{2} \phi}(x_1)
             e^{-\frac{1}{2} \phi}(x_2) \rangle \n
   & \times \langle
      i k_\lambda \bar{j}^{\lambda \nu} (\bar{z}) \psi^\mu (z) S^a (x_1) S^b (x_2)
     \rangle .
\label{BNN1}
\end{align}
The ghost operator part and picture operator part are calculated as
 \begin{align}
   & \langle c(z) \bar{c} (\bar{z}) c(x_1) \rangle
   = ( z-\bar{z} ) (z-x_1) (\bar{z}-x_1) , \n
   & \langle e^{-\phi}(z) e^{-\frac{1}{2} \phi}(x_1) e^{-\frac{1}{2} \phi}(x_2) \rangle
   = (z-x_1)^{-\frac{1}{2}} (z-x_2)^{-\frac{1}{2}}
     (x_1-x_2)^{-\frac{1}{4}} .
 \label{BNNgp}
\end{align}
The spin field part is
 \begin{align}
   & \langle
       \bar{j}^{\lambda \nu} (\bar{z}) \psi^\mu (z) S^a (x_1) S^b (x_2)
     \rangle \n
 = & \sum_{i}
      \frac{M^{\lambda \nu} (i)}{\bar{z}-x_i}
      \frac{ (\gamma^\mu)^{ab} }{(z-x_1)^{ \frac{1}{2} } (z-x_2)^{ \frac{1}{2} } (x_1-x_2)^{\frac{3}{4}}  } ,
 \label{jpss1}
\end{align}
where $M^{\lambda \nu}(i)$ is the Lorentz generator which acts on
vector and spinor indices.
$(i)$ denotes the fields and the summations are taken over all fields in the correlator.
For vector indices, $M^{\mu \nu}$ acts as,
\begin{equation}
 iM^{\mu \nu} v^\rho = i ( g^{\rho \nu} v^\mu - g^{\mu \rho} v^\nu ) .
\end{equation}
For spinor indices, it acts like
\begin{equation}
 iM^{\mu \nu} \psi_a = \frac{i}{2} (\gamma^{\mu \nu} \psi)_a .
\end{equation}
Then (\ref{jpss1}) becomes
 \begin{align}
    & \langle
        \bar{j}^{\lambda \nu} (\bar{z}) \psi^\mu (z) S^a (x_1) S^b (x_2)
      \rangle \n
  = & \bigg( \frac{1}{\bar{z} - z }
               \left(  g^{\mu \nu} (\gamma^\lambda)^{a b}
                     - g^{\lambda \mu} (\gamma^\nu)^{a b} \right) \n
    & + \frac{ \frac{1}{2} }{ \bar{z} - x_1 }
     \left( \gamma^{\lambda \nu } \gamma^\mu \right)^{a b}
      - \frac{ \frac{1}{2} }{ \bar{z} - x_2 }
     \left( \gamma^\mu \gamma^{\lambda \nu} \right)^{a b} \bigg) \n
    & \times \frac{ 1 }
                  {(z-x_1)^{ \frac{1}{2} } (z-x_2)^{ \frac{1}{2} }
                   (x_1-x_2)^{\frac{3}{4}}  } \n
  = & \bigg( \frac{1}{\bar{z} - z }
               \left(  g^{\mu \nu} (\gamma^\lambda)^{a b}
                     - g^{\lambda \mu} (\gamma^\nu)^{a b} \right) \n
    & + \frac{ \frac{1}{2} }{ \bar{z} - x_1 }
     \left( \gamma^{\lambda \nu \mu}
           - g^{\lambda \mu} \gamma^{\nu} + g^{\mu \nu} \gamma^{\lambda} \right)^{a b} \n
    & - \frac{ \frac{1}{2} }{ \bar{z} - x_2 }
     \left( \gamma^{\lambda \nu \mu}
           + g^{\lambda \mu} \gamma^{\nu} - g^{\mu \nu} \gamma^{\lambda} \right)^{a b}
      \bigg) \n
    & \times \frac{ 1 }
                  {(z-x_1)^{ \frac{1}{2} } (z-x_2)^{ \frac{1}{2} }
                   (x_1-x_2)^{\frac{3}{4}}  } .
 \label{jpss2}
\end{align}
Substituting (\ref{BNNgp}) and (\ref{jpss2}) into (\ref{BNN1}),
we obtain
\begin{align}
   & \langle B^{ \mathrm{NN} } \rangle \n
 = & \zeta^{ \mathrm{NN} }_{\mu \nu} i k_{\lambda}
     Tr \mathbf{P}
        \exp
          \left(
           i \int d \tau k \cdot A
          \right)
     \epsilon_a \epsilon_b \n
   & \times
     \int dx_2
     \left[ ( z-\bar{z} ) (z-x_1) (\bar{z}-x_1) \right]
     \left[ (z-x_1)^{-\frac{1}{2}} (z-x_2)^{-\frac{1}{2}}
            (x_1-x_2)^{-\frac{1}{4}} \right] \n
   & \times   \bigg( \frac{1}{\bar{z} - z }
                \left(  g^{\mu \nu} (\gamma^\lambda)^{a b}
                      - g^{\lambda \mu} (\gamma^\nu)^{a b} \right)
            + \frac{ \frac{1}{2} }{ \bar{z} - x_1 }
                \left( \gamma^{\lambda \nu \mu}
                - g^{\lambda \mu} \gamma^{\nu} + g^{\mu \nu} \gamma^{\lambda} \right)^{a b} \n
    & \qquad - \frac{ \frac{1}{2} }{ \bar{z} - x_2 }
     \left( \gamma^{\lambda \nu \mu}
           + g^{\lambda \mu} \gamma^{\nu} - g^{\mu \nu} \gamma^{\lambda} \right)^{a b}
      \bigg) \n
    & \times \frac{ 1 }
                  {(z-x_1)^{ \frac{1}{2} } (z-x_2)^{ \frac{1}{2} }
                   (x_1-x_2)^{\frac{3}{4}}  } \n
 = & i \zeta^{ \mathrm{NN} }_{\mu \nu} k_\lambda
     Tr \mathbf{P}
        \exp
          \left(
           i \int d \tau k \cdot A
          \right)
     \left[ \epsilon_a (\gamma^{ \lambda \nu \mu })^{a b} \epsilon_b \right] \n
   & \times
     \int d x_2
     \left[ (z-\bar{z})(z-x_1)^{\frac{1}{2}} (\bar{z}-x_1)
            (z-x_2)^{- \frac{1}{2}}(x_1-x_2)^{-\frac{1}{4}} \right] \n
   & \times \frac{1}{2} \left( \frac{x_1 - x_2}{(\bar{z}-x_1)(\bar{z}-x_2)} \right)
     \left( \frac{1}
            { (z-x_1)^{\frac{1}{2}} (z-x_2)^{ \frac{1}{2}}
              (x_1-x_2)^{\frac{3}{4}} } \right) \n
 = &  \frac{i}{2} k_\lambda \zeta^{ \mathrm{NN} }_{\mu \nu}
     STr
        \exp
          \left(
           i k \cdot A
          \right)
     \left[ \epsilon_a (\gamma^{ \lambda \nu \mu })^{a b} \epsilon_b \right]
     \int d x_2 \left[ \frac{z-\bar{z}}{(z-x_2)(\bar{z}-x_2)} \right] \n
 = & - \pi k_\lambda \zeta^{ \mathrm{NN} }_{\mu \nu}
     STr
        \exp
          \left(
           i k \cdot A
          \right)
     \left[ \epsilon_a (\gamma^{ \lambda \mu \nu })^{a b} \epsilon_b
 \right] ,
\end{align}
where we used the fact that for Majorana-Weyl bispinor matrix in symmetrized trace,
\begin{equation}
STr\exp
          \left(
           i k \cdot A
          \right) [ \bar{\epsilon} \gamma^{\mu_1 \mu_2 \dots \mu_n} \epsilon ] = 0
\label{g-bi-spinor}
\end{equation}
in $10$-dimensional theory unless $ n = 3$ or $7$.
Ignoring the normalization ambiguity, the disk amplitude for B-field for NN part is derived as
\begin{equation}
  \langle
   B^{\mathrm{NN}}
  \rangle
 \sim  k_\lambda \zeta^{ \mathrm{NN} }_{\mu \nu} STr 
 \exp
          \left(
           i k \cdot A
          \right)
   \left[ \epsilon_a (\gamma^{ \lambda \mu \nu })^{a b} \epsilon_b \right] .
\end{equation}

For RR part,
\begin{align}
   & \langle B^{ \mathrm{RR} } \rangle \n
 = & \langle
       c(z) \bar{c}(\bar{z})
       V^{\mathrm{RR}}_{(- \frac{1}{2} , - \frac{3}{2} )} (z, \bar{z})
     Tr \mathbf{P}
     \exp
          \left(
           i \int dt W^{(0)} (t)
          \right)
       c(x_1) F_{ -\frac{1}{2} } (x_1)
       \int d x_2 F_{ \frac{1}{2} } (x_2)
     \rangle \n
 = & \langle
       c(z) \bar{c}(\bar{z})
        \zeta^{ \mathrm{RR} }_{\mu \nu}
        e^{- \frac{1}{2} \phi}(z)
        S^a (z) (\gamma^{\mu \nu})_{a b} S^b(\bar{z})
        e^{- \frac{3}{2} \bar{\phi} } (\bar{z})
        e^{i k X}(z,\bar{z})
     Tr \mathbf{P}
     \exp
          \left(
           i \int dt \Phi^\rho i g_{\rho \sigma} \partial_\perp X^\sigma (t)
          \right)\n
   & \times
       c(x_1) \epsilon_c S^c (x_1) e^{- \frac{1}{2} \phi} (x_1)
       \int d x_2 \epsilon_f (\gamma^\rho)^f_{ \ d}
                  S^d (x_2) \partial_\perp X_\rho (x_2)
     \rangle \n
 = & \zeta^{\mathrm{RR}}_{\mu \nu}
      Tr \mathbf{P}
        \exp
          \left(
           i \int d \tau k \cdot A
          \right)
     \epsilon_c \epsilon_f (\gamma^{\mu \nu})_{a b} \n
   & \times
     \int d x_2
     \langle c(z) c (\bar{z}) c(x_1) \rangle
     \langle e^{-\frac{1}{2} \phi}(z)
             e^{-\frac{3}{2} \bar{\phi}}(\bar{z})
             e^{-\frac{1}{2} \phi} (x_1)  e^{-\frac{1}{2} \phi} (x_2)
     \rangle \n
   & \times \langle S^a (z) S^b (\bar{z}) S^c (x_1) S^d (x_2) \rangle
     \left[ - i (\gamma^\rho)^f_{ \ d} k^{\sigma} g_{\sigma \rho}
     \partial_\perp \ln \left| \frac{z-x_2}{\bar{z}-x_2} \right| \right] \n
 = & - \zeta^{\mathrm{RR}}_{\mu \nu}
      Tr \mathbf{P}
        \exp
          \left(
           i \int d \tau k \cdot A
          \right)
     \epsilon_c \epsilon_f (\gamma^{\mu \nu})_{a b} \int d x_2
     \left[ ( z-\bar{z} )( z-x_1 )( \bar{z}-x_1 ) \right] \n
   & \times
     \left[ (z-\bar{z})^{ -\frac{3}{4} }
            (z-x_1)^{-\frac{1}{4}}
            (z-x_2)^{\frac{1}{4}}
            (\bar{z} - x_1)^{-\frac{3}{4}}
            (\bar{z} - x_2)^{ \frac{3}{4} }
            (x_1 - x_2)^{\frac{1}{4}} \right] \n
   & \times
     \left[ \frac{   (z-x_2) (\bar{z}-x_1) (\gamma_\tau)^{a b} (\gamma^\tau)^{c d}
                   - (z-\bar{z})(x_1-x_2)  (\gamma_\tau)^{a d} (\gamma^\tau)^{b c} }
            { \left[ (z-\bar{z})(z-x_1)(z-x_2)
                     (\bar{z}-x_1)(\bar{z}-x_2)(x_1-x_2)
              \right]^{ \frac{3}{4} } }
     \right] \n
   & \times \left[ (\gamma^\rho)^f_{ \ d} k_\rho \frac{ z - \bar{z} }{(z-x_2)(\bar{z}-x_2)} \right] \n
 = & - \zeta^{\mathrm{RR}}_{\mu \nu}
      Tr \mathbf{P}
        \exp
          \left(
           i \int d \tau k \cdot A
          \right)
     \epsilon_c \epsilon_f \n
   & \times \int d x_2
     \left[ (z-\bar{z})^{ \frac{1}{4} }
            (z-x_1)^{\frac{3}{4}}
            (z-x_2)^{\frac{1}{4}}
            (\bar{z} - x_1)^{ \frac{1}{4}}
            (\bar{z} - x_2)^{ \frac{3}{4} }
            (x_1 - x_2)^{\frac{1}{4}} \right] \n
   & \times
     \left[ \frac{   (z-x_2) (\bar{z}-x_1)
                        tr \left( \gamma^{\mu \nu} \gamma_\tau \right)
                        (\gamma^\tau \gamma^\rho)^{c f}
                   - (z-\bar{z})(x_1-x_2)
                        (\gamma_\tau \gamma^{\mu \nu} \gamma^\tau)^{c d}
                        (\gamma^\rho)^f_{ \ d} }
            { \left[ (z-\bar{z})(z-x_1)(z-x_2)
                     (\bar{z}-x_1)(\bar{z}-x_2)(x_1-x_2)
              \right]^{ \frac{3}{4} } }
     \right] \n
   & \times \left[  k_\rho \frac{ z - \bar{z} }{(z-x_2)(\bar{z}-x_2)} \right] ,
\end{align}
where we used (\ref{4spinfield}).
Then
 \begin{align}
  & \epsilon_c \epsilon_f
  \left[
  (z-x_2) (\bar{z}-x_1)
    tr \left( \gamma^{\mu \nu} \gamma_\tau \right)
    (\gamma^\tau \gamma^\rho)^{c f}
  - (z-\bar{z})(x_1-x_2)
    (\gamma_\tau \gamma_{\mu \nu} \gamma^\tau)^{c d}
    (\gamma^\rho)^f_{ \ d}
  \right] \n
 = & (z-x_2) (\bar{z}-x_1)
       tr \left( \gamma^{\mu \nu} \gamma_\tau \right)
       [ \epsilon_c ((\gamma^{\tau \rho})^{c f} + \eta^{\tau \rho} \delta^{c f}) \epsilon_f ] \n
   & - (z-\bar{z})(x_1-x_2)
       [ \epsilon_c (6 \gamma^{\mu \nu} \gamma^\rho)^{c f} \epsilon_f ] \n
 = & - 6 (z-\bar{z})(x_1-x_2)
       [ \epsilon_c (\gamma^{\mu \nu \rho}
         - g^{\mu \rho} \gamma^\nu
         + g^{\rho \nu} \gamma^\mu )^{c f} \epsilon_f ] \n
 = & - 6 (z-\bar{z})(x_1-x_2)
     [ \epsilon_c (\gamma^{\mu \nu \rho})^{c f} \epsilon_f] ,
 \end{align}
where $\gamma_\tau \gamma^{\mu \nu} \gamma^{\tau} = 6 \gamma^{\mu \nu}$ 
and (\ref{g-bi-spinor}) hold.
Therefore
\begin{equation}
   \langle B^{\mathrm{RR}} \rangle
 =  6 \zeta^{\mathrm{RR}}_{\mu \nu} k_\rho
      STr \mathbf{P}
        \exp
          \left(
           i \int d \tau k \cdot A
          \right)
     [ \epsilon_c (\gamma^{\mu \nu \rho})^{c f} \epsilon_f]
     \int d x_2 \left[ \frac{(z-\bar{z})^{\frac{3}{2}}
                             (x_1-x_2)^{\frac{1}{2}}}
                            {(z-x_2)^{\frac{3}{2}}
                             (\bar{z}-x_1)^{\frac{1}{2}}
                             (\bar{z}-x_2)
                            } \right] .
\end{equation}
This result is calculated up to constant.
Therefore taking the limit $x_1 \rightarrow z$,
\begin{align}
   \langle B^{ \mathrm{RR} } \rangle
 = &   6  \zeta^{\mathrm{RR}}_{\mu \nu} k_\rho
      STr \mathbf{P}
        \exp
          \left(
           i \int d \tau k \cdot A
          \right)
     [ \epsilon_c (\gamma^{\mu \nu \rho})^{c f} \epsilon_f]
     \int d x_2 \left[ \frac{ -i(z-\bar{z}) }{(z-x_2)(\bar{z}-x_2)} \right] \n
 = & - 6 i \zeta^{\mathrm{RR}}_{\mu \nu} k_\rho STr
 \exp
          \left(
           i k \cdot A
          \right)
     [ \epsilon_c (\gamma^{\mu \nu \rho})^{c f} \epsilon_f] (2 \pi i) \n
 = &  12 \pi \zeta^{\mathrm{RR}}_{\mu \nu} k_\rho
      STr
        \exp
          \left(
           i k \cdot A
          \right)
     [ \epsilon_c (\gamma^{\mu \nu \rho})^{c f} \epsilon_f] .
\end{align}
Ignoring the normalization ambiguity, the disk amplitude of B-field for RR part is derived as
\begin{equation}
    i \langle B^{ \mathrm{RR} } \rangle
 \sim  i \zeta^{\mathrm{RR}}_{\mu \nu} k_\lambda
       STr
        \exp
          \left(
           i k \cdot A
          \right)
     [ \epsilon_c (\gamma^{\lambda \mu \nu})^{c f} \epsilon_f] .
\end{equation}
Therefore the disk amplitudes for B-field are derived as
\begin{equation}
  \langle B^{\mathrm{NN}} \rangle + i \langle B^{\mathrm{RR}} \rangle
 \sim (\zeta^{\mathrm{NN}}_{\mu \nu} + i \zeta^{\mathrm{RR}}_{\mu \nu} )
   STr
     \exp
        \left(
          i k \cdot A
        \right)
\left[ \epsilon_a k_\lambda (\gamma^{\lambda \mu \nu})^{a b} \epsilon_b \right] .
\end{equation}
This result coincides with the following matrix model vertex operator:
\begin{equation}
  B^{\mu \nu}(\lambda) V^B (A, \epsilon)
= B^{\mu \nu}(\lambda) \left( STr \left[ e^{i k \cdot A} \frac{1}{16}
  \left(\bar{\epsilon} \cdot \Slash{k} \gamma_{\mu \nu} \epsilon \right) \right] \right) .
\end{equation}

\subsection{B-field coupling to one bosonic string}
Instead of two fermionic open strings,
B-field can couples to one bosonic open string.
Now the disk amplitude in which NS-NS B-field couples to one bosonic
string is :
 \begin{align}
   &  Tr \mathbf{P}
      \sum_{n=0}^{\infty}  2 \pi i
       \int_{-\infty}^{+\infty} dt_2 dt_3  \dots
       dt_{n-1} dt_n \n
   & \times
     \langle
        c (z) \bar{c} (\bar{z}) V^{\mathrm{NN}}_{(-1,-1)}(z,\bar{z})
        c (t_1) \frac{1}{n!} \prod_{a=1}^n \left( i W^{(0)} (t_a) \right)
        \left(i \int d x D^{(0)}(x) \right)
      \rangle \n
 \sim
   &  Tr \mathbf{P}
      \sum_{n=0}^{\infty}  2 \pi i
       \frac{1}{n!} \int_{-\infty}^{+\infty} dt_2 dt_3  \dots
       dt_{n-1} dt_n \n
   & \times
     \langle
      c (z) \bar{c} (\bar{z})
     \zeta_{\mu \nu}^{\mathrm{NN}}
      e^{-\phi(z)} \psi^\mu (z)  e^{-\bar{\phi}(\bar{z})} \bar{\psi}^\nu (\bar{z})
     e^{i k X}(z,\bar{z}) \n
   & \times
     c (t_1)
     \prod_{a=1}^n [ i \Phi^\rho g_{\rho \sigma} i
       \partial_\perp (-i g^{\tau \sigma} k_{\tau} \ln \left| \frac{ z - t_a}{ \bar{z} - t_a } \right| )]
     \left( \int dx
            (g_{\alpha \gamma} g_{\beta \delta}
            [\Phi^{\gamma},\Phi^{\delta}]
            \Psi^\alpha \Psi^\beta )(x)
            ] \right)
    \rangle \n
 \sim
   &  Tr \mathbf{P}
      \sum_{n=0}^{\infty}  2 \pi i
       \frac{1}{n!} \int_{-\infty}^{+\infty} dt_2  dt_3  \dots
       dt_{n-1} dt_n \n
   & \times
     \langle
      c (z) \bar{c} (\bar{z}) c (t_1)
     \rangle
     \langle
      e^{-\phi(z)} e^{-\bar{\phi}(\bar{z})}
     \rangle
     \langle
     \zeta_{\mu \nu}^{\mathrm{NN}} \psi^\mu (z) \bar{\psi}^\nu (\bar{z})
     e^{i k X}(z,\bar{z}) \n
 & \times
     \prod_{a=1}^n [ (\Phi \cdot k)
       i \partial_\perp \ln \left| \frac{z - t_a}{\bar{z}-t_a}  \right|]
     \left( \frac{1}{(2 \pi)^2} \int dx
            [ F_{\alpha \beta}
            \Psi^\alpha \Psi^\beta (x)
            ] \right)
    \rangle .
 \end{align}
Contributions from the ghosts are given as:
 \begin{align}
 &
 \langle
 c(z) \bar{c} (\bar{z}) c (t_1)
 \rangle
 =(z-\bar{z}) (z-t_1) (\bar{z}-t_1) , \n
 &
 \langle
 e^{-\phi(z)} e^{-\bar{\phi} (\bar{z}) }
 \rangle
 =\frac{1}{z-\bar{z}} .
\end{align}
Contributions from the fermions are given as :
\begin{align}
   & \langle
      \psi^\mu (z) \bar{\psi}^\nu (\bar{z}) \Psi^\alpha \Psi^\beta(x)
     \rangle \n
 = & \langle
     j^{\alpha \beta}(x) \psi^\mu (z) \bar{\psi}^\nu (\bar{z})
     \rangle \n
 = & \sum_i \frac{ M^{\alpha \beta}(i) }{ x-z_i }
     \langle
     \psi^\mu (z) \bar{\psi}^\nu (\bar{z})
     \rangle \n
 = & \sum_i \frac{ M^{\alpha \beta}(i) }{ x-z_i } \frac{g^{\mu \nu}}{z-\bar{z}} \n
 = &   \frac{1}{(x-z)(z-\bar{z})}
       (g^{\mu \beta} g^{\alpha \nu} - g^{\alpha \mu} g^{\beta \nu})
     + \frac{1}{(x-\bar{z})(z-\bar{z})}
       (g^{\nu \beta} g^{\mu \alpha} - g^{\alpha \nu} g^{\mu \beta}) .
 \end{align}
Therefore the result is
 \begin{align}
   & \frac{1}{(2 \pi)^2} \zeta_{\mu \nu}^{\mathrm{NN}} Tr \mathbf{P}
      \sum_{n=0}^{\infty}  2 \pi i \frac{1}{n!}
       \int_{-\infty}^{+\infty} dt_2  dt_3  \dots
       dt_{n-1} dt_n \n
   & \times [(z-\bar{z}) (z-t_1) (\bar{z}-t_1)] \frac{1}{z-\bar{z}} \n
 & \times
      (\Phi \cdot k)
       \frac{z-\bar{z}}{(z-t_1)(\bar{z}-t_1)}
     \prod_{a=2}^n [ (\Phi \cdot k)
       2 \pi i \frac{\partial \tau (t_a,z)}{\partial t_a} ] \n
 & \times F_{\alpha \beta} \left(  \int dx
            [
            \frac{1}{(x-z)(z-\bar{z})}
            (g^{\mu \beta} g^{\alpha \nu} - g^{\alpha \mu} g^{\beta \nu})
            + \frac{1}{(x-\bar{z})(z-\bar{z})}
            (g^{\nu \beta} g^{\mu \alpha} - g^{\alpha \nu} g^{\mu \beta})
            ] \right) \n
 = & \frac{1}{(2 \pi)^2} Tr \mathbf{P} \sum_{n=0}^{\infty} \frac{1}{n!}
      (i A \cdot k)
     \prod_{a=2}^n [ \int_0^1 d \tau_a(i A \cdot k) ] \n
 & \times \left(  \int dx
            [
            \frac{1}{x-z}
            (\zeta_{\mu \nu}^{\mathrm{NN}} F^{\nu \mu} - \zeta_{\mu \nu}^{\mathrm{NN}} F^{\mu \nu})
            + \frac{1}{x-\bar{z}}
            (\zeta_{\mu \nu}^{\mathrm{NN}} F^{\mu \nu} - \zeta_{\mu \nu}^{\mathrm{NN}} F^{\mu \nu})
            ] \right) \n
 = &   \frac{1}{(2 \pi)^2} Tr \mathbf{P} \exp \left(\int_0^1 d \tau (i A \cdot k ) \right)
       \left(- \zeta_{\mu \nu}^{\mathrm{NN}} F^{\mu \nu}
       \int dx (\frac{1}{x-z} - \frac{1}{x-\bar{z}}) \right) \n
 = &   - \frac{1}{(2 \pi)^2} \zeta_{\mu \nu}^{\mathrm{NN}}
       STr \exp \left(i A \cdot k \right) F^{\mu \nu}
       \left( 2 \pi i \right) \n
 = & - \frac{i}{2 \pi} \zeta_{\mu \nu}^{\mathrm{NN}} STr[
     \exp \left(i k \cdot A \right) F^{\mu \nu} ] ,
 \end{align}
where we used the fact that $F^{\mu \nu}$ is the antisymmetric tensor.

Next we will consider RR part.
 \begin{align}
   &  \sum_{n=0}^{\infty}  2 \pi i
       \int_{-\infty}^{+\infty} dt_2 dt_3  \dots
       dt_{n-1} dt_n \n
   & \times
     \langle
        c (z) \bar{c} (\bar{z}) V^{\mathrm{RR}}_{(-\frac{1}{2},-\frac{3}{2})}(z,\bar{z})
        c (t_1) Tr \mathbf{P} \frac{1}{n!} \prod_{a=1}^n
        \left( i W^{(0)} (t_a) \right)
        \left(
           i \int d x D^{(0)}(x)
        \right)
      \rangle \n
 \sim & Tr \mathbf{P}
      \sum_{n=0}^{\infty}  2 \pi i \frac{1}{n!}
       \int_{-\infty}^{+\infty} dt_2 dt_3  \dots
       dt_{n-1} dt_n \n
   & \times
     \langle
        c (z) \bar{c} (\bar{z}) c (t_1)
     \rangle
     \langle
        \zeta_{\mu \nu}^{\mathrm{RR}}
        e^{-\frac{1}{2} \phi(z)}
        S^a(z) \gamma^{\mu \nu}_{a b} S^b(\bar{z})
        e^{-\frac{3}{2} \bar{\phi}(\bar{z})}
        e^{i k X} \n
   & \times
     \prod_{a=1}^n [ (\Phi \cdot k)
       i \partial_\perp \ln \left| \frac{z - t_a}{\bar{z}-t_a}  \right|]
     \left( \frac{1}{(2 \pi)^2} \int dx
            [ F_{\alpha \beta}
            \Psi^\alpha \Psi^\beta (x)
            ] \right)
    \rangle .
\end{align}
Contributions from the ghosts are given as:
 \begin{align}
 &
 \langle
 c(z) \bar{c} (\bar{z}) c (t_1)
 \rangle
 =(z-\bar{z}) (z-t_1) (\bar{z}-t_1) , \n
 &
 \langle
 e^{- \frac{1}{2} \phi(z)} e^{- \frac{3}{2} \bar{\phi} (\bar{z}) }
 \rangle
 =\frac{1}{ (z-\bar{z})^{\frac{3}{4}} } .
\end{align}
Contributions from the spin fields and fermions are given as:
\begin{align}
   & \langle
      S^a (z) S^b (\bar{z}) \Psi^\alpha \Psi^\beta(x)
     \rangle \n
 = & \langle
     j^{\alpha \beta}(x) S^a (z) S^b (\bar{z})
     \rangle \n
 = & \sum_i \frac{ M^{\alpha \beta}(i) }{ x-z_i }
     \langle
     S^a (z) S^b (\bar{z})
     \rangle \n
 = & \sum_i \frac{ M^{\alpha \beta}(i) }{ x-z_i }
            \frac{- \delta^{a b} } { (z-\bar{z})^{\frac{5}{4}} } \n
 = &   \frac{ \frac{1}{2} }{ (x-z)(z-\bar{z})^{ \frac{5}{4} } }
       (\gamma^{\alpha \beta})^{a b}
     - \frac{ \frac{1}{2} }{ (x-\bar{z})(z-\bar{z})^{ \frac{5}{4} } }
       (\gamma^{\alpha \beta})^{a b} \n
 = & \frac{1}{2} (\gamma^{\alpha \beta})^{a b}
     \left( \frac{1}{ (x-z)(z-\bar{z})^{ \frac{5}{4} } }
          - \frac{1}{(x-\bar{z})(z-\bar{z})^{ \frac{5}{4} } }
     \right)
 .
 \end{align}
Therefore the amplitude becomes
\begin{align}
 & \frac{1}{(2 \pi)^2}
   \zeta_{\mu \nu}^{\mathrm{RR}}
   (\gamma^{\mu \nu})_{a b}
      Tr \mathbf{P}
      \sum_{n=0}^{\infty} \frac{1}{n!}
       \int_{-\infty}^{+\infty} dt_2 dt_3  \dots
       dt_{n-1} dt_n \n
   & \times
      [(z-\bar{z}) (z-t_1) (\bar{z}-t_1)] \frac{1}{ (z-\bar{z})^{\frac{3}{4}} } \n
   & \times
      (i k \cdot A) \frac{z-\bar{z}}{(z-t_1)(\bar{z}-t_1)}
     \prod_{a=2}^n  (\Phi \cdot k)
       2 \pi i \frac{\partial \tau(t_a,z)}{\partial t_a} \n
   & \times F_{\alpha \beta} \left(  \int dx
               \frac{1}{2} (\gamma^{\alpha \beta})^{a b}
                \left( \frac{1}{ (x-z)(z-\bar{z})^{ \frac{5}{4} } }
                     - \frac{1}{(x-\bar{z})(z-\bar{z})^{ \frac{5}{4} } }
                \right)
             \right) \n
 = & \frac{1}{2} \frac{1}{(2 \pi)^2}
   \zeta_{\mu \nu}^{\mathrm{RR}}
   \mathrm{tr}(\gamma^{\mu \nu} \gamma^{\alpha \beta} )
      Tr
      \sum_{n=0}^{\infty} \frac{1}{n!}
      \mathbf{P} (i k \cdot A)
      \prod_{a=2}^n \int_0^1 d \tau_a \left( i k \cdot A \right) \n
   & \times F_{\alpha \beta} \left(  \int dx
              \frac{z-\bar{z}}{(x-z)(x-\bar{z})}
             \right) .
\end{align}

Here we can use the equation (\ref{g2g2}).
Substituting this into the above equation,
\begin{align}
   & \frac{1}{2} \frac{1}{(2 \pi)^2}
     \zeta_{\mu \nu}^{\mathrm{RR}}
   STr [ \exp \left( i k \cdot A \right) \n
   & \times F_{\alpha \beta} (2 \pi i)
            \left(
                  - \frac{1}{4} 2^{\frac{D}{2}}
                  [   g^{\mu \alpha} g^{\nu \beta}
                    - g^{\mu \beta} g^{\nu \alpha}
                    - g^{\nu \alpha} g^{\mu \beta}
                    + g^{\nu \beta} g^{\mu \alpha} ]
             \right) ] \n
 = & \frac{ 2^{ \frac{D}{2} } }{8} \frac{-i}{2 \pi}
    STr \exp \left( i k \cdot A \right)
        \left[  \zeta_{\mu \nu}^{\mathrm{RR}} F^{\mu \nu}
           - \zeta_{\mu \nu}^{\mathrm{RR}} F^{\nu \mu}
           - \zeta_{\mu \nu}^{\mathrm{RR}} F^{\nu \mu}
           + \zeta_{\mu \nu}^{\mathrm{RR}} F^{\mu \nu}
        \right] \n
 = & - \frac{2^{ \frac{D}{2} } i }{4 \pi} \zeta_{\mu \nu}^{\mathrm{RR}}
   STr \exp [\left( i k \cdot A \right)]
   F^{\mu \nu} .
\end{align}
The result for type IIB superstring multiplet is
\begin{equation}
       \langle
         B^{\mathrm{NN}}
       \rangle
   + i \langle
         B^{\mathrm{RR}}
       \rangle
 = (\zeta_{\mu \nu}^{\mathrm{NN}} + i \zeta_{\mu \nu}^{\mathrm{RR}} )
    STr \exp [\left( i k \cdot A \right) F^{\mu \nu} ].
\end{equation}
The corresponding part of the vertex operator in the matrix model is
\begin{equation}
   B^{\mu \nu}(\lambda) V_{\mu \nu}^{B}(A,\epsilon)
 = B^{\mu \nu}(\lambda)
   STr e^{i k \cdot A}
   \left(
     -\frac{i}{2}[A_\mu, A_\nu]
   \right) .
\end{equation}
They coincide up to normalization coefficients.
\section{Gravitino}
Now we calculate the disk amplitudes of the gravitino, which is written like:
\begin{equation}
   \langle \mathbf{\Psi} \rangle
 =   \langle \mathbf{\Psi}^{\mathrm{RN}} \rangle
  +i \langle \mathbf{\Psi}^{\mathrm{NR}} \rangle .
\end{equation}
\subsection{Gravitino coupling to three fermionic strings}
Since the gravitino is the $3$ times SUSY transformed field in type IIB
massless superstring multiplet, the closed string vertex operator for the gravitino
can couples to $3$ fermionic open strings.
Firstly, RN part becomes as
 \begin{align}
   & \langle \mathbf{\Psi}^{\mathrm{RN}} \rangle \n
 = & \langle
      c(z) \bar{c} (\bar{z})
      V^{\mathrm{RN}}_{(-\frac{1}{2},0)} (z,\bar{z})
     Tr \mathbf{P}
     \exp
          \left(
           i \int dt W^{(0)} (t)
          \right) \n
   & \times
      c(x_1) F_{-\frac{1}{2}}(x_1)
      \int d x_2 F_{-\frac{1}{2}}(x_2)
      \int d x_3 F_{-\frac{1}{2}}(x_3)
     \rangle \n
 = & \langle c(z) \bar{c} (\bar{z})
      \zeta^{\mathrm{RN}}_{a \mu}
       e^{-\frac{1}{2} \phi}(z) S^a(z)
       (\bar{\partial} X^\mu + i k_\rho j^{\rho \mu} )(\bar{z})
      e^{i k X}(z,\bar{z})
     Tr \mathbf{P}
     \exp
          \left(
           i \int dt \Phi^\rho i g_{\rho \sigma} \partial_\perp X^\sigma (t)
          \right) \n
   & \times
      c(x_1) \epsilon_b S^b(x_1) e^{-\frac{1}{2} \phi}(x_1)
      \int d x_2 \epsilon_c S^c(x_2) e^{-\frac{1}{2} \phi}(x_2)
      \int d x_3 \epsilon_d S^d(x_3) e^{-\frac{1}{2} \phi}(x_3)
     \rangle \n
 = & i \zeta^{\mathrm{RN}}_{a \mu} k_\rho
     Tr \mathbf{P}
     \exp
          \left(
           i \int d \tau k \cdot A
          \right)
     \epsilon_b \epsilon_c \epsilon_d
     \int d x_2 \int d x_3
     \langle c(z) \bar{c}(\bar{z}) c(x_1) \rangle
     \langle e^{- \frac{1}{2} \phi} (z)
             e^{- \frac{1}{2} \phi} (x_1)
             e^{- \frac{1}{2} \phi} (x_2)
             e^{- \frac{1}{2} \phi} (x_3)
     \rangle \n
   & \times \langle j^{\rho \mu} (\bar{z})
                    S^a (z) S^b (x_1) S^c (x_2) S^d (x_3)
            \rangle \n
 = & i \zeta^{\mathrm{RN}}_{a \mu} k_\rho
     Tr \mathbf{P}
     \exp
          \left(
           i \int d \tau k \cdot A
          \right)
     \epsilon_b \epsilon_c \epsilon_d
     \int d x_2 \int d x_3
     \left[ (z-\bar{z})(z-x_1)(\bar{z}-x_1)  \right] \n
   & \times
     \left[
            (z - x_1)^{- \frac{1}{4}}
            (z - x_2)^{- \frac{1}{4}}
            (z - x_3)^{- \frac{1}{4}}
            (x_1 - x_2)^{- \frac{1}{4}}
            (x_1 - x_3)^{- \frac{1}{4}}
            (x_2 - x_3)^{- \frac{1}{4}}
     \right] \n
   & \times \sum_i \frac{M^{\rho \mu}(i)}{\bar{z}-z_i}
                    \frac{  (z-x_3)(x_1-x_2)(\gamma^\tau)^{a b}(\gamma_\tau)^{cd}
                          - (z-x_1)(x_2-x_3)(\gamma^\tau)^{a d}(\gamma_\tau)^{bc}
                         }
                  { \left[ ( z-x_1 )
                           ( z-x_2 )
                           ( z-x_3 )
                           ( x_1-x_2 )
                           ( x_1-x_3 )
                           ( x_2-x_3 )
                    \right]^{ \frac{3}{4} } } \n
 = & i \zeta^{\mathrm{RN}}_{a \mu} k_\rho
     Tr \mathbf{P}
     \exp
          \left(
           i \int d \tau k \cdot A
          \right) \n
   & \times
     \epsilon_b \epsilon_c \epsilon_d
     \int d x_2 \int d x_3
     [ (z-\bar{z}) (z-x_1)^{\frac{3}{4}}
       (z-x_2)^{-\frac{1}{4}}(z-x_3)^{-\frac{1}{4}} (\bar{z}-x_1) \n
   & \times (x_1-x_2)^{-\frac{1}{4}} (x_1-x_3)^{-\frac{1}{4}} (x_2-x_3)^{-\frac{1}{4}} ] \n
   & \times \sum_i \frac{M^{\rho \mu}(i)}{\bar{z}-z_i}
                    \frac{  (z-x_3)(x_1-x_2)(\gamma_\tau)^{a b}(\gamma^\tau)^{cd}
                          - (z-x_1)(x_2-x_3)(\gamma_\tau)^{a d}(\gamma^\tau)^{bc}
                         }
                  { \left[ ( z-x_1 )
                           ( z-x_2 )
                           ( z-x_3 )
                           ( x_1-x_2 )
                           ( x_1-x_3 )
                           ( x_2-x_3 )
                    \right]^{ \frac{3}{4} } } .
 \end{align}
Because (\ref{g-bi-spinor}) holds, $\epsilon_c (\gamma^{\mu_1 \mu_2
\cdots \mu_m})^{ab} \epsilon_d$ and
$\epsilon_b (\gamma^{\nu_1 \nu_2 \cdots \nu_n})^{bc} \epsilon_c$ vanish unless $m$ and
$n$ are $3$ or $7$.
This is why the factors on which $M^{\mu \rho}$ acts are constrained.
Therefore in the above equation, the only contributing factors are as follows:
\begin{align}
& i \zeta^{\mathrm{RN}}_{a \mu} k_\rho
     Tr \mathbf{P}
     \exp
          \left(
           i \int d \tau k \cdot A
          \right) \n
&    \times
     \epsilon_b \epsilon_c \epsilon_d
     \int d x_2 \int d x_3 [
     \frac{1}{2} (\gamma_\tau)^{ab} (\gamma^{\rho \mu \tau})^{cd}
     \left( \frac{1}{\bar{z}-x_2} - \frac{1}{\bar{z}-x_3} \right)
     \left[ \frac{(z-\bar{z})(\bar{z}-x_1)}{(z-x_2)(x_1-x_3)(x_2-x_3)}\right] \n
& -  \frac{1}{2} (\gamma_\tau)^{ad} (\gamma^{\rho \mu \tau})^{bc}
     \left( \frac{1}{\bar{z}-x_1} - \frac{1}{\bar{z}-x_2} \right)
     \left[ \frac{(z-\bar{z})(z-x_1)(\bar{z}-x_1)}
          {(z-x_2)(z-x_3)(x_1-x_2)(x_1-x_3)}\right] ] \n
= &  \frac{i}{2}
     Tr \mathbf{P}
     \exp
          \left(
           i \int d \tau k \cdot A
          \right) \n
&    \times
     \left[ \zeta^{\mathrm{RN}}_{a \mu} (\gamma_\tau)^{ab} \epsilon_b \right]
     \left[ \epsilon_c (k_\rho \gamma^{\rho \mu \tau})^{cd} \epsilon_d \right]
     \int d x_2 \int d x_3
     \left[ \frac{z-\bar{z}}{(z-x_2)(\bar{z}-x_2)}
            \frac{\bar{z}-x_1}{(x_1-x_3)(\bar{z}-x_3)}
     \right] \n
 & -\frac{i}{2} Tr \mathbf{P}
     \exp
          \left(
           i \int d \tau k \cdot A
          \right) \n
 &   \times
     \left[ \zeta^{\mathrm{RN}}_{a \mu} (\gamma_\tau)^{ab} \epsilon_b \right]
     \left[ \epsilon_c (k_\rho \gamma^{\rho \mu \tau})^{cd} \epsilon_d \right]
     \int d x_2 \int d x_3
     \left[ \frac{z-\bar{z}}{(z-x_2)(\bar{z}-x_2)}
            \frac{z-x_1}{(z-x_3)(x_1-x_3)} \right] .
\label{gravitinoRN1}
\end{align}
Here we have two possibilities.
We can take the limit $x_1 \rightarrow z$ or $x_1 \rightarrow \bar{z}$.
Taking the former limit, only the first term in (\ref{gravitinoRN1})
contributes and it becomes:
\begin{align}
& - \frac{i}{2}
     STr  \exp
          \left(
           i k \cdot A
          \right)
     \left[ \zeta^{\mathrm{RN}}_{a \mu} (\gamma_\tau)^{ab} \epsilon_b \right]
     \left[ \epsilon_c (k_\rho \gamma^{\rho \mu \tau})^{cd} \epsilon_d \right]
     (2 \pi i)(2 \pi i) \n
= &  2 \pi^2 i
     STr  \exp
          \left(
           i k \cdot A
          \right)
     \left[ \zeta^{\mathrm{RN}}_{a \mu} (\gamma_\tau)^{ab} \epsilon_b \right]
     \left[ \epsilon_c (k_\rho \gamma^{\rho \mu \tau})^{cd} \epsilon_d \right] .
\label{gr:first}
\end{align}
For the second case, taking the limit $x_1 \rightarrow \bar{z}$, 
only the second term in (\ref{gravitinoRN1}) contributes and 
it takes exactly the same form as (\ref{gr:first}).
Therefore in the both cases, this result is essentially regarded as
\begin{equation}
     STr  \exp
          \left(
           i k \cdot A
          \right)
   \left[ \zeta^{\mathrm{RN}}_{a \mu} (\gamma_\tau)^{ab} \epsilon_b \right]
   \left[ \epsilon_c (k_\rho \gamma^{\rho \mu \tau})^{cd} \epsilon_d \right] .
\label{gravitinoRNs}
\end{equation}

The NR contribution can be evaluated in an analogous way 
and the identical expression is obtained.
In this way the disk amplitude for the gravitino becomes
\begin{equation}
   \langle \mathbf{\Psi} \rangle
 = (\zeta^{\mathrm{RN}}_{a \mu} + i \zeta^{NR}_{\mu a})
   STr
     \exp
          \left(
           i k \cdot A
          \right)
   \left[(\gamma_\tau)^{ab} \epsilon_b \right]
   \left[ \epsilon_c (k_\rho \gamma^{\rho \mu \tau})^{cd} \epsilon_d \right] .
\end{equation}
It agrees with the following vertex operator in type IIB matrix model:
\begin{equation}
  \mathbf{\Psi}_{\mu} (\lambda) V^{\mathbf{\Psi}}(A,\epsilon)
 = \mathbf{\Psi}_{\mu} (\lambda) STr \left[ e^{i k \cdot A}
   \left( \bar{\epsilon} \cdot \Slash{k} \gamma^{\mu \nu} \epsilon \right)
   \cdot \bar{\epsilon} \gamma_{\nu} \right] .
\end{equation}

\subsection{Gravitino coupling to one fermionic string and one bosonic string}
We also consider the case in which the vertex operator of the gravitino couples to
the vertex operator of one bosonic and one fermionic string.
\begin{align}
  & \langle
      c (z) \bar{c} (\bar{z}) V^{\mathrm{RN}}_{(-\frac{1}{2}, -1)}(z,\bar{z})
     Tr \mathbf{P}
     \exp
          \left(
           i \int dx W^{(0)} (x)
          \right)
         \left(i \int dt D^{(0)} (t)
     \right)
      c (x_1) F_{-\frac{1}{2}}(x_1)
    \rangle \n
 = & \langle c (z) \bar{c}(\bar{z})
             \zeta^{\mathrm{RN}}_{a \mu}
                        e^{-\frac{1}{2}\phi}(z) S^{a}(z) 
                        e^{-\bar{\phi}}(\bar{z}) \bar{\psi}^{\mu}(\bar{z})
             e^{i k X}(z,\bar{z})
     Tr \mathbf{P}
     \exp
          \left(
           i \int dx \Phi^\rho i g_{\rho \sigma} \partial_\perp X^\sigma (x)
          \right) \n
   & \times \left( i  \int dt
                        (- i g_{\alpha \gamma} g_{\beta \delta}
                         [\Phi^\alpha, \Phi^\beta]
                          \Psi^{\gamma} \Psi^{\delta}(t) ) \right) \n
   & \times  c(x_1) \epsilon_b S^b (x_1) e^{-\frac{1}{2} \phi}(x_1)
     \rangle \n
 = & - \zeta^{\mathrm{RN}}_{a \mu} \frac{1}{(2 \pi)^2}
     Tr \mathbf{P}
     \exp
          \left(
           i \int d \tau k \cdot A
          \right) \n
   & \times \left[2 \pi \Phi_\gamma, 2 \pi \Phi_\delta \right] \epsilon_b
             \int dt \langle c(z) \bar{c}(\bar{z}) c(x_1) \rangle
             \langle  e^{-\frac{1}{2} \phi }(z) e^{-\bar{\phi}}(\bar{z}) e^{-\frac{1}{2} \phi}(x_1) \rangle
             \langle  j^{\gamma \delta}(t) \bar{\psi}^\mu(\bar{z}) S^a(z) S^b  (x_1) \rangle \n
 = & \frac{i}{(2 \pi)^2} \zeta^{\mathrm{RN}}_{a \mu}
     Tr \mathbf{P}
     \exp
          \left(
           i \int d \tau k \cdot A
          \right) \n
   & \times F_{\gamma \delta} \epsilon_b
       \int dt [(z-\bar{z})(z-x_1)(\bar{z}-x_1)]
         [(z-\bar{z})^{-\frac{1}{2}}
          (z-x_1)^{-\frac{1}{4}}
          (\bar{z}-x_1)^{-\frac{1}{2}}] \n
 & \times \sum_i \frac{M^{\gamma \delta}(i)}{t-z_i}
          \frac{(\gamma^\mu)^{ab}}
               {(z-\bar{z})^{\frac{1}{2}}
                (\bar{z}-x_1)^{\frac{1}{2}}
                (z-x_1)^{\frac{3}{4}}
               } \n
 = &  \frac{i}{(2 \pi)^2} \zeta^{\mathrm{RN}}_{a \mu}
     Tr \mathbf{P}
     \exp
          \left(
           i \int d \tau k \cdot A
          \right)
       F_{\gamma \delta} \epsilon_b \n
   &   \int dt [  \frac{1}{t-\bar{z}}
                  (g^{\mu \delta} \gamma^\gamma - g^{\gamma \mu} \gamma^{\delta} )^{a b}
                + \frac{ \frac{1}{2} }{t-z}
                  (\gamma^{\gamma \delta} \gamma^\mu)^{a b}
                - \frac{ \frac{1}{2} }{t-x_1} (\gamma^\mu \gamma^{\gamma \delta})^{a b}] \n
 = &  \frac{i}{(2 \pi)^2} \zeta^{\mathrm{RN}}_{a \mu}
     Tr \mathbf{P}
     \exp
          \left(
           i \int d \tau k \cdot A
          \right)
       F_{\gamma \delta} \epsilon_b
       \int dt [  \frac{1}{t-\bar{z}}
                  (g^{\mu \delta} \gamma^\gamma - g^{\gamma \mu} \gamma^{\delta} )^{a b} \n
   &            + \frac{ \frac{1}{2} }{t-z}
                  (\gamma^{\gamma \delta \mu}
                    - g^{\gamma \mu} \gamma^{\delta}
                    + g^{\delta \mu} \gamma^{\gamma})^{a b}
                - \frac{ \frac{1}{2} }{t-x_1}
                  (\gamma^{\mu \gamma \delta}
                    + g^{\gamma \mu} \gamma^{\delta}
                    - g^{\delta \mu} \gamma^{\gamma} )^{a b}] .
\end{align}
Taking the limit $x_1 \rightarrow z$, it becomes like :
\begin{align}
   & \frac{-i}{(2 \pi)^2} \zeta^{\mathrm{RN}}_{a \mu}
     Tr \mathbf{P}
     \exp
          \left(
           i \int d \tau k \cdot A
          \right)
       F_{\gamma \delta} \epsilon_b (g^{\mu \delta} \gamma^\gamma - g^{\gamma \mu} \gamma^{\delta} )^{a b}
       \int dt \frac{z-\bar{z}}{(t-z)(t-\bar{z})} \n
 = & \frac{1}{2 \pi} \zeta^{\mathrm{RN}}_{a \mu}
     Tr \mathbf{P}
     \exp
          \left(
           i \int d \tau k \cdot A
          \right)
       F_{\gamma \delta}
      (g^{\mu \delta} \gamma^\gamma - g^{\gamma \mu} \gamma^{\delta} )^{a b}
       \epsilon_b .
\label{Gravitinobf1}
\end{align}

The NR contribution can be evaluated in an analogous way.
In this way, we find the other term of the gravitino vertex operator as follows:
  \begin{align}
      &  (\zeta^{\mathrm{RN}}_{a \mu} + i \zeta^{\mathrm{NR}}_{\mu a})
     STr
     \exp
          \left(
           i \int d \tau k \cdot A
          \right)
      F_{\gamma \delta}
      (g^{\mu \delta} \gamma^\gamma - g^{\gamma \mu} \gamma^{\delta} )^{a b}
      \epsilon_b \n
    = & (\zeta^{\mathrm{RN}}_{\mu} + i \zeta^{\mathrm{NR}}_{\mu})
     STr
     \exp
          \left(
           i \int d \tau k \cdot A
          \right)
      \left[ (- 2 F^{\mu \nu}) \cdot \bar{\epsilon} \gamma_\nu \right] .
  \end{align}
The corresponding part of the vertex operator in the matrix model is
\begin{equation}
   \mathbf{\Psi}_{\mu}(\lambda) V^{\Psi}(A,\epsilon)
 = \mathbf{\Psi}_{\mu}(\lambda)
   \left[ STr e^{i k \cdot A} \left(- 2 F^{\mu \nu} \right)
          \cdot \bar{\epsilon} \gamma_\nu  \right] .
\end{equation}
Here again they coincide.
\subsection{Gravitino coupling to $C_\frac{1}{2}$}
Because the closed string vertex operator for gravitino
couples to $3$ fermionic open strings,
it can also couple to $C_{\frac{1}{2}}$ in (\ref{C(1/2)}).
In what follows we show that such a coupling vanishes for gravitino
case.

R-NS part is given as follows.
We can consider the following disk amplitude :
\begin{align}
   & \langle
      c(z) \bar{c}(\bar{z}) V_{(-\frac{3}{2},-1)}^{\mathrm{RN}}(z,\bar{z})
      Tr \mathbf{P}
      \exp \left( i \int dt W^{(0)} (t) \right)
      c(x) C_{\frac{1}{2}}(x)
     \rangle \n
 = & \langle c(z) \bar{c}(\bar{z}) c(x) \rangle
     Tr \mathbf{P} \exp \left( \int d \tau i k \cdot A \right) \n
   & \times
     \langle
        \zeta_{a \mu} e^{ - \frac{3}{2} \phi(z) } S^a (z)
        e^{ - \bar{\phi} (\bar{z}) } \bar{\psi}^\mu (\bar{z})
        e^{i k X}(z,\bar{z})
        [\epsilon^c,\Phi_\alpha] (\gamma_\beta)_{c b}
        \Psi^\alpha \Psi^\beta (x) S^b (x)
        e^{-\frac{1}{2} \phi(x) } \rangle \n
 = & - \zeta_{a \mu}
     Tr \mathbf{P} \exp \left( \int d \tau i k \cdot A \right)
     [\epsilon^c, \Phi_\alpha] (\gamma_\beta)_{c b} \n
   & \times
     \langle
       c(z) \bar{c}(\bar{z}) c(x)
     \rangle
     \langle
       e^{-\frac{3}{2}\phi(z)} e^{- \bar{\phi}(\bar{z})} e^{-\frac{1}{2}\phi(x)}
     \rangle
     \langle
      j^{\alpha \beta} (x) \bar{\psi}^\mu (\bar{z}) S^a (z) S^b(x)
     \rangle
\label{grC}
\end{align}
Contribution from the ghost terms are given as :
\begin{align}
 &   \langle c(z) \bar{c}(\bar{z}) c(x) \rangle
   = (z-\bar{z})(z-x)(\bar{z}-x) \n
 &   \langle
      e^{-\frac{3}{2}\phi(z)} e^{-\bar{\phi}(\bar{z})}  e^{\frac{1}{2} \phi(x)}
     \rangle
   = (z-\bar{z})^{-\frac{3}{2}} (z-x)^{\frac{3}{4}} (\bar{z}-x)^{\frac{1}{2}} .
\label{ghost,grC}
\end{align}
Contributions from the fermion fields and spin fields are given as:
\begin{align}
  & \langle
     j^{\alpha \beta} (x) \bar{\psi}^\mu (\bar{z}) S^a (z) S^b (x)
    \rangle \n
 = & \frac{1}{x-\bar{z}}
   \frac{(g^{\alpha \mu} \gamma^\beta - g^{\mu \beta} \gamma^\alpha)^{a b}}
        {
         (z-\bar{z})^{\frac{1}{2}}
         (\bar{z}-x)^{\frac{1}{2}}
         (z-x)^{\frac{3}{4}}
        } \n
   & - \frac{1}{x-z}
       \frac{[\gamma^\alpha, \gamma^\beta]^b_{\ d} (\gamma^\mu)^{a d}}
            {(z-\bar{z})^{\frac{1}{2}}
             (\bar{z}-x)^{\frac{1}{2}}
             (z-x)^{\frac{3}{4}}} .
\label{jjpp,grC}
\end{align}
Substituting (\ref{ghost,grC}) and (\ref{jjpp,grC}) into (\ref{grC}), we get :
\begin{equation}
 - \zeta_{a \mu}
     Tr \mathbf{P} \exp \left( \int d \tau i k \cdot A \right)
     [\epsilon^c, \Phi_\alpha] (\gamma_\beta)_{c b}
     \left(
       (g^{\alpha \mu} \gamma^{\beta} - g^{\mu \beta} \gamma^\alpha)^{a b}
       \frac{z-x}{z-\bar{z}}
       -
       [\gamma^\alpha,\gamma^\beta]^b_{\ d} (\gamma^\mu)^{a d}
       \frac{\bar{z}-x}{z-\bar{z}}
     \right) .
\end{equation}
In this equation, we take the value of $x$ arbitrarily.
Thus we take the limit $x \rightarrow z$, then we obtain:
\begin{align}
   &- \zeta_{a \mu} Tr \mathbf{P} \exp \left( \int d \tau i k \cdot A \right)
     [\epsilon^c, \Phi_\alpha] (\gamma_\beta)_{c b}
     (2 (\gamma^{\alpha} \gamma^\beta)^b_{\ d} - 2 g^{\alpha \beta} \delta^b_d ) (\gamma^\mu)^{a d} \n
 = &- \zeta_{a \mu} Tr \mathbf{P} \exp \left( \int d \tau i k \cdot A \right)
     [\epsilon^c, \Phi_\alpha] \n
   & \times
     \left( 2 (\gamma_\beta \gamma^{\alpha} \gamma^\beta)_{c d} (\gamma^\mu)^{a d}
            - 2 (\gamma^\alpha)_{c b}(\gamma^\mu)^{a b}
     \right)\n
 = & 18 \zeta_{a \mu} Tr \mathbf{P} \exp \left( \int d \tau i k \cdot A \right)
     [\epsilon^c, \Phi_\alpha]( \gamma^{\alpha})_{c d} (\gamma^\mu)^{a d}.
\label{grC2}
\end{align}
From IIB matrix model action (\ref{action}),
we find the following equation of motion
\begin{equation}
 (\gamma^\mu)^{a b} [A_\mu, \epsilon_b] = 0.
 \label{eom}
\end{equation}
Therefore (\ref{grC2}) vanishes due to the equation of motion. 
NS-R part also vanishes in the similar way.
Therefore, this coupling does not contribute to this vertex operator.

\section{Graviton}
Graviton is the $4$ times SUSY transformed field in
type IIB supergravity multiplet and the vertex operator satisfy the
the NS-NS boundary condition.
The matrix model vertex operator consists of 4 terms as shown in Appendix A.5.
We reproduce each of them in the following subsections.

\subsection{Graviton coupling to four fermionic open strings}
We calculate the disk amplitude where the vertex operator of the
graviton field couples to four fermionic open strings.
The disk amplitude is :
\begin{align}
   & \langle h^{ \mathrm{NN} } \rangle \n
 = & \langle c (z) \bar{c} (\bar{z})
      V^{ \mathrm{NN} }_{(0,0)} (z, \bar{z})
     Tr \mathbf{P}
     \exp
          \left(
           i \int dt W^{(0)} (t)
          \right) \n
   & \times
      c(x_1) F_{-\frac{1}{2}} (x_1)
      \int dx_2 F_{-\frac{1}{2}} (x_2)
      \int dx_3 F_{-\frac{1}{2}} (x_3)
      \int dx_4 F_{-\frac{1}{2}} (x_4)
     \rangle \n
 = & \langle
      c(z) \bar{c}(\bar{z})
      \zeta^{\mathrm{NN}}_{\mu \nu}
      (\partial X^\mu + i k_{\rho} j^{\mu \rho})(z)
      (\bar{\partial} X^\nu + i k_{\lambda} \bar{j}^{\nu \lambda})(\bar{z})
      e^{ikX}(z,\bar{z})
     Tr \mathbf{P}
     \exp
          \left(
           i \int dt \Phi^\rho i g_{\rho \sigma} \partial_\perp X (t)
          \right) \n
   &  \times
      c(x_1) \epsilon_a S^a (x_1) e^{-\frac{1}{2} \phi} (x_1)
      \int d x_2 \epsilon_b S^b (x_2) e^{-\frac{1}{2} \phi} (x_2)
      \int d x_3 \epsilon_b S^c (x_3) e^{-\frac{1}{2} \phi} (x_3)
      \int d x_4 \epsilon_b S^d (x_4) e^{-\frac{1}{2} \phi} (x_4)
   \rangle \n
 = & - \zeta^{\mathrm{NN}}_{\mu \nu}
     k_\rho k_\lambda
     Tr \mathbf{P}
     \exp
          \left(
           i \int d \tau k \cdot A
          \right) \n
   & \times
     \epsilon_a \epsilon_b \epsilon_c \epsilon_d
     \int d x_2 \int d x_3 \int d x_4
     \langle c(z)c(\bar{z})c(x_1) \rangle
     \langle e^{-\frac{1}{2} \phi}(x_1)
             e^{-\frac{1}{2} \phi}(x_2)
             e^{-\frac{1}{2} \phi}(x_3)
             e^{-\frac{1}{2} \phi}(x_4)
     \rangle \n
 & \times
   \langle
    j^{\rho \mu}(z) \bar{j}^{\lambda \nu}(\bar{z})
    S^a(x_1) S^b (x_2) S^c(x_3) S^d(x_4)
   \rangle .
\end{align}
Contributions from the ghosts are:
\begin{equation}
  \langle c(z)c(\bar{z})c(x_1) \rangle = (z-\bar{z})(z-x_1)(\bar{z}-x_1)
\end{equation}
and
\begin{align}
   &  \langle e^{-\frac{1}{2} \phi}(x_1)
            e^{-\frac{1}{2} \phi}(x_2)
            e^{-\frac{1}{2} \phi}(x_3)
            e^{-\frac{1}{2} \phi}(x_4)
      \rangle \n
 = &  (x_1-x_2)^{-\frac{1}{4}}
      (x_1-x_3)^{-\frac{1}{4}}
      (x_1-x_4)^{-\frac{1}{4}}
      (x_2-x_3)^{-\frac{1}{4}}
      (x_2-x_4)^{-\frac{1}{4}}
      (x_3-x_4)^{-\frac{1}{4}}.
\end{align}
Contributions from the spin fields are:
\begin{align}
   & \langle j^{\rho \mu}(z) \bar{j}^{\lambda \nu}(\bar{z})
           S^a(x_1) S^b(x_2) S^c(x_3) S^d(x_4)
     \rangle \n
 = & \langle
      g_{\tau \sigma}
      \sum_m \frac{ M^{\rho \mu}(m) }{z-x_m}
      \sum_n \frac{ M^{\lambda \nu}(n) }{\bar{z}-x_n}
      \frac{ (x_1-x_4)(x_2-x_3)(\gamma^\tau)^{a b}(\gamma^\sigma)^{c d}
            -(x_1-x_2)(x_3-x_4)(\gamma^\tau)^{a d}(\gamma^\sigma)^{b c}}
           { [(x_1-x_2)(x_1-x_3)(x_1-x_4)(x_2-x_3)(x_2-x_4)(x_3-x_4)]^{\frac{3}{4}} }
     \rangle .
\label{MMSSSS}
\end{align}
Because (\ref{g-bi-spinor}) holds, considering the Majorana-Weyl fermion
bispinors, the factors which $M^{\rho \mu}(m)$ or
$M^{\lambda \nu}(n)$ acts on are constrained.

 Here we consider the case when $M^{\rho \mu}(m)$ acts on
$(\gamma^\tau)^{ab}$ and $M^{\lambda \nu}(n)$ acts on $(\gamma^\sigma)^{cd}$.
\begin{align}
 & - \zeta^{\mathrm{NN}}_{\mu \nu}
     k_\rho k_\lambda
     Tr \mathbf{P}
     \exp
          \left(
           i \int d \tau k \cdot A
          \right)
     \int d x_2 \int d x_3 \int d x_4
     \left[ (z-\bar{z})(z-x_1)(\bar{z}-x_1) \right] \n
 & \times
     \left[
      (x_1-x_2)^{-\frac{1}{4}}
      (x_1-x_3)^{-\frac{1}{4}}
      (x_1-x_4)^{-\frac{1}{4}}
      (x_2-x_3)^{-\frac{1}{4}}
      (x_2-x_4)^{-\frac{1}{4}}
      (x_3-x_4)^{-\frac{1}{4}}
     \right] \n
 & \times
     \left( \frac{1}{z-x_1} - \frac{1}{z-x_2} \right)
     \left( \frac{1}{\bar{z}-x_3} - \frac{1}{\bar{z}-x_4} \right) \n
 & \times
      \frac{ g_{\tau \sigma}(x_1-x_4)(x_2-x_3)
                \frac{i}{2}
                \left[ \epsilon_a (\gamma^{\rho \mu} \gamma^\tau)^{a b} \epsilon_b \right]
                \frac{i}{2}
                \left[ \epsilon_c (\gamma^{\lambda \nu} \gamma^\sigma)^{c d} \epsilon_d \right]}
                {[(x_1-x_2)(x_1-x_3)(x_1-x_4)(x_2-x_3)(x_2-x_4)(x_3-x_4)]^{\frac{3}{4}}} \n
 = & \frac{1}{4} g_{\tau \sigma}
       \zeta^{\mathrm{NN}}_{\mu \nu}
       k_\rho k_\lambda
     Tr \mathbf{P}
     \exp
          \left(
           i \int d \tau k \cdot A
          \right)
       \left[ \epsilon_a (\gamma^{\rho \mu \tau})^{a b} \epsilon_b \right]
       \left[ \epsilon_c (\gamma^{\lambda \nu \sigma})^{c d} \epsilon_d \right] \n
   & \times \int d x_2 \int d x_3 \int d x_4
     \frac{(z-\bar{z})(z-x_1)(\bar{z}-x_1)}{x_{12} x_{13} x_{24} x_{34}}
     \left[ \frac{(x_1-x_2)(x_3-x_4)}
                 {(z-x_1)(z-x_2)(\bar{z}-x_3)(\bar{z}-x_4)} \right] \n
 = & \frac{1}{4} g_{\tau \sigma}
       \zeta^{\mathrm{NN}}_{\mu \nu}
       k_\rho k_\lambda
     Tr \mathbf{P}
     \exp
          \left(
           i \int d \tau k \cdot A
          \right)
       \left[ \epsilon_a (\gamma^{\rho \mu \tau})^{a b} \epsilon_b \right]
       \left[ \epsilon_c (\gamma^{\lambda \nu \sigma})^{c d} \epsilon_d \right] \n
   & \times \int d x_2 \int d x_3 \int d x_4
     \frac{(z-\bar{z})(\bar{z}-x_1)}
          {x_{13} x_{24}(z-x_2)(\bar{z}-x_3)(\bar{z}-x_4)} \n
 = & \frac{1}{4} g_{\tau \sigma}
       \zeta^{\mathrm{NN}}_{\mu \nu}
       k_\rho k_\lambda
     Tr \mathbf{P}
     \exp
          \left(
           i \int d \tau k \cdot A
          \right)
       \left[ \epsilon_a (\gamma^{\rho \mu \tau})^{a b} \epsilon_b \right]
       \left[ \epsilon_c (\gamma^{\lambda \nu \sigma})^{c d} \epsilon_d \right] \n
   & \times \int d x_2 \int d x_3 \int d x_4
     \frac{z-\bar{z}}{(z-x_2)(\bar{z}-x_2)} \cdot
     \frac{x_2-\bar{z}}{(x_2-x_4)(\bar{z}-x_4)} \cdot
     \frac{x_1-\bar{z}}{(x_1-x_3)(\bar{z}-x_3)} .
\end{align}
To calculate this expression, we take the limit as $x_1 \rightarrow z$.
We also consider $x_2$ is closer to $z$ than to $x_4$. That is,
on integrating over $x_4$, we regard that $x_2$ is on the upper half plane.
In this way it becomes like :
\begin{align}
  & \frac{1}{4} g_{\tau \sigma}
       \zeta^{\mathrm{NN}}_{\mu \nu}
       k_\rho k_\lambda
     STr
     \exp
          \left(
           i k \cdot A
          \right)
       \left[ \epsilon_a (\gamma^{\rho \mu \tau})^{a b} \epsilon_b \right]
       \left[ \epsilon_c (\gamma^{\lambda \nu \sigma})^{c d} \epsilon_d \right]
    (2 \pi i)^3 \n
= & - 2 \pi^3 i g_{\tau \sigma}
       \zeta^{\mathrm{NN}}_{\mu \nu}
       k_\rho k_\lambda
     STr
     \exp
          \left(
           i k \cdot A
          \right)
       \left[ \epsilon_a (\gamma^{\rho \mu \tau})^{a b} \epsilon_b \right]
       \left[ \epsilon_c (\gamma^{\lambda \nu \sigma})^{c d} \epsilon_d \right] .
\end{align}
This formula can be regarded as :
\begin{equation}
       \zeta^{\mathrm{NN}}_{\mu \nu}
     STr
     \exp
          \left(
           i k \cdot A
          \right)
       k_\rho k_\lambda
       \left[ \epsilon_a (\gamma^{\rho \mu \tau})^{a b} \epsilon_b \right]
       \left[ \epsilon_c (\gamma^{\lambda \nu \sigma})^{c d} \epsilon_d
      \right] .
 \label{Gravitonf4:result}
\end{equation}
Of course, $M^{\rho \mu}(m)$ and $M^{\lambda \nu}(n)$ also act on
other factors in the (\ref{MMSSSS}).
Such cases are considered in Appendix.A.3.
The conclusion is that they give the same result.
The corresponding term of the type IIB matrix model vertex operator is :
\begin{equation}
    h_{\mu \nu}(\lambda) V^h(A,\epsilon)
  = h_{\mu \nu}(\lambda)
    \left[
    STr e^{i k \cdot A}
    \left( - \frac{1}{96}
           k_\rho k_\sigma
           (\bar{\epsilon} \cdot \gamma^{\mu \rho \beta} \epsilon ) \cdot
           (\bar{\epsilon} \cdot {\gamma^{\nu \sigma}}_{\beta} \epsilon)
    \right)
    \right].
\end{equation}
We find again they coincide.
\subsection{Graviton coupling to one bosonic string and two fermionic strings}
We calculate the disk amplitude where the vertex operator of the
graviton field couples to two fermionic open strings and
one bosonic open string.
We calculate the following disk amplitude:
\begin{align}
   & \langle
       c (z) \bar{c} (\bar{z}) V^{\mathrm{NN}}_{(0,-1)} (z,\bar{z})
     Tr \mathbf{P}
     \exp
          \left(
           i \int dx W^{(0)} (x)
          \right) \n
   & \times \left(i \int_{- \infty}^{+ \infty} dt D^{(0)} (t)
         \right)
       c (x_1) F_{-\frac{1}{2}}(x_1) \int d x_2 F_{-\frac{1}{2}}(x_2)
     \rangle \n
 \sim & \langle
       c(z) \bar{c}(\bar{z})
       \zeta^{\mathrm{NN}}_{\mu \nu}
         (\partial X^\mu + i k_{\rho} j^{\mu \rho})(z)
         \bar{\psi}^\nu (\bar{z}) e^{-\bar{\phi}}(\bar{z})
         e^{i k X}(z,\bar{z})
     Tr \mathbf{P}
     \exp
          \left(
           i \int dx \Phi^\rho i g_{\rho \sigma} \partial_\perp X (x)
          \right) \n
 & \times \left( i \int_{- \infty}^{+ \infty} dt
                (- i g_{\alpha \gamma} g_{\beta \delta}
                   [\Phi^\alpha , \Phi^\beta] \Psi^\gamma \Psi^\delta (t) )
          \right)
       c (x_1)
       \epsilon_a S^a(x_1) e^{-\frac{1}{2} \phi}(x_1)
       \int d x_2 \epsilon_b S^b(x_2) e^{-\frac{1}{2} \phi}(x_2)
     \rangle \n
 = & \frac{i k_{\rho}}{(2 \pi)^2}
     \zeta^{\mathrm{NN}}_{\mu \nu}
     Tr \mathbf{P}
     \exp
          \left(
           i \int d \tau k \cdot A
          \right)
      F_{\gamma \delta}
      \epsilon_a \epsilon_b
       \int d t \int d x_2
        \langle
            c(z) \bar{c}(\bar{z})  c (x_1)
        \rangle
      \langle
        e^{-\bar{\phi}}(\bar{z})
        e^{-\frac{1}{2} \phi}(x_1)
        e^{-\frac{1}{2} \phi}(x_2)
      \rangle \n
 & \times
      \langle
         j^{\mu \rho} (z) j^{\gamma \delta} (t)
         \bar{\psi}^{\nu} (\bar{z})
         S^a(x_1)  S^b(x_2)
      \rangle \n
 = & \frac{ik_{\rho}}{(2 \pi)^2}
     \zeta^{\mathrm{NN}}_{\mu \nu}
     Tr \mathbf{P}
     \exp
          \left(
           i \int d \tau k \cdot A
          \right)
      F_{\gamma \delta}
      \epsilon_a \epsilon_b \n
   &  \times
      \int d t \int d x_2
      \left[ (z-\bar{z})(z-x_1)(\bar{z}-x_1) \right]
      \left[ (\bar{z}-x_1)^{-\frac{1}{2}}
             (\bar{z}-x_2)^{-\frac{1}{2}}
             (x_1-x_2)^{-\frac{1}{4}} \right] \n
  & \times
      \left[ \sum_i \frac{M^{\mu \rho}(i)}{z-x_i}
             \sum_j \frac{M^{\gamma \delta}(j)}{t-x_j}
             \frac{(\gamma^\nu)^{a b}}
                  {(\bar{z}-x_1)^{\frac{1}{2}}
                   (\bar{z}-x_2)^{\frac{1}{2}}
                   (x_1-x_2)^{\frac{3}{4}} }
      \right] \n
 = & \frac{ik_{\rho}}{(2 \pi)^2}
     \zeta^{\mathrm{NN}}_{\mu \nu}
     Tr \mathbf{P}
     \exp
          \left(
           i \int d \tau k \cdot A
          \right)
      F_{\gamma \delta}
      \epsilon_a \epsilon_b
       \int d t \int d x_2
        \frac{(z-\bar{z})(z-x_1)}{(\bar{z}-x_2)(x_1-x_2)} \sum_i \frac{M^{\mu \rho}(i)}{z-x_i} \n
  & \times
       [\frac{1}{t-\bar{z}}
         (g^{\nu \delta} \gamma^{\gamma} - g^{\gamma \nu} \gamma^{\delta})^{a b}
        +\frac{\frac{1}{2}}
              {t-x_1}
         (\gamma^{\gamma \delta \nu}
          - g^{\nu \delta} \gamma^\gamma + g^{\gamma \nu} \gamma^{\delta})^{a b}
       -\frac{\frac{1}{2}}
              {t-x_2}
         (\gamma^{\nu \gamma \delta}
          + g^{\nu \delta} \gamma^\gamma - g^{\gamma \nu}
      \gamma^{\delta})^{a b} ] .
\end{align}
When we integrate over $t$, we concentrate on the terms in the square bracket,
and take the limit $x_1 \rightarrow z$ and $x_2 \rightarrow \bar{z}$.
That is, to define the integration over $t$ well,
we consider the case when $x_1$ and $x_2$
locate in the upper side of the complex plane
than $t$ .
Then we obtain
\begin{align}
 & \int dt
        [\frac{1}{t-\bar{z}}
         (g^{\nu \delta} \gamma^{\gamma} - g^{\gamma \nu} \gamma^{\delta})^{a b}
        +\frac{\frac{1}{2}}
              {t-x_1}
         (\gamma^{\gamma \delta \nu}
          - g^{\nu \delta} \gamma^\gamma + g^{\gamma \nu} \gamma^{\delta})^{a b}
       -\frac{\frac{1}{2}}
              {t-x_2}
         (\gamma^{\nu \gamma \delta}
          + g^{\nu \delta} \gamma^\gamma - g^{\gamma \nu} \gamma^{\delta})^{a b} ] \n
 \rightarrow & - \int dt
   \frac{z - \bar{z}}{(t-z)(t-\bar{z})}
   (g^{\nu \delta} \gamma^{\gamma} - g^{\gamma \nu} \gamma^{\delta})^{a b} \n
 = & - 2 \pi i (g^{\nu \delta} \gamma^{\gamma} - g^{\gamma \nu} \gamma^{\delta})^{a b} .
\end{align}
Using (\ref{g-bi-spinor}), we can obtain the following equation :
\begin{align}
   & \epsilon_a \epsilon_b
     \sum_i \frac{M^{\mu \rho}(i)}{z-x_i}
     (g^{\nu \delta} \gamma^{\gamma} - g^{\gamma \nu} \gamma^{\delta})^{a b} \n
 = & \frac{1}{2} (\frac{1}{z-x_1} - \frac{1}{z-x_2})
     \epsilon_a
     (  g^{\nu \delta} \gamma^{\mu \rho \gamma}
     - g^{\gamma \nu} \gamma^{\mu \rho \delta})^{a b} \epsilon_b .
\end{align}
Consequently the disk amplitude becomes as follows:
 \begin{align}
  & \frac{k_{\rho}}{(2 \pi)^2}
     \zeta^{\mathrm{NN}}_{\mu \nu}
     Tr \mathbf{P}
     \exp
          \left(
           i \int d \tau k \cdot A
          \right)
      F_{\gamma \delta} \pi i \n
   & \times
      \int d x_2 \frac{(z-\bar{z})(z-x_1)}{(\bar{z}-x_2)(x_1-x_2)}
    \frac{x_1-x_2}{(z-x_1)(z-x_2)}
    \epsilon_a
     (  g^{\nu \delta} \gamma^{\mu \rho \gamma}
     - g^{\gamma \nu} \gamma^{\mu \rho \delta})^{a b} \epsilon_b \n
 = & \frac{k_{\rho}}{(2 \pi)^2}
     \zeta^{\mathrm{NN}}_{\mu \nu}
     Tr \mathbf{P}
     \exp
          \left(
           i \int d \tau k \cdot A
          \right)
      F_{\gamma \delta} \pi i
    \epsilon_a
     (  g^{\nu \delta} \gamma^{\mu \rho \gamma}
     - g^{\gamma \nu} \gamma^{\mu \rho \delta})^{a b} \epsilon_b
      \int d x_2 \frac{(z-\bar{z})}{(z-x_2)(\bar{z}-x_2)} \n
 = & \frac{k_{\rho}}{(2 \pi)^2} \pi i
     \zeta^{\mathrm{NN}}_{\mu \nu}
     STr
     \exp
          \left(
           i k \cdot A
          \right)
      F_{\gamma \delta}
    \epsilon_a
     (  g^{\nu \delta} \gamma^{\mu \rho \gamma}
     - g^{\gamma \nu} \gamma^{\mu \rho \delta})^{a b} \epsilon_b
    (2 \pi i) \n
 = & -\frac{k_{\rho}}{2}
     \zeta^{\mathrm{NN}}_{\mu \nu}
     STr
     \exp
          \left(
           i k \cdot A
          \right)
     (  {F_{\gamma}}^{\nu} \epsilon_a \gamma^{\mu \rho \gamma}
     -  {F^\nu}_{\delta} \epsilon_a \gamma^{\mu \rho \delta})^{a b} \epsilon_b \n
 = & -\zeta^{\mathrm{NN}}_{\mu \nu}
     STr e^{i k \cdot A}
     \left( k_\rho
            \epsilon_a (\gamma^{\rho \gamma \mu})^{ab} \epsilon_b
            F^{\nu}_{\ \ \gamma} \right).
 \end{align}
We compare this to the result from the type IIB matrix model, which is
given as:
\begin{equation}
   h_{\mu \nu}(\lambda) V^{h}(A,\epsilon)
 = h_{\mu \nu}(\lambda)
    \left[
     STr e^{i k \cdot A}
      \left(
       - \frac{i}{4} k^\alpha
         \bar{\epsilon} \cdot
         {\gamma_{\alpha \beta}}^{(\mu}
         \epsilon \cdot F^{\nu) \beta}
      \right)
    \right].
\end{equation}
They coincide up to normalization coefficients.

\subsection{Graviton coupling to $C_{\frac{1}{2}}$ and one fermionic open string}
We also need to consider the coupling through $C_{\frac{1}{2}}$ type
vertex operator.
The disk amplitude is:
\begin{align}
   & \langle c(z) \bar{c}(\bar{z})
          V_{(-1,-1)}^{NN}(z,\bar{z})
          Tr \mathbf{P}
          \exp \left(
                 i \int dt
                 W^{(0)}(t)
               \right)
          c(x_1) C_{\frac{1}{2}} (x_1) \int d x_2 F_{-\frac{1}{2}} (x_2)
    \rangle \n
 = & \langle c(z) \bar{c}(\bar{z}) c(x_1) \rangle
     \langle \zeta_{\mu \nu}^{\mathrm{NN}}
              e^{-\phi(z)} \psi^\mu (z)
              e^{-\bar{\phi}}(\bar{z}) \bar{\psi}^\nu (\bar{z}) e^{i k X (z,\bar{z})}
          Tr \mathbf{P}
     \exp \left( i \int d \tau k \cdot A \right) \n
   & \times c (x_1) [\epsilon^a, \Phi_\alpha] (\gamma_\beta)_{a b}
     \Psi^\alpha \Psi^\beta (x_1) S^b(x) e^{+ \frac{1}{2} \phi(x_1)}
     \int d x_2 \epsilon_c S^c (x_2) e^{-\frac{1}{2} \phi(x_2)} \rangle \n
 = & \left( (z-\bar{z}) (z-x_1) (\bar{z}-x_1) \right)
     \zeta_{\mu \nu}^{\mathrm{NN}}
     Tr \mathbf{P}
     \exp \left( i \int d \tau k \cdot A \right)
     [\epsilon^a, \Phi_\alpha] (\gamma_\beta)_{a b} \epsilon_c \n
   & \times \int d x_2
            (z-\bar{z})^{-1} (z-x_1)^{\frac{1}{2}} (z-x_2)^{-\frac{1}{2}}
            (\bar{z}-x_1)^{\frac{1}{2}} (\bar{z}-x_2)^{-\frac{1}{2}}
            (x_1 - x_2)^{\frac{1}{4}} \n
   & \times \langle
              j^{\alpha \beta} (x_1)
              \psi^\mu (z) \bar{\psi}^\nu (\bar{z})
              S^b(x_1) S^c (x_2)
            \rangle .
\label{hCF}
\end{align}
To calculate the OPE :
\begin{equation}
 \langle
 j^{\alpha \beta} (x_1)
 \psi^\mu (z) \bar{\psi}^\nu (\bar{z})
 S^b(x_1) S^c (x_2)
 \rangle ,
 \label{jjpp}
\end{equation}
we specify the U(1) charges of bosonized $S^b$ and $S^c$ as in Table.\ref{jppSS:chirality}.
\begin{table}
\caption{The U(1) charges for spin fields to calculate (\ref{jjpp}).}
\begin{center}
 \begin{tabular}{cc}  \hline
  $ S^b$ & $S^c$  \\  \hline
   $+$   & $+$    \\
   $+$   & $+$    \\
   $+$   & $-$    \\
   $+$   & $-$    \\
   $+$   & $-$    \\ \hline
 \end{tabular}
\end{center}
\label{jppSS:chirality}
\end{table}
Recalling the OPE of the 2-point function of the spin fields in  (\ref{2spin}),
we can show :
 \begin{align}
& \langle
   j^{\alpha \beta}(x_1) \psi^\mu (z) \bar{\psi}^\nu (\bar{z}) S^b(x_1) S^c(x_2)
  \rangle \n
 \sim
& \langle
   j^{\alpha \beta}(x_1)
   \frac{ j^{\gamma \delta}(x_2)
           \left( \gamma_{\gamma \delta} \right)^{b c}
        }
        {(x_1-x_2)^{ \frac{1}{4} } }
  \psi^\mu (z) \bar{\psi}^\nu (\bar{z})
  \rangle \n
 = & \frac{\left( \gamma_{\gamma \delta} \right)^{b c}}
          {(x_1-x_2)^{\frac{1}{4} }}
     \langle
       j^{\alpha \beta}(x_1)
       j^{\gamma \delta}(x_2)
       \psi^\mu (z) \bar{\psi}^\nu (\bar{z})
     \rangle .
 \label{jjpp:L1}
\end{align}
(\ref{jjpp:L1}) becomes :
\begin{align}
   & \langle
     j^{\alpha \beta}(x_1) j^{\gamma \delta} (x_2)
     \psi^\mu (z) \bar{\psi}^\nu (\bar{z})
     \rangle \n
 = & \langle
      \Psi^\alpha \Psi^\beta (x_1) \Psi^\gamma \Psi^\delta(x_2) \psi^\mu(z) \bar{\psi}^\nu(\bar{z})
     \rangle .
\label{jjpp11}
\end{align}
This gives
 \begin{align}
  & -\frac{g^{\alpha \delta} g^{\gamma \mu} g^{\beta \nu}}{(x_1-z)(x_2-z)(z-\bar{z})}
    +(-(\alpha \leftrightarrow \beta),-(\gamma \leftrightarrow \delta)) \n
  & -\frac{ g^{\alpha \nu} g^{\gamma \mu} g^{\beta \delta}}{(x_1-\bar{z})(x_2-z)(z-\bar{z})}
    +(-(\alpha \leftrightarrow \beta),-(\gamma \leftrightarrow \delta)) \n
  & +\frac{ g^{\alpha \delta} g^{\gamma \nu} g^{\beta \mu}}{(x_1-z)(x_2-\bar{z})(z-\bar{z})}
    +(-(\alpha \leftrightarrow \beta),-(\gamma \leftrightarrow \delta)) \n
  & +\frac{ g^{\alpha \mu} g^{\gamma \nu} g^{\beta \delta}}{(x_1-\bar{z})(x_2-\bar{z})(z-\bar{z})}
    +(-(\alpha \leftrightarrow \beta),-(\gamma \leftrightarrow \delta))
\label{jjpp1a}
\end{align}
and the following terms
 \begin{align}
  & -\frac{g^{\alpha \gamma} g^{\beta \mu} g^{\delta \nu}}{(x_1-x_2)(x_2-z)(z-\bar{z})}
    +(-(\alpha \leftrightarrow \beta),-(\gamma \leftrightarrow \delta)) \n
  & -\frac{ g^{\alpha \delta} g^{\beta \nu} g^{\gamma \mu}}{(x_1-x_2)(x_2-z)(z-\bar{z})}
    +(-(\alpha \leftrightarrow \beta),-(\gamma \leftrightarrow \delta)) \n
  & +\frac{ g^{\alpha \gamma} g^{\beta \mu} g^{\delta \nu}}{(x_1-x_2)(x_2-\bar{z})(z-\bar{z})}
    +(-(\alpha \leftrightarrow \beta),-(\gamma \leftrightarrow \delta)) \n
  & +\frac{ g^{\alpha \delta} g^{\beta \nu} g^{\gamma \mu} }{(x_1-x_2)(x_2-\bar{z})(z-\bar{z})}
    +(-(\alpha \leftrightarrow \beta),-(\gamma \leftrightarrow \delta)) .
\label{jjpp1b}
\end{align}
This expression contains the poles on the real axis.
Thus (\ref{hCF}) becomes :
\begin{align}
 & \zeta_{\mu \nu}^{\mathrm{NN}}
     Tr \mathbf{P}
     \exp \left( i \int d \tau k \cdot A \right) \n
 & \times (z-x_1)(\bar{z}-x_1)
        \left[
           \frac{ (z-x_1)^{ \frac{1}{2} }
                  (\bar{z}-x_1)^{ \frac{1}{2} }
                  (x_1-x_2)^{ \frac{1}{4} }
                }
                { (z-x_2)^{\frac{1}{2}}
                  (\bar{z}-x_2)^{\frac{1}{2}}
                }
        \right] \n
 &   \times \bigg([\epsilon^a, \Phi_\alpha] (\gamma^\nu)_{a b}
                   \epsilon_c (\gamma^{\alpha \mu} )^{bc}
                   \frac{z-\bar{z}}
                  {(x_1-x_2)^{ \frac{1}{4}} (x_1-z)(x_1-\bar{z})(x_2-z)(z-\bar{z})} \n
 &   \ \qquad -  [\epsilon^a, \Phi_\alpha] (\gamma^\mu)_{a b}
                   \epsilon_c (\gamma^{\alpha \nu})^{bc}
                   \frac{z-\bar{z}}
                  {(x_1-x_2)^{ \frac{1}{4}} (x_1-z)(x_1-\bar{z})(x_2-\bar{z})(z-\bar{z})} \n
 &   \ \qquad +  [\epsilon^a, \Phi^\nu] (\gamma_\beta)_{a b}
                   \epsilon_c (\gamma^{\mu \beta} )^{bc}
                   \frac{z-\bar{z}}
                  {(x_1-x_2)^{ \frac{1}{4}} (x_1-z)(x_1-\bar{z})(x_2-z)(z-\bar{z})} \n
 &   \ \qquad -  [\epsilon^a, \Phi^\mu] (\gamma_\beta)_{a b}
                   \epsilon_c (\gamma^{\nu \beta} )^{bc}
                   \frac{z-\bar{z}}
                   {(x_1-x_2)^{ \frac{1}{4}} (x_1-z)(x_1-\bar{z})(x_2-\bar{z})(z-\bar{z})}
            \bigg) .
\end{align}
We have used the symmetry of  $\zeta_{\mu \nu}^{\mathrm{NN}}$ under the exchange of
$\mu$ and $\nu$.
We find that the contribution (\ref{jjpp1b}) vanishes identically.
Taking the limit $x_1 \rightarrow x_2$, this integration can be done.
Using the equation of motion (\ref{eom}), we obtain
\begin{equation}
 \zeta^{\mathrm{NN}}_{\mu \nu}
     STr \exp \left( i k \cdot A \right)
       \left( \bar{\epsilon} \cdot
       \gamma^{(\mu}
         \left[ A^{\nu)},\epsilon \right] \right) .
\end{equation}
It agrees with the corresponding matrix model vertex operator:
\begin{equation}
   h_{\mu \nu}(\lambda) V^{h}(A,\epsilon)
 = h_{\mu \nu}(\lambda)
   \left[
    STr e^{i k \cdot A}
    \left(
    \frac{1}{2}
     \bar{\epsilon} \cdot
       \gamma^{(\mu}
         \left[ A^{\nu)},\epsilon \right]
    \right)
   \right] .
 \label{Gbf2-matrix}
\end{equation}

\subsection{Graviton coupling to bosonic strings}
We consider the case where the vertex operator of the graviton
couples to only bosonic open strings for completeness.
It couples two bosonic open strings.
The disk amplitude is :
 \begin{align}
   &  Tr \mathbf{P}
      \sum_{n=0}^{\infty}  2 \pi i
       \int_{-\infty}^{+\infty} dt_2  dt_3  \dots
       dt_{n-1} dt_n \n
   & \times
     \langle
        c (z) \bar{c} (\bar{z}) V^{\mathrm{NN}}_{(-1,-1)}(z,\bar{z})
        c (t_1) \frac{1}{n!} \prod_{a=1}^n \left( i W^{(0)} (t_a) \right)
         \left( i \int d x D^{(0)} (x) \right)
         \left( i \int d y D^{(0)}(y)  \right)
      \rangle \n
 =
   &  Tr \mathbf{P}
      \sum_{n=0}^{\infty}  2 \pi i
       \int_{-\infty}^{+\infty} dt_2  dt_3  \dots
       dt_{n-1} dt_n \n
   & \times
     \langle
        c (z) \bar{c} (\bar{z}) c (t_1)
     \rangle
     \langle
        \zeta_{\mu \nu}^{\mathrm{NN}}
        e^{-\phi(z)} \psi^\mu (z)  e^{-\bar{\phi}(\bar{z})} \bar{\psi}^\nu (\bar{z})
        e^{i k X}(z,\bar{z})
     \frac{1}{n!} \prod_{a=1}^n 
     \left( i \Phi^\rho g_{\rho \sigma} \partial_\perp X^\sigma (t_a) \right) \n
   & \times
         [ i \int d x
            (- i g_{\alpha_1 \gamma_1} g_{\beta_1 \delta_1}
            [\Phi^{\gamma_1},\Phi^{\delta_1}]
            \Psi^{\alpha_1} \Psi^{\beta_1} )(x)
         ]
         [ i \int d y
            (- i g_{\alpha_2 \gamma_2} g_{\beta_2 \delta_2}
            [\Phi^{\gamma_2},\Phi^{\delta_2}]
            \Psi^{\alpha_2} \Psi^{\beta_2} )(y)
         ]
      \rangle \n
 \sim
   &  \frac{1}{(2 \pi)^4} \zeta_{\mu \nu}^{\mathrm{NN}}
     Tr \mathbf{P}
      \sum_{n=0}^{\infty}
       \frac{1}{n!} \int_{-\infty}^{+\infty} dt_2 dt_3  \dots
       dt_{n-1} dt_n \n
   & \times
     [(z-\bar{z})(z-t_1)(\bar{z}-t_1)] \frac{1}{z-\bar{z}}
     \left(
       2 \pi i (\Phi \cdot k) \frac{z-\bar{z}}{(z-t_1)(\bar{z}-t_1)}
       \prod_{a=2}^n [ (\Phi \cdot k) 2 \pi i \frac{\partial \tau(t_a,z)}{\partial t_a}]
     \right) \n
   & \times
       F_{\alpha_1 \beta_1} F_{\alpha_2 \beta_2}
     \int dx \int d y
     \langle
      j^{\alpha_1 \beta_1}(x) j^{\alpha_2 \beta_2}(y)
      \psi^\mu (z)   \bar{\psi}^\nu (\bar{z})
     \rangle \n
 = & \frac{1}{(2 \pi)^4} \zeta_{\mu \nu}^{\mathrm{NN}}
     Tr \mathbf{P}
     \exp \left( \int_0^1 d \tau (i k \cdot A) \right)
       F_{\alpha_1 \beta_1} F_{\alpha_2 \beta_2} (z-\bar{z}) \n
   & \times \int dx \int d y
     \langle
      j^{\alpha_1 \beta_1}(x) j^{\alpha_2 \beta_2}(y)
      \psi^\mu (z)   \bar{\psi}^\nu (\bar{z})
     \rangle .
\label{GravitonBoson}
\end{align}
The OPE
\begin{equation}
     \langle
      j^{\alpha_1 \beta_1}(x) j^{\alpha_2 \beta_2}(y)
      \psi^\mu (z) \bar{\psi}^\nu (\bar{z})
     \rangle
\end{equation}
is given in (\ref{jjpp1a}) and (\ref{jjpp1b}).
Using the fact that
$\zeta_{\mu \nu}^{\mathrm{NN}}$ is the symmetric tensor and
$F$ is the antisymmetric tensor, the disk amplitude becomes
 \begin{align}
 & \frac{1}{(2 \pi)^4} \zeta_{\mu \nu}^{\mathrm{NN}}
   Tr \mathbf{P} \exp \bigg( \int_0^1 d \tau (i k \cdot A) \bigg) \n
 & \times \int dx \int dy
 \bigg(
          \frac{(z-\bar{z})^2}{(x - z)(x-\bar{z})(y-z)(y-\bar{z})}
          (- 4 F^{\alpha_1 \mu} {F^{\nu}}_{\alpha_1})
 \bigg) \n
 = & \frac{1}{\pi^2}
     \zeta_{\mu \nu}^{\mathrm{NN}}
     STr \bigg[
               \exp (i k \cdot A)
               F^{\alpha \mu} {F^\nu}_{\alpha}
         \bigg] .
\label{Gbb}
\end{align}
Up to the normalization coefficient, (\ref{Gbb}) gives :
\begin{equation}
     \zeta_{\mu \nu}^{\mathrm{NN}}
     STr \bigg[
               \exp (i k \cdot A)
               F^{\mu \alpha} {F^\nu}_{\alpha}
         \bigg] .
\end{equation}
At the same time, the result from the type IIB matrix model is as follows :
\begin{equation}
   h_{\mu \nu}(\lambda) V^{h}(A,\epsilon)
 = h_{\mu \nu}(\lambda) STr
   \left( e^{i k \cdot A} F^{\mu \rho} \cdot {F^{\nu}}_{\rho} \right) .
\end{equation}
They agree with each other.

\section{4-th rank antisymmetric tensor}
Finally we consider the fourth-rank antisymmetric tensor $A_{\mu \nu 
\rho \sigma}$.
Similarly to the graviton field, 
it is the $4$ times SUSY transformed part of the IIB supergravity multiplet.
The difference is that the vertex operator satisfies the R-R boundary condition.
Just like the graviton case, the matrix model vertex operator consists of 4 terms 
as shown in Appendix A.5.
We reproduce each of them in the following subsections.

\subsection{$A_{\mu \nu \rho \sigma}$ coupling to $4$ fermionic open strings}
We calculate the disk amplitude, in which the fourth-rank antisymmetric 
tensor couples to $4$ fermionic open strings.
It is given as :
\begin{align}
   & \langle A_{\mu_1 \mu_2 \mu_3 \mu_4} \rangle \n
 = & \langle c(z) \bar{c} (\bar{z})
         V^{\mathrm{RR}}_{(-\frac{1}{2},-\frac{3}{2})} (z, \bar{z})
     Tr \mathbf{P}
     \exp
          \left(
           i \int dt W^{(0)} (t)
          \right) \n
   & \times c(x_1) F_{\frac{1}{2}}(x_1)
         \int d x_2 F_{\frac{1}{2}}(x_2)
         \int d x_3 F_{- \frac{1}{2}}(x_3)
         \int d x_4 F_{- \frac{1}{2}}(x_4) \rangle \n
 \sim & \langle c(z) \bar{c} (\bar{z})
         \zeta^{\mathrm{RR}}_{\mu_1 \mu_2 \mu_3 \mu_4}
           e^{-\frac{1}{2} \phi}(z)
           S^a (z)
           (\gamma^{\mu_1 \mu_2 \mu_3 \mu_4})_{a b}
           S^b (\bar{z})
           e^{-\frac{3}{2} \bar{\phi}}(\bar{z})
           e^{i k X}(z,\bar{z}) \n
 & \times  Tr \mathbf{P}
              \exp
               \left(
                  i \int d \tau k \cdot A
               \right)
          c(x_1) \epsilon^g (\gamma^{\mu})_{g c} S^c (x_1)
                      \partial_\perp X_\mu e^{+\frac{1}{2} \phi}(x_1) \n
 & \times  \int d x_2 \epsilon^h (\gamma^{\nu})_{h d} S^d (x_2)
                      \partial_\perp X_\nu e^{+\frac{1}{2} \phi}(x_2)
           \int d x_3 \epsilon_e S^e (x_3) e^{-\frac{1}{2} \phi}(x_3)
           \int d x_4 \epsilon_f S^f (x_4) e^{-\frac{1}{2} \phi}(x_4) \rangle .
\end{align}
Ghost terms give :
\begin{equation}
    \langle c(z) \bar{c} (\bar{z}) c (x_1) \rangle
    = (z-\bar{z})(z-x_1)(\bar{z}-x_1) ,
\end{equation}
and
\begin{align}
  & \langle e^{-\frac{1}{2} \phi}(z)
              e^{-\frac{3}{2} \bar{\phi}}(\bar{z})
              e^{\frac{1}{2} \phi}(x_1)
              e^{\frac{1}{2} \phi}(x_2)
              e^{-\frac{1}{2} \phi}(x_3)
              e^{-\frac{1}{2} \phi}(x_4) \rangle \n
  = & (z-\bar{z})^{-\frac{3}{4}}
     (z-x_1)^{+\frac{1}{4}}
     (z-x_2)^{+\frac{1}{4}}
     (z-x_3)^{-\frac{1}{4}}
     (z-x_4)^{-\frac{1}{4}} \n
    & \times
     (\bar{z}-x_1)^{+\frac{3}{4}}
     (\bar{z}-x_2)^{+\frac{3}{4}}
     (\bar{z}-x_3)^{-\frac{3}{4}}
     (\bar{z}-x_4)^{-\frac{3}{4}} \n
    & \times
     (x_1-x_2)^{-\frac{1}{4}}
     (x_1-x_3)^{+\frac{1}{4}}
     (x_1-x_4)^{+\frac{1}{4}} \n
    & \times
     (x_2-x_3)^{+\frac{1}{4}}
     (x_2-x_4)^{+\frac{1}{4}}
     (x_3-x_4)^{-\frac{1}{4}} .
 \end{align}
Spin field $S$ can be shown in the bosonized form as follows :
\begin{equation}
 S = e^{\frac{i}{2}s_i H^i}.
\end{equation}
Here $s^i$  take value :  $(\pm1, \pm1, \pm1, \pm1 ,\pm1)$ and
$H^i$ are scalar fields.
Here we want to consider the $6$ point function of the spin field given
as :
\begin{equation}
   \langle
     S^a(z) (\gamma^{\mu_1 \mu_2 \mu_3 \mu_4})_{ab} S^b(\bar{z})
     \epsilon_g {(\gamma^\mu)^g}_c S^c(x_1)
     \epsilon_h {(\gamma^\nu)^h}_d S^d(x_2)
     \epsilon_e S^e(x_3) \epsilon_f S^f(x_4)
    \rangle .
\label{6spin}
\end{equation}
Here we take a specific configuration of $s^i$'s as in the Table.\ref{spinfield:chirality}.
Then (\ref{6spin}) is calculated as :
\begin{align}
  & \langle
     S^a(z) (\gamma^{\mu_1 \mu_2 \mu_3 \mu_4})_{ab} S^b(\bar{z})
     \epsilon_g {(\gamma^\mu)^g}_c S^c(x_1)
     \epsilon_h {(\gamma^\nu)^h}_d S^d(x_2)
     \epsilon_e S^e(x_3) \epsilon_f S^f(x_4)
    \rangle \n
\sim
  & \epsilon_g \epsilon_h \epsilon_e \epsilon_f
    (z-\bar{z})^{\frac{3}{4}}
    (z-x_1)^{-\frac{1}{4}}
    (z-x_2)^{-\frac{1}{4}}
    (z-x_3)^{-\frac{3}{4}}
    (z-x_4)^{-\frac{3}{4}} \n
  & \times
    (\bar{z}-x_1)^{-\frac{3}{4}}
    (\bar{z}-x_2)^{-\frac{3}{4}}
    (\bar{z}-x_3)^{-\frac{1}{4}}
    (\bar{z}-x_4)^{-\frac{1}{4}} \n
  & \times
    (x_1-x_2)^{\frac{1}{4}}
    (x_1-x_3)^{-\frac{1}{4}}
    (x_1-x_4)^{-\frac{1}{4}}
    (x_2-x_3)^{-\frac{1}{4}}
    (x_2-x_4)^{-\frac{1}{4}}
    (x_3-x_4)^{\frac{1}{4}} \n
  & \times
    (\gamma^{\mu_1 \mu_2 \mu_3 \mu_4} \gamma^\mu \gamma^\nu )^{g h e f} .
\end{align}

\begin{table}
\caption{The U(1) charges for spin fields to calculate (\ref{6spin}).}
\begin{center}
 \begin{tabular}{cccccc}      \hline
  $ S^a$ & $S^b$  & $S^c$ & $S^d$ & $S^e$ & $S^f$  \\  \hline
   $+$   & $-$    & $+$   & $+$   & $-$   & $-$    \\
   $+$   & $+$    & $-$   & $+$   & $-$   & $-$    \\
   $+$   & $+$    & $-$   & $-$   & $+$   & $-$    \\
   $+$   & $+$    & $-$   & $-$   & $-$   & $+$    \\
   $+$   & $+$    & $+$   & $-$   & $-$   & $-$    \\  \hline
 \end{tabular}
\end{center}
\label{spinfield:chirality}
\end{table}

Therefore the disk amplitude of the $A_{\mu_1 \mu_2 \mu_3 \mu_4}$ becomes
as follows :
\begin{align}
   & \langle A_{\mu_1 \mu_2 \mu_3 \mu_4} \rangle \n
 = & \zeta^{\mathrm{RR}}_{\mu_1 \mu_2 \mu_3 \mu_4}
     Tr \mathbf{P}
              \exp
               \left(
                  i \int d \tau k \cdot A
               \right) \n
   & \times
     \int d x_2 \int d x_3 \int d x_4
     [(z-\bar{z})(z-x_1)(\bar{z}-x_1)]
     k_\mu \left[ \frac{1}{z-x_1} - \frac{1}{\bar{z}-x_1} \right]
     k_\nu \left[ \frac{1}{z-x_2} - \frac{1}{\bar{z}-x_2} \right] \n
   & \times
     \frac{1}{(z-x_3)(\bar{z}-x_3)(z-x_4)(\bar{z}-x_4)}
     \left[ \epsilon_g (\gamma^{\nu [\mu_1 \mu_2})^{gh}  \epsilon_h \right]
     \left[ \epsilon_e (\gamma^{\mu_3 \mu_4] \mu })^{ef} \epsilon_f \right] \n
 = & \zeta^{\mathrm{RR}}_{\mu_1 \mu_2 \mu_3 \mu_4}
     Tr \mathbf{P}
              \exp
               \left(
                  i \int d \tau k \cdot A
               \right)
      k_\mu k_\nu
     \left[ \epsilon_g (\gamma^{\nu [\mu_1 \mu_2})^{gh}  \epsilon_h \right]
     \left[ \epsilon_e (\gamma^{mu_3 \mu_4] \mu})^{ef} \epsilon_f \right] \n
   & \times \int dx_2
      \frac{z-\bar{z}}{(z-x_2)(\bar{z}-x_2)}
     \int dx_3
      \frac{z-\bar{z}}{(z-x_3)(\bar{z}-x_3)}
     \int dx_4
      \frac{z-\bar{z}}{(z-x_4)(\bar{z}-x_4)} \n
 = &  - 8 \pi^3 i
       \zeta^{\mathrm{RR}}_{\mu_1 \mu_2 \mu_3 \mu_4}
      STr \left( e^{i k \cdot A}
      k_\mu k_\nu
     \left[ \epsilon_g (\gamma^{\nu [\mu_1 \mu_2})^{gh}  \epsilon_h \right]
     \left[ \epsilon_e (\gamma^{\mu_3 \mu_4] \mu})^{ef} \epsilon_f \right] \right) .
\end{align}
The result from type IIB matrix model is
\begin{equation}
   A^{\mu \nu \rho \sigma}(\lambda) V^{A} (A, \epsilon)
 = A^{\mu \nu \rho \sigma}(\lambda) STr e^{i k \cdot A}
   \frac{i}{8 \cdot 4 !}
   k_\alpha k_\gamma
   \left[
     \bar{\epsilon} \cdot {\gamma_{[\mu \nu}}^{\alpha}
     \epsilon
   \right] \cdot
  \left[
    \bar{\epsilon} \cdot {\gamma_{\rho \sigma]}}^{\gamma}
    \epsilon
  \right] .
\end{equation}
They coincide up to normalization coefficients.

\subsection{$A_{\mu \nu \rho \sigma}$ coupling to two fermionic strings and one bosonic string}
We calculate the disk amplitude where the vertex operator of the
4-th rank anti-symmetric tensor field couples to two fermionic open strings and
one open string.
The corresponding disk amplitude is :
\begin{align}
   & \langle A_{\mu_1 \mu_2 \mu_3 \mu_4} \rangle \n
 = & \langle c(z) \bar{c} (\bar{z})
         V^{\mathrm{RR}}_{(-\frac{1}{2},-\frac{3}{2})} (z, \bar{z})
     Tr \mathbf{P}
     \exp
          \left(
           i \int dt W^{(0)} (t)
          \right) \n
   & \times c(x_1) 
         \left(
           iD^{(0)} (x_1)
         \right)
         \int d x_2 F_{-\frac{1}{2}}(x_2)
         \int d x_3 F_{- \frac{1}{2}}(x_3) \rangle \n
 = & \langle c(z) \bar{c} (\bar{z}) c(x_1) \rangle
     \langle
      \zeta_{\mu \nu \rho \sigma}^{\mathrm{RR}}
       e^{-\frac{1}{2} \phi(z)} 
       S^a (z) (\gamma^{\mu \nu \rho \sigma})_{a b}
       S^b(\bar{z}) e^{-\frac{3}{2} \bar{\phi}(\bar{z})}
       e^{i k X(z,\bar{z})} \n
   & \times   Tr \mathbf{P}
        \exp
         \left(i \int d \tau k \cdot A \right)
        [\Phi_\alpha, \Phi_\beta] \Psi^\alpha \Psi^\beta (x_1) \n
   & \times \int d x_2 \epsilon^e (\gamma^\gamma)_{e c} 
        \partial_\perp X_\gamma S^c (x_2) e^{\frac{1}{2}\phi(x_2)}
       \int d x_3 \epsilon_d S^d (x_3) e^{-\frac{1}{2}\phi(x_3)}
     \rangle \n
 = & \int d x_2 \int d x_3 \langle c(z) \bar{c} (\bar{z}) c(x_1) \rangle
     \langle
      e^{-\frac{1}{2} \phi(z)} e^{-\frac{3}{2} \bar{\phi}(\bar{z})}
      e^{\frac{1}{2}\phi(x_2)} e^{-\frac{1}{2}\phi(x_3)}
     \rangle \n 
   & \times
      \frac{1}{4 \pi^2} \zeta_{\mu \nu \rho \sigma}^{\mathrm{RR}}
       Tr \mathbf{P}
        \exp
         \left(i \int d \tau k \cdot A \right)
       [A_\alpha, A_\beta] \epsilon^e \epsilon_d \n
   & \times k_\gamma \left( \frac{1}{z-x_2} - \frac{1}{\bar{z}-x_2} \right)
       (\gamma^{\mu \nu \rho \sigma})_{a b} (\gamma^\gamma)_{e c}
     \langle
       j^{\alpha \beta} (x_1) S^a (z) S^b(\bar{z}) S^c (x_2) S^d (x_3) 
     \rangle .
\label{AUFF}
\end{align}
Ghost contributions give
\begin{equation}
   \langle
     c(z) \bar{c} (\bar{z}) c(x_1)
   \rangle
  = (z-\bar{z})(z-x_1)(\bar{z}-x_1),
\end{equation}
and
 \begin{align}
   &     \langle
           e^{-\frac{1}{2}\phi(z)} e^{-\frac{3}{2} \bar{\phi}(\bar{z})}
           e^{\frac{1}{2}\phi(x_2)} e^{-\frac{1}{2} \phi(x_3)}
         \rangle \n
  =&
         (z - \bar{z})^{-\frac{3}{4}} 
         (z - x_2)^{\frac{1}{4}} 
         (z - x_3)^{-\frac{1}{4}} 
         (\bar{z} - x_2)^{\frac{3}{4}} 
         (\bar{z} - x_3)^{-\frac{3}{4}} 
         (x_2 - x_3)^{\frac{1}{4}} .
 \end{align}
The OPE for the fermions and spin fields are :
\begin{align}
  &  \langle
       j^{\alpha \beta} (x_1) S^a (z) S^b(\bar{z}) S^c (x_2) S^d (x_3)
     \rangle \n
 = & \sum_i \frac{M^{\alpha \beta}(i)}{x_1 - z_i}
     \frac{(z-x_3)(\bar{z}-x_2)
           (\gamma^\tau)^{ab}(\gamma_\tau)^{cd}
           -(z-\bar{z})(x_2-x_3)
           (\gamma^\tau)^{ad}(\gamma_\tau)^{bc}}
           {(z-\bar{z})^{\frac{3}{4}}
           (z-x_2)^{\frac{3}{4}}
           (z-x_3)^{\frac{3}{4}}
           (\bar{z}-x_2)^{\frac{3}{4}}
           (\bar{z}-x_3)^{\frac{3}{4}}
           (x_2-x_3)^{\frac{3}{4}}
          } .
\end{align} 
Thus (\ref{AUFF}) becomes :
\begin{align}
& - \frac{1}{16 \pi^2} \zeta_{\mu \nu \rho \sigma}^{\mathrm{RR}}
       Tr \mathbf{P}
        \exp
         \left(i \int d \tau k \cdot A \right)
       F_{\alpha \beta} \epsilon^e \epsilon_d \n
& \times k_\gamma \int d x_2 \int d x_3 \frac{z-\bar{z}}{(z-x_2)(\bar{z}-x_2)}
       (\gamma^{\mu \nu \rho \sigma})_{a b} (\gamma^\gamma)_{e c} \n
& \times 
  \left(
        \frac{(\gamma^{\alpha \beta \tau})^{ad}(\gamma_\tau)^{bc}}{x_1-z}
      - \frac{(\gamma^{\tau \alpha \beta})^{ad}(\gamma_\tau)^{bc}}{x_1-x_3}
      - \frac{(\gamma^{\tau})^{ad}(\gamma^{\alpha \beta}_{\ \ \ \tau})^{bc}}{x_1-\bar{z}}
      + \frac{(\gamma^{\tau})^{ad}(\gamma_{\tau}^{\ \alpha \beta})^{bc}}{x_1-x_2}
\right) \n
& \times
\frac{
      (z-\bar{z})^\frac{1}{2}(z-x_1)(\bar{z}-x_1)(x_2 - x_3)^{\frac{1}{2}}
     }
     {(z - x_2)^\frac{1}{2}
      (z-x_3)
      (\bar{z}-x_3)^{\frac{3}{2}}} \n
= & - \frac{1}{16 \pi^2} \zeta_{\mu \nu \rho \sigma}^{\mathrm{RR}}
       Tr \mathbf{P}
        \exp
         \left(i \int d \tau k \cdot A \right)
       F_{\alpha \beta} \epsilon^e \epsilon_d \n
& \times k_\gamma \int d x_2 \int d x_3 
  \frac{z-\bar{z}}{(z-x_2)(\bar{z}-x_2)} \n
& \times
 \left[
    (\gamma^{\gamma}  \gamma_\tau \gamma^{\alpha \beta} \gamma^{\mu \nu \rho \sigma} \gamma^\tau )_{e}^{\ d}
    \left( \frac{z-\bar{z}}{(z-x_1)(\bar{z}-x_1)} + \frac{x_2-x_3}{(x_1-x_2)(x_1-x_3)} \right)
 \right] \n
& \times
\frac{
      (z-\bar{z})^\frac{1}{2}(z-x_1)(\bar{z}-x_1)(x_2 - x_3)^{\frac{1}{2}}
     }
     {(z - x_2)^\frac{1}{2}
      (z-x_3)
      (\bar{z}-x_3)^{\frac{3}{2}}} \n
= & - \frac{1}{16 \pi^2} \zeta_{\mu \nu \rho \sigma}^{\mathrm{RR}}
       Tr \mathbf{P}
        \exp
         \left(i \int d \tau k \cdot A \right)
       k_\gamma F_{\alpha \beta}
       \epsilon^e
      (\gamma^{\gamma}  \gamma_\tau \gamma^{\alpha \beta}
       \gamma^{\mu \nu \rho \sigma} \gamma^\tau )_{e}^{\ d}
      \epsilon_d \n
   & \times
     \int d x_2 
      \frac{z-\bar{z}}{(z-x_2)(\bar{z}-x_2)} \n
   & \times
      \int d x_3
      \left[ 
       \frac{
             (z-\bar{z})^\frac{3}{2} (x_2 - x_3)^{\frac{1}{2}}
            }
            {(z - x_2)^\frac{1}{2}
             (z-x_3)
             (\bar{z}-x_3)^{\frac{3}{2}}
            }
     +        \frac{
             (z-\bar{z})^\frac{1}{2}(z-x_1)(\bar{z}-x_1)(x_2 - x_3)^{\frac{3}{2}}
            }
            {(x_1-x_2)(x_1-x_3)
             (z - x_2)^\frac{1}{2}
             (z-x_3)
             (\bar{z}-x_3)^{\frac{3}{2}}
            }
      \right] .
\label{DAtotalAUFF}
\end{align}
When we integrate over $x_3$ in the last line of (\ref{DAtotalAUFF}),
we can take the limit $x_2 \rightarrow \bar{z}$.
The first term in the integrand become
\begin{equation}
 \frac{z-\bar{z}}{(z-x_3)(\bar{z}-x_3)}.
\end{equation}
The second term becomes
\begin{equation}
 - \frac{ z-x_1 }
        {(x_1-x_3)
         (z-x_3)}
\end{equation}
and we can take any value for $x_1$.
In particular taking $x_1 \rightarrow z$, it vanishes.
Ignoring the normalization ambiguity, (\ref{DAtotalAUFF}) can be 
evaluated as :
\begin{equation}
   \zeta_{\mu \nu \rho \sigma}^{\mathrm{RR}}
       STr 
        \left(
          e^{i k \cdot A }
           F^{[\mu \nu} k_\gamma
        \left(
        \bar{\epsilon}
           \gamma^{ \rho \sigma] \gamma}
        \epsilon
        \right)
       \right) .
\end{equation}
The corresponding result from the matrix model is given as
\begin{equation}
A^{\mu \nu \rho \sigma}(\lambda) V^A(A,\epsilon)
=
A^{\mu \nu \rho \sigma}(\lambda)
STr
\left( e^{ik \cdot A}
\frac{1}{4}
          F_{[\mu\nu} \cdot (\bar{\epsilon} \cdot
          {\gamma_{\rho\sigma]}}^{\beta}\epsilon)k_{\beta}
\right) .
\end{equation}
Therefore they coincide.
\subsection{$A_{\mu \nu \rho \sigma}$ coupling to $C_{\frac{1}{2}}$ and one fermionic open string}
Same as the graviton field,
$A_{\mu \nu \rho \sigma}$ can couple to $C_{\frac{1}{2}}$ and one 
fermionic open string .
The corresponding disk amplitude is :
\begin{align}
   & \langle A_{\mu_1 \mu_2 \mu_3 \mu_4} \rangle \n
 = & \langle c(z) \bar{c} (\bar{z})
         V^{\mathrm{RR}}_{(-\frac{1}{2},-\frac{3}{2})} (z, \bar{z})
     Tr \mathbf{P}
     \exp
          \left(
           i \int dt W^{(0)} (t)
          \right) \n
   & \times c(x_1) C_{\frac{1}{2}} (x_1)
         \int d x_2 F_{- \frac{1}{2}}(x_2) \rangle \n
 = & \langle c(z) \bar{c} (\bar{z})
       \zeta_{\mu \nu \rho \sigma}^{\mathrm{RR}} e^{-\frac{1}{2}\phi(z)}
       S^a (\gamma^{\mu \nu \rho \sigma})_{a b} S^b
       e^{ - \frac{3}{2} \bar{\phi} ( \bar{z} ) } e^{i k X}
     Tr \mathbf{P}
     \exp
          \left(
           i \int d \tau k \cdot A
          \right) \n
   & \times c(x_1) [\epsilon^e, \Phi_\alpha]
           (\gamma_\beta)_{e c} \Psi^\alpha \Psi^\beta(x_1)
           S^c (x_1) e^{\frac{1}{2}\phi(x_1)}
         \int d x_2 \epsilon_d S^d (x_2) e^{-\frac{1}{2}\phi(x_2)} \rangle \n
 = & \int d x_2 \langle c(z) \bar{c} (\bar{z}) c(x_1) \rangle
   \langle 
     e^{-\frac{1}{2}\phi(z)} e^{ -\frac{3}{2} \bar{\phi} (\bar{z}) }
     e^{\frac{1}{2}\phi(x_1)} e^{-\frac{1}{2} \phi(x_2)}
   \rangle
   \zeta_{\mu \nu \rho \sigma}^{\mathrm{RR}}
     Tr \mathbf{P}
     \exp
          \left(
            i \int d \tau k \cdot A
          \right) \n
   & \times [\epsilon^e, \Phi_\alpha]
     (\gamma^{\mu \nu \rho \sigma})_{ab} (\gamma_\beta)_{e c} \epsilon_d
     \n
 & \times
   \langle
     j^{\alpha \beta}(x_1) S^a(z) S^b(\bar{z}) S^c(x_1) S^d (x_2)
   \rangle .
\end{align}
Ghost contributions give
\begin{equation}
   \langle
     c(z) \bar{c} (\bar{z}) c(x_1)
   \rangle
  = (z-\bar{z})(z-x_1)(\bar{z}-x_1),
\end{equation}
and
 \begin{align}
   &     \langle
           e^{-\frac{1}{2}\phi(z)} e^{-\frac{3}{2} \bar{\phi}(\bar{z})}
           e^{\frac{1}{2}\phi(x_1)} e^{-\frac{1}{2} \phi(x_2)}
         \rangle \n
  =&
         (z - \bar{z})^{-\frac{3}{4}} 
         (z - x_1)^{\frac{1}{4}} 
         (z - x_2)^{-\frac{1}{4}} 
         (\bar{z} - x_1)^{\frac{3}{4}} 
         (\bar{z} - x_2)^{-\frac{3}{4}} 
         (x_1 - x_2)^{\frac{1}{4}} .
 \end{align}
The OPE for the fermions and spin fields are :
\begin{align}
 & \langle
     j^{\alpha \beta}(x_1) S^a(z) S^b(\bar{z}) S^c(x_1) S^d (x_2)
   \rangle \n
 = & \sum_{ \{i | z_i \neq x_1 \} } 
     \frac{M^{\alpha \beta}(i)}{x_1-z_i}
     \frac{   (z-x_2)(\bar{z}-x_1)(\gamma^\tau)^{ab}(\gamma_\tau)^{cd}
            - (z-\bar{z})(x_1-x_2)(\gamma^\tau)^{ad}(\gamma_\tau)^{bc}}
          {(z-\bar{z})^{\frac{3}{4}}
           (z-x_1)^{\frac{3}{4}}
           (z-x_2)^{\frac{3}{4}}
           (\bar{z}-x_1)^{\frac{3}{4}}
           (\bar{z}-x_2)^{\frac{3}{4}}
           (x_1-x_2)^{\frac{3}{4}}
          } .
\end{align}
Thus the disk amplitude becomes
\begin{align}
 & - \zeta_{\mu \nu \rho \sigma}^{\mathrm{RR}}
     Tr \mathbf{P}
     \exp
          \left(
            i \int d \tau k \cdot A
          \right) [\epsilon^e, \Phi_\alpha]
     (\gamma^{\mu \nu \rho \sigma})_{ab} (\gamma_\beta)_{e c} \epsilon_d \n
 & \times
   \int d x_2 
   \left( 
        \frac{(\gamma^{\alpha \beta \tau})^{ad}(\gamma_\tau)^{bc}}{x_1-z}
      - \frac{(\gamma^{\tau})^{ad}(\gamma^{\alpha \beta}_{\ \ \ \tau})^{bc}}{x_1-\bar{z}}
      - \frac{(\gamma^{\tau \alpha \beta})^{ad}(\gamma_\tau)^{bc}}{x_1-x_2}
   \right) \n
 & \times \frac{   (z-\bar{z})^{\frac{1}{2}}
                   (z-x_1)^{\frac{1}{2}}
                   (\bar{z}-x_1)
                   (x_1-x_2)^{\frac{1}{2}}
               }
               {(z-x_2)
                (\bar{z}-x_2)^{\frac{3}{2}}
               } \n
= & - \zeta_{\mu \nu \rho \sigma}^{\mathrm{RR}}
     Tr \mathbf{P}
     \exp
          \left(
            i \int d \tau k \cdot A
          \right)
     [\epsilon^e, \Phi_\alpha] 
      (\gamma_\beta \gamma_\tau 
       \gamma^{\mu \nu \rho \sigma} 
       \gamma^{\alpha \beta \tau})_e^{\ d}
     \epsilon_d    \n
 & \times    \int d x_2
   \left( 
        \frac{z-\bar{z}}{(z-x_1)(\bar{z}-x_1)}
      - \frac{1}{x_1-x_2}
   \right) \n
 & \times \frac{   (z-\bar{z})^{\frac{1}{2}}
                   (z-x_1)^{\frac{1}{2}}
                   (\bar{z}-x_1)
                   (x_1-x_2)^{\frac{1}{2}}
               }
               {(z-x_2)
                (\bar{z}-x_2)^{\frac{3}{2}}
               } \n
= & - \zeta_{\mu \nu \rho \sigma}^{\mathrm{RR}}
     Tr \mathbf{P}
     \exp
          \left(
            i \int d \tau k \cdot A
          \right)
     [\epsilon^e, \Phi_\alpha] 
      (\gamma_\beta \gamma_\tau 
       \gamma^{\mu \nu \rho \sigma} 
       \gamma^{\alpha \beta \tau})_e^{\ d}
     \epsilon_d    \n
 & \times
   \int d x_2
   \left( 
        \frac{ (z-\bar{z})^{\frac{3}{2}} (x_1-x_2)^{\frac{1}{2}} }
             { (z-x_1)^{\frac{1}{2}} (z-x_2) (\bar{z}-x_2)^{\frac{3}{2}} }
      - \frac{   (z-\bar{z})^{\frac{1}{2}}
                   (z-x_1)^{\frac{1}{2}}
                   (\bar{z}-x_1)
               }
               {(z-x_2)
                (\bar{z}-x_2)^{\frac{3}{2}}
                (x_1-x_2)^{\frac{1}{2}}
               }
   \right) .
\end{align}
We can take the limit $x_1 \rightarrow \bar{z}$.
We obtain
\begin{equation}
- 2 \pi i \zeta_{\mu \nu \rho \sigma}^{\mathrm{RR}}
     Tr \mathbf{P}
     \exp
          \left(
            i \int d \tau k \cdot A
          \right)
     [\epsilon^e, \Phi_\alpha] 
      (\gamma_\beta \gamma_\tau 
       \gamma^{\mu \nu \rho \sigma} 
       \gamma^{\alpha \beta \tau})_e^{\ d}
     \epsilon_d .
\end{equation}
Ignoring the normalization ambiguity, we obtain :
\begin{equation}
 \zeta_{\mu \nu \rho \sigma}^{\mathrm{RR}}
     STr e^{i k \cdot A}
     \epsilon \gamma^{[\nu \rho \sigma} 
     [\epsilon, \Phi^{\mu]}] .
\end{equation}
The corresponding result from the matrix model is given as :
\begin{equation}
A^{\mu \nu \rho \sigma}(\lambda) V^A(A,\epsilon)
=
A^{\mu \nu \rho \sigma}(\lambda)
STr
\left( e^{ik \cdot A}
          \frac{i}{3}\bar{\epsilon}
          \cdot
          \gamma_{[\nu\rho\sigma}[\epsilon,A_{\mu]}]
\right).
\end{equation}
We find an agreement in this case again.
\subsection{$A_{\mu \nu \rho \sigma}$ coupling to bosonic open strings}
Lastly we consider the case where the vertex operator of the 4-th rank 
antisymmetric tensor field couples to two bosonic open strings.
The disk amplitude is :
\begin{align}
   & \langle A_{\mu_1 \mu_2 \mu_3 \mu_4} \rangle \n
 = & \langle c(z) \bar{c} (\bar{z})
         V^{\mathrm{RR}}_{(-\frac{1}{2},-\frac{3}{2})} (z, \bar{z})
     Tr \mathbf{P}
     \exp
          \left(
           i \int dt W^{(0)} (t)
          \right) \n
   & \times c(x_1) 
      i \left( D^{(0)} (x_1) \right)
      \int d x_2 
      i \left( D^{(0)} (x_2) \right)
     \rangle \n
 = & \langle c(z) \bar{c}(\bar{z}) c(x_1)
       \zeta_{\mu \nu \rho \sigma}^{\mathrm{RR}}
       e^{- \frac{1}{2} \phi(z) } S^a(z)
       (\gamma^{\mu \nu \rho \sigma})_{a b} S^b(\bar{z})
       e^{-\frac{3}{2} \bar{\phi}(\bar{z})} e^{i k X} \n
   & \times
     Tr \mathbf{P}
     \exp
          \left(
           i \int d \tau k \cdot A
          \right)
       \frac{1}{16 \pi^4} [A_\alpha, A_\beta][A_\gamma,A_\delta]
       j^{\alpha \beta}(x_1) j^{\gamma \delta}(x_2)
     \rangle \n
 = & \langle c(z) \bar{c}(\bar{z}) c(x_1) \rangle
     \langle 
       e^{- \frac{1}{2} \phi(z) } e^{-\frac{3}{2} \bar{\phi}(\bar{z})}
     \rangle
     \frac{1}{16 \pi^4} \zeta_{\mu \nu \rho \sigma}^{\mathrm{RR}}
     Tr \mathbf{P}
     \exp
          \left(
           i \int d \tau k \cdot A
          \right) \n
   & \times
       F_{\alpha \beta} F_{\gamma \delta}
       (\gamma^{\mu \nu \rho \sigma})_{a b}
     \langle
       j^{\alpha \beta}(x_1) j^{\gamma \delta}(x_2) S^a(z) S^b(\bar{z})
     \rangle .
\label{AUU}
\end{align}
Ghost contributions give
\begin{align}
 & \langle c(z)\bar{c}(\bar{z})c(x_1) \rangle =(z-\bar{z})(z-x_1)(\bar{z}-x_1), \n
 & \langle e^{-\frac{1}{2}\phi(z)} e^{-\frac{3}{2}\bar{\phi}(\bar{z})} \rangle
   = (z-\bar{z})^{-\frac{3}{4}}.
\end{align}
The OPE from the fermions and spin fields gives
\begin{align}
 & \langle
    j^{\alpha \beta}(x_1) j^{\gamma \delta}(x_2) S^a(z) S^b(\bar{z})
   \rangle \n
 = & \left(\frac{1}{x_1-z} - \frac{1}{x_1-\bar{z}} \right)
     \left(\frac{1}{x_2-z} - \frac{1}{x_2-\bar{z}} \right)
     \frac{(\gamma^{\alpha \beta} \gamma^{\gamma \delta})^{ab}}
          {(z-\bar{z})^{\frac{5}{4}}} .
\end{align}
Substituting these formulae into (\ref{AUU}), the disk amplitude becomes
\begin{align}
 & \frac{1}{16 \pi^4} \zeta_{\mu \nu \rho \sigma}^{\mathrm{RR}}
     Tr \mathbf{P}
     \exp
          \left(
           i \int d \tau k \cdot A
          \right)
     F_{\alpha \beta} F_{\gamma \delta}
     tr(\gamma^{\mu \nu \rho \sigma} \gamma^{\alpha \beta} \gamma^{\gamma \delta}) \n
 & \times 
   \int d x_2 \frac{z-\bar{z}}{(z-x_2)(\bar{z}-x_2) }.
\end{align}
Ignoring the normalization ambiguity, this formula can be simplified as
\begin{equation}
    \zeta_{\mu \nu \rho \sigma}^{\mathrm{RR}}
     STr
     e^{i k \cdot A}
     (- i F^{[\mu \nu} F^{\rho \sigma]}).
\end{equation}
It agrees with corresponding result from the matrix model 
\begin{equation}
A^{\mu \nu \rho \sigma}(\lambda) V^A(A,\epsilon)
=
A^{\mu \nu \rho \sigma}(\lambda)
STr
e^{ik \cdot A}
\left(
       -iF_{[\mu\nu} \cdot F_{\rho \sigma]}
\right).
\end{equation}
\section{Conclusion}
We have investigated the vertex operators of the supergravity
multiplet in the type IIB matrix model from the first principle by
using conformal field theory. The vertex operators couple closed
strings to open strings that are introduced by the existence of the
D-branes. We have generalized a single D instanton calculation
\cite{GreenGutperle,Gutperle} into that for multiple D instantons by
introducing matrix Majorana-Weyl spinor fields. Although Okawa and Ooguri
considered multiple D-branes, they only investigated the couplings to bosonic open
strings. In this respect, we have investigated the most generic case
in which the both bosonic and fermionic open strings are involved.
Our results are consistent with the previous results based on the
BPS nature of the supergravity multiplet. We have explicitly carried
out the conformal field theory calculation up to the 4-th rank antisymmetric tensor field.

Our investigation thus put our understandings of the supergravity
vertex operators in IIB matrix model on a very firm basis. It is
very gratifying that the symmetry arguments are confirmed by the
first principle calculations in string perturbation theory. Our
investigations have thus justified the basic assumptions in the
previous symmetry arguments.
To be precise, we have confirmed the existence of each term of the matrix model vertex operators
from the first principle, namely conformal field theory.
The symmetry arguments need to assume the existence of these operators.
Once their existence is assured, we can trust the symmetry arguments to determine
the exact structure of the vertex operators including the numerical coefficients. 

The vertex operators enables us to compute the correlation functions
in IIB matrix model. In fact, this problem has been investigated in
\cite{NK} and perturbative superstring amplitudes are
reproduced in a matrix string like background.

The supergravity multiplet are very important to understand the
dynamics of IIB matrix model as they are expected to control the low
energy and long distance physics. Let us consider a block diagonal matrix
configuration whose center of mass are widely separated. It has
been found that the effective action for such a configuration is
given by the supergravity which couples to the vertex operators of
the respective matrix configuration. We thus expect that the long
distance dynamics should be investigated by supergravity. We
hope that matrix configurations could be self-consistently
determined in such an analysis.

Another issue is a possible relationship to gauge/gravity duality.
Since we have argued that the effective theory of IIB matrix model is IIB supergravity,
a consistent background of it must be a solution of supergravity.
In fact non-commutative backgrounds are argued to be dual to supergravity solutions with
various fluxes\cite{duality}. It is conceivable that such a correspondence can be better understood
from our point of view.

\section*{Acknowledgment}
We are very grateful to useful discussion with Satoshi Iso, Shun'ya
Mizoguchi and Osamu Saito.
This work is supported in part by the Grant-in-Aid for Scientific Research from
the Ministry of Education, Science and Culture of Japan.

\newpage
\setcounter{section}{0}
\renewcommand{\theequation}{A.\arabic{equation}}
\renewcommand{\thesection}{\Alph{section}}
\section{Appendix}
\subsection{Gamma matrices and traces}
We give some useful fomulae for calculating traces of $\gamma$ - matrices.
Mathematically, $\gamma$ - matrices are given by the Clifford algebra:
\begin{equation}
 \{ \gamma^\mu,\gamma^\nu \} = 2 g^{\mu \nu}.
\end{equation}
Thus,
 \begin{align}
   \mathrm{tr} (\gamma^\mu \gamma^\nu)
 = & \mathrm{tr} (2 g^{\mu \nu} - \gamma^\nu \gamma^\mu) \n
 = &   2 g^{\mu \nu} \mathrm{tr} 1
     - \mathrm{tr} \gamma^\nu \gamma^\mu \n
 = &    2 g^{\mu \nu} 2^{\frac{D}{2}}
      - \mathrm{tr} (\gamma^\mu \gamma^\nu) .
 \end{align}
\begin{equation}
  \therefore
      \mathrm{tr} (\gamma^\mu \gamma^\nu)
    = g^{\mu \nu} 2^{\frac{D}{2}} .
\end{equation}
Therefore
 \begin{align}
   \mathrm{tr} (\gamma^\mu \gamma^\nu \gamma^\rho \gamma^\sigma)
 = & \mathrm{tr}
       [(2 g^{\mu \nu} - \gamma^\nu \gamma^\mu)
        \gamma^\rho \gamma^\sigma] \n
 = &   2 g^{\mu \nu} \mathrm{tr} [\gamma^\rho \gamma^\sigma]
     - \mathrm{tr} \gamma^\nu \gamma^\mu \gamma^\rho \gamma^\sigma \n
 = &    2 g^{\mu \nu} g^{\rho \sigma} 2^{\frac{D}{2}}
      - \mathrm{tr}
         [\gamma^\nu (2 g^{\mu \rho} - \gamma^\rho \gamma^\mu) \gamma^\sigma] \n
 = &   2 \cdot 2^{\frac{D}{2}} g^{\mu \nu} g^{\rho \sigma}
     - 2 g^{\mu \rho} \cdot 2^{\frac{D}{2}} g^{\nu \sigma}
     + \mathrm{tr} [\gamma^\nu \gamma^\rho \gamma^\mu \gamma^\sigma] ,
 \end{align}
where
 \begin{align}
      \mathrm{tr} [\gamma^\nu \gamma^\rho \gamma^\mu \gamma^\sigma]
  = & \mathrm{tr}
       [\gamma^\nu \gamma^\rho (2 g^{\mu \sigma} - \gamma^\sigma \gamma^\mu)] \n
  = &   2 g^{\mu \sigma} \cdot 2^{\frac{D}{2}} g^{\nu \rho}
      - \mathrm{tr} [\gamma^\mu \gamma^\nu \gamma^\rho \gamma^\sigma] .
 \end{align}
\begin{equation}
   \therefore
     \mathrm{tr} (\gamma^\mu \gamma^\nu \gamma^\rho \gamma^\sigma)
 = 2^{\frac{D}{2}}
    [  g^{\mu \nu} g^{\rho \sigma}
     - g^{\mu \rho} g^{\nu \sigma}
     + g^{\mu \sigma} g^{\nu \rho}
    ] .
\end{equation}
For example, in the case for $ D = 4$,
\begin{equation}
   \mathrm{tr} (\gamma^\mu \gamma^\nu \gamma^\rho \gamma^\sigma)
 = 4 (  g^{\mu \nu} g^{\rho \sigma}
      - g^{\mu \rho} g^{\nu \sigma}
      + g^{\mu \sigma} g^{\nu \rho}
     ) .
\end{equation}
Therefore
\begin{align}
   & \mathrm{tr} (\gamma^{\mu \nu} \gamma^{\rho \sigma}) \n
 = & \frac{1}{4}
     \mathrm{tr}
     \left[
           (\gamma^\mu \gamma^\nu - \gamma^\nu \gamma^\mu)
           (\gamma^\rho \gamma^\sigma - \gamma^\sigma \gamma^\rho)
     \right] \n
 = &  \frac{1}{4} \left[
      \mathrm{tr} [\gamma^\mu \gamma^\nu \gamma^\rho \gamma^\sigma]
    - \mathrm{tr} [\gamma^\mu \gamma^\nu \gamma^\sigma \gamma^\rho]
    - \mathrm{tr} [\gamma^\nu \gamma^\mu \gamma^\rho \gamma^\sigma]
    + \mathrm{tr} [\gamma^\nu \gamma^\mu \gamma^\sigma \gamma^\rho] \right] \n
 = & \frac{1}{4}
     \bigg[
       2^{\frac{D}{2}}
        [  g^{\mu \nu} g^{\rho \sigma}
         - g^{\mu \rho} g^{\nu \sigma}
         + g^{\mu \sigma} g^{\nu \rho}
        ] \n
   & - 2^{\frac{D}{2}}
        [
           g^{\mu \nu} g^{\sigma \rho}
         - g^{\mu \sigma} g^{\nu \rho}
         + g^{\mu \rho} g^{\nu \sigma}
        ] \n
   & - 2^{\frac{D}{2}}
        [
           g^{\nu \mu} g^{\rho \sigma}
         - g^{\nu \rho} g^{\mu \sigma}
         + g^{\nu \sigma} g^{\mu \rho}
        ] \n
   & + 2^{\frac{D}{2}}
        [
           g^{\nu \mu} g^{\sigma \rho}
         - g^{\nu \sigma} g^{\mu \rho}
         + g^{\nu \rho} g^{\mu \sigma}
        ]
     \bigg] \n
 = & - \frac{1}{4} 2^{\frac{D}{2}}
       [   g^{\mu \rho} g^{\nu \sigma}
         - g^{\mu \sigma} g^{\nu \rho}
         - g^{\nu \rho} g^{\mu \sigma}
         + g^{\nu \sigma} g^{\mu \rho} ] .
\label{g2g2}
\end{align}

\subsection{Spin field}
The OPE for 2-point function of the spin fields is given in \cite{CFQS} :
\begin{equation}
    S^a (z) S^b (w)
  \sim   \frac{- \delta^{a b}}{  (z-w)^{ \frac{5}{4} }  }
  +  \frac{ (\gamma_\mu)^{a b} \psi^\mu (z) }{  (z-w)^{ \frac{3}{4} } }
  +  \frac{ (\gamma_{\mu \nu})^{a b} j^{\mu \nu} (z) }{ (z-w)^\frac{1}{4} } ,
\label{2spin}
\end{equation}
where $j^{\mu \nu}(z) = \psi^\mu (z) \psi^\nu (z)$.

To calculate the disk amplitude,
we need the OPE of 4-point function for the spin fields.
The result is given as
\begin{equation}
   \langle S^a (z_1) S^b (z_2) S^c(z_3) S^d(z_4) \rangle
 = \frac{   z_{14} z_{23} (\gamma^{ \rho})^{ a b } (\gamma_{\rho})^{ c d }
          - z_{12} z_{34} (\gamma^{ \rho})^{ a d } (\gamma_{\rho})^{ b c } }
        { (z_{12} z_{13} z_{14} z_{23} z_{24} z_{34} )^{ \frac{3}{4}} } ,
 \label{4spinfield}
\end{equation}
where $z_{i j} \equiv z_i - z_j$.
When we consider SCFT on the 2-dimensional world sheet,
there is the $SL(2,\mathbb{C})$ invariance.
And using the anharmonic quotient
\begin{equation}
 z \equiv \frac{ z_{12} z_{34} }{ z_{13} z_{24} } \ ,
 \quad 1-z = \frac{ z_{14} z_{23} }{ z_{13} z_{24} } \qquad \quad ,
\end{equation}
the following equation holds for the same 4-operators $\mathcal{O}$ :
\begin{equation}
  \langle
   \mathcal{O}^a (z_1) \mathcal{O}^b (z_2) \mathcal{O}^c (z_3) \mathcal{O}^d (z_4)
  \rangle
= (z_{13} z_{24})^{- 2 \Delta}
          \langle \mathcal{O}^a (+\infty) \mathcal{O}^b (1) \mathcal{O}^c (z) \mathcal{O}^d (0)
          \rangle ,
\end{equation}
where $\Delta$ is the conformal dimension of $\mathcal{O}$.
For the spin field $S(z)$, $\Delta = \frac{5}{8}$.
Therefore
 \begin{align}
    & \langle S^a (z_1) S^b (z_2) S^c(z_3) S^d(z_4) \rangle \n
  = & (z_{13} z_{24})^{ - 2 \times \frac{5}{8}}
      \langle S^a (+ \infty) S^b (1) S^c (z) S^d (0) \rangle .
\end{align}
Following \cite{CFQS},
\begin{equation}
   \langle S^a (+ \infty) S^b (1) S^c (z) S^d (0) \rangle
 = \left[ z(1-z) \right]^{ -\frac{3}{4} }
      \left[ (1-z) (\gamma^{ \rho })^{ a b } (\gamma_{ \rho })^{c d}
            - z (\gamma^{ \rho })^{ a d } (\gamma_{ \rho })^{b c}
      \right] .
\end{equation}
Substituting this into the above equation,
we get :
\begin{align}
    & \langle S^a (z_1) S^b (z_2) S^c(z_3) S^d(z_4) \rangle \n
  = & (z_{13} z_{24})^{- \frac{5}{4}}
      \left[ z(1-z) \right]^{ -\frac{3}{4} }
      \left[ (1-z) (\gamma^{ \rho })^{ a b } (\gamma_{ \rho })^{c d}
            - z (\gamma^{ \rho })^{ a d } (\gamma_{ \rho })^{b c}
      \right] \n
  = & (z_{13} z_{24})^{ - \frac{5}{4} }
       \left[ \frac{ z_{12} z_{34} }{ z_{13} z_{24} } \cdot
              \frac{ z_{14} z_{23} }{ z_{13} z_{24} }
       \right]^{-\frac{3}{4}}
      \left[ \frac{ z_{14} z_{23} }{ z_{13} z_{24} }
                (\gamma^{ \rho})^{ a b } (\gamma_{ \rho })^{c d}
            - \frac{ z_{12} z_{34} }{ z_{13} z_{24} }
                (\gamma^{ \rho})^{ a d } (\gamma_{ \rho })^{b c}
      \right] \n
  = & (z_{13} z_{24})^{ - \frac{5}{4} } \cdot
      \frac{ (z_{13} z_{24})^{ 2 \cdot \frac{3}{4} } }
           { (z_{12} z_{14} z_{23} z_{34})^{ \frac{3}{4} } }
      \left[ \frac{ z_{14} z_{23} }{ z_{13} z_{24} }
                (\gamma^{ \rho })^{ a b } (\gamma_{ \rho })^{c d}
            - \frac{ z_{12} z_{34} }{ z_{13} z_{24} }
                (\gamma^{ \rho })^{ a d } (\gamma_{ \rho })^{b c}
      \right] \n
 = & \frac{   z_{14} z_{23} (\gamma^\rho)^{ab} (\gamma_{\rho})^{c d}
          - z_{12} z_{34} (\gamma^\rho)^{ad} (\gamma_{\rho})^{b c} }
        { (z_{12} z_{13} z_{14} z_{23} z_{24} z_{34} )^{\frac{3}{4}} } .
\end{align}
Therefore we got (\ref{4spinfield}).
\subsection{Disk amplitude for the graviton field}
In the main text, we calculated (\ref{MMSSSS}) for the case
when $M^{\rho \mu}(m)$ acts on
$(\gamma^\tau)^{ab}$ and $M^{\lambda \nu}(n)$ acts on $(\gamma^\sigma)^{cd}$.
Here we consider the case when they
act on the other part the of the $\gamma$ matrices in (\ref{MMSSSS}).

 Firstly, we consider the case when $M^{\rho \mu}(m)$ acts on
$(\gamma^\sigma)^{cd}$ and $M^{\lambda \nu}(n)$ acts on $(\gamma^\tau)^{ab}$.
\begin{align}
 & - \zeta^{\mathrm{NN}}_{\mu \nu}
     Tr \mathbf{P}
     \exp
          \left(
           i \int d \tau k \cdot A
          \right)
     k_\rho k_\lambda
     \int d x_2 \int d x_3 \int d x_4
     \left[ (z-\bar{z})(z-x_1)(\bar{z}-x_1) \right] \n
 & \times
     \left[
      (x_1-x_2)^{-\frac{1}{4}}
      (x_1-x_3)^{-\frac{1}{4}}
      (x_1-x_4)^{-\frac{1}{4}}
      (x_2-x_3)^{-\frac{1}{4}}
      (x_2-x_4)^{-\frac{1}{4}}
      (x_3-x_4)^{-\frac{1}{4}}
     \right] \n
 & \times
     \left( \frac{1}{z-x_3} - \frac{1}{z-x_4} \right)
     \left( \frac{1}{\bar{z}-x_1} - \frac{1}{\bar{z}-x_2} \right) \n
 & \times
      \frac{ g_{\tau \sigma}(x_1-x_4)(x_2-x_3)
                \frac{i}{2}
                \left[ \epsilon_a (\gamma^{\lambda \nu} \gamma^\sigma)^{a b} \epsilon_b \right]
                \frac{i}{2}
                \left[ \epsilon_c (\gamma^{\rho \mu} \gamma^\tau)^{c d} \epsilon_d \right]}
                {[(x_1-x_2)(x_1-x_3)(x_1-x_4)(x_2-x_3)(x_2-x_4)(x_3-x_4)]^{\frac{3}{4}}} \n
 = & \frac{1}{4} g_{\tau \sigma}
       \zeta^{\mathrm{NN}}_{\mu \nu}
     Tr \mathbf{P}
     \exp
          \left(
           i \int d \tau k \cdot A
          \right)
       k_\rho k_\lambda
       \left[ \epsilon_c (\gamma^{\lambda \nu \sigma})^{c d} \epsilon_d \right]
       \left[ \epsilon_a (\gamma^{\rho \mu \tau})^{a b} \epsilon_b \right] \n
   & \times \int d x_2 \int d x_3 \int d x_4
     \frac{(z-\bar{z})(z-x_1)(\bar{z}-x_1)}{x_{12} x_{13} x_{24} x_{34}}
     \left[ \frac{(x_3-x_4)(x_1-x_2)}
                 {(z-x_3)(z-x_4)(\bar{z}-x_1)(\bar{z}-x_2)} \right] \n
 = & \frac{1}{4} g_{\tau \sigma}
       \zeta^{\mathrm{NN}}_{\mu \nu}
     Tr \mathbf{P}
     \exp
          \left(
           i \int d \tau k \cdot A
          \right)
       k_\rho k_\lambda
       \left[ \epsilon_c (\gamma^{\lambda \nu \sigma})^{c d} \epsilon_d \right]
       \left[ \epsilon_a (\gamma^{\rho \mu \tau})^{a b} \epsilon_b \right] \n
   & \times \int d x_2 \int d x_3 \int d x_4
     \frac{(z-\bar{z})(z-x_1)}
          {x_{13} x_{24}(z-x_3)(z-x_4)(\bar{z}-x_2)} \n
 = & \frac{1}{4} g_{\tau \sigma}
       \zeta^{\mathrm{NN}}_{\mu \nu}
     Tr \mathbf{P}
     \exp
          \left(
           i \int d \tau k \cdot A
          \right)
       k_\rho k_\lambda
       \left[ \epsilon_c (\gamma^{\lambda \nu \sigma})^{c d} \epsilon_d \right]
       \left[ \epsilon_a (\gamma^{\rho \mu \tau})^{a b} \epsilon_b \right] \n
   & \times \int d x_2 \int d x_3 \int d x_4
     \frac{z-\bar{z}}{(z-x_2)(\bar{z}-x_2)} \cdot
     \frac{x_2-z}{(x_2-x_4)(z-x_4)} \cdot
     \frac{x_1-z}{(x_1-x_3)(z-x_3)} .
\end{align}
Taking the limit $x_1 \rightarrow \bar{z}$ and $x_2 \rightarrow \bar{z}$,
it becomes as:
\begin{align}
  & \frac{1}{4} g_{\tau \sigma}
       \zeta^{\mathrm{NN}}_{\mu \nu}
     STr
     \exp
          \left(
           i k \cdot A
          \right)
       k_\rho k_\lambda
       \left[ \epsilon_c (\gamma^{\rho \mu \sigma})^{c d} \epsilon_d \right]
       \left[ \epsilon_a (\gamma^{\lambda \nu \tau})^{a b} \epsilon_b \right]
    (2 \pi i)^3 \n
= & - 2 \pi^3 i g_{\tau \sigma}
       \zeta^{\mathrm{NN}}_{\mu \nu}
     STr
     \exp
          \left(
           i k \cdot A
          \right)
       k_\rho k_\lambda
       \left[ \epsilon_c (\gamma^{\rho \mu \tau})^{c d} \epsilon_d \right]
       \left[ \epsilon_a (\gamma^{\lambda \nu \sigma})^{a b} \epsilon_b  \right] .
\label{AS41}
\end{align}

  Secondly, we consider the case when $M^{\rho \mu}(m)$ acts on
$(\gamma^\tau)^{a d}$ and $M^{\lambda \nu}(n)$ acts on $(\gamma^\sigma)^{b c}$.
\begin{align}
 & - \zeta^{\mathrm{NN}}_{\mu \nu}
     Tr \mathbf{P}
     \exp
          \left(
           i \int d \tau k \cdot A
          \right)
     k_\rho k_\lambda
     \int d x_2 \int d x_3 \int d x_4
     \left[ (z-\bar{z})(z-x_1)(\bar{z}-x_1) \right] \n
 & \times
     \left[
      (x_1-x_2)^{-\frac{1}{4}}
      (x_1-x_3)^{-\frac{1}{4}}
      (x_1-x_4)^{-\frac{1}{4}}
      (x_2-x_3)^{-\frac{1}{4}}
      (x_2-x_4)^{-\frac{1}{4}}
      (x_3-x_4)^{-\frac{1}{4}}
     \right] \n
 & \times
     \left( \frac{1}{z-x_1} - \frac{1}{z-x_4} \right)
     \left( \frac{1}{\bar{z}-x_2} - \frac{1}{\bar{z}-x_3} \right) \n
 & \times
      \frac{- g_{\tau \sigma}(x_1-x_2)(x_3-x_4)
                \frac{i}{2}
                \left[ \epsilon_a (\gamma^{\rho \mu} \gamma^\tau)^{a d} \epsilon_d \right]
                \frac{i}{2}
                \left[ \epsilon_b (\gamma^{\lambda \nu} \gamma^\sigma)^{b c} \epsilon_c \right]}
                {[(x_1-x_2)(x_1-x_3)(x_1-x_4)(x_2-x_3)(x_2-x_4)(x_3-x_4)]^{\frac{3}{4}}} \n
 = &- \frac{1}{4} g_{\tau \sigma}
       \zeta^{\mathrm{NN}}_{\mu \nu}
     Tr \mathbf{P}
     \exp
          \left(
           i \int d \tau k \cdot A
          \right)
       k_\rho k_\lambda
       \left[ \epsilon_a (\gamma^{\rho \mu \tau})^{a d} \epsilon_d \right]
       \left[ \epsilon_b (\gamma^{\lambda \nu \sigma})^{b c} \epsilon_c \right] \n
   & \times \int d x_2 \int d x_3 \int d x_4
     \frac{(z-\bar{z})(z-x_1)(\bar{z}-x_1)}{x_{13} x_{14} x_{23} x_{24}}
     \left[ \frac{(x_1-x_4)(x_2-x_3)}
                 {(z-x_1)(z-x_4)(\bar{z}-x_2)(\bar{z}-x_3)} \right] \n
 = & - \frac{1}{4} g_{\tau \sigma}
       \zeta^{\mathrm{NN}}_{\mu \nu}
     Tr \mathbf{P}
     \exp
          \left(
           i \int d \tau k \cdot A
          \right)
       k_\rho k_\lambda
       \left[ \epsilon_a (\gamma^{\rho \mu \tau})^{a d} \epsilon_d \right]
       \left[ \epsilon_b (\gamma^{\lambda \nu \sigma})^{b c} \epsilon_c \right] \n
   & \times \int d x_2 \int d x_3 \int d x_4
     \frac{(z-\bar{z})(\bar{z}-x_1)}
          {x_{13} x_{24}(z-x_4)(\bar{z}-x_2)(\bar{z}-x_3)} \n
 = & \frac{1}{4} g_{\tau \sigma}
       \zeta^{\mathrm{NN}}_{\mu \nu}
     Tr \mathbf{P}
     \exp
          \left(
           i \int d \tau k \cdot A
          \right)
       k_\rho k_\lambda
       \left[ \epsilon_a (\gamma^{\rho \mu \tau})^{a d} \epsilon_d \right]
       \left[ \epsilon_b (\gamma^{\lambda \nu \sigma})^{b c} \epsilon_c \right] \n
   & \times \int d x_2 \int d x_3 \int d x_4
     \frac{z-\bar{z}}{(z-x_2)(\bar{z}-x_2)} \cdot
     \frac{\bar{z}-x_1}{(x_1-x_3)(\bar{z}-x_3)} \cdot
     \frac{z-x_2}{(z-x_4)(x_2-x_4)} .
\end{align}
Taking the limit $x_1 \rightarrow z$ and $x_2 \rightarrow \bar{z}$,
it becomes as:
\begin{align}
  & \frac{1}{4} g_{\tau \sigma}
       \zeta^{\mathrm{NN}}_{\mu \nu}
     STr
     \exp
          \left(
           i k \cdot A
          \right)
       k_\rho k_\lambda
       \left[ \epsilon_a (\gamma^{\rho \mu \tau})^{a d} \epsilon_d \right]
       \left[ \epsilon_c (\gamma^{\lambda \nu \sigma})^{b c} \epsilon_c \right]
    (2 \pi i)^3 \n
= & - 2 \pi^3 i g_{\tau \sigma}
       \zeta^{\mathrm{NN}}_{\mu \nu}
     STr
     \exp
          \left(
           i k \cdot A
          \right)
       k_\rho k_\lambda
       \left[ \epsilon_a (\gamma^{\rho \mu \tau})^{a d} \epsilon_d \right]
       \left[ \epsilon_b (\gamma^{\lambda \nu \sigma})^{b c} \epsilon_c \right] .
\label{AS42}
\end{align}

 Thirdly, we consider the case
when $M^{\rho \mu}(m)$ acts on $(\gamma^\sigma)^{b c}$
and $M^{\lambda \nu}(n)$ acts on $(\gamma^\tau)^{a d}$.
\begin{align}
 & - \zeta^{\mathrm{NN}}_{\mu \nu}
     Tr \mathbf{P}
     \exp
          \left(
           i \int d \tau k \cdot A
          \right)
     k_\rho k_\lambda
     \int d x_2 \int d x_3 \int d x_4
     \left[ (z-\bar{z})(z-x_1)(\bar{z}-x_1) \right] \n
 & \times
     \left[
      (x_1-x_2)^{-\frac{1}{4}}
      (x_1-x_3)^{-\frac{1}{4}}
      (x_1-x_4)^{-\frac{1}{4}}
      (x_2-x_3)^{-\frac{1}{4}}
      (x_2-x_4)^{-\frac{1}{4}}
      (x_3-x_4)^{-\frac{1}{4}}
     \right] \n
 & \times
     \left( \frac{1}{z-x_2} - \frac{1}{z-x_3} \right)
     \left( \frac{1}{\bar{z}-x_1} - \frac{1}{\bar{z}-x_4} \right) \n
 & \times
      \frac{- g_{\tau \sigma}(x_1-x_2)(x_3-x_4)
                \frac{i}{2}
                \left[ \epsilon_a (\gamma^{\lambda \nu} \gamma^\tau)^{a d} \epsilon_d \right]
                \frac{i}{2}
                \left[ \epsilon_b (\gamma^{\rho \mu} \gamma^\sigma)^{b c} \epsilon_c \right]}
                {[(x_1-x_2)(x_1-x_3)(x_1-x_4)(x_2-x_3)(x_2-x_4)(x_3-x_4)]^{\frac{3}{4}}} \n
 = &- \frac{1}{4} g_{\tau \sigma}
       \zeta^{\mathrm{NN}}_{\mu \nu}
     Tr \mathbf{P}
     \exp
          \left(
           i \int d \tau k \cdot A
          \right)
       k_\rho k_\lambda
       \left[ \epsilon_a (\gamma^{\lambda \nu \tau})^{a d} \epsilon_d \right]
       \left[ \epsilon_b (\gamma^{\rho \mu \sigma})^{b c} \epsilon_c \right] \n
   & \times \int d x_2 \int d x_3 \int d x_4
     \frac{(z-\bar{z})(z-x_1)(\bar{z}-x_1)}{x_{13} x_{14} x_{23} x_{24}}
     \left[ \frac{(x_2-x_3)(x_1-x_4)}
                 {(z-x_2)(z-x_3)(\bar{z}-x_1)(\bar{z}-x_4)} \right] \n
 = & - \frac{1}{4} g_{\tau \sigma}
       \zeta^{\mathrm{NN}}_{\mu \nu}
     Tr \mathbf{P}
     \exp
          \left(
           i \int d \tau k \cdot A
          \right)
       k_\rho k_\lambda
       \left[ \epsilon_a (\gamma^{\lambda \nu \tau})^{a d} \epsilon_d \right]
       \left[ \epsilon_b (\gamma^{\rho \mu \sigma})^{b c} \epsilon_c \right] \n
   & \times \int d x_2 \int d x_3 \int d x_4
     \frac{(z-\bar{z})(z-x_1)}
          {x_{13} x_{24}(z-x_2)(z-x_3)(\bar{z}-x_4)} \n
 = & \frac{1}{4} g_{\tau \sigma}
       \zeta^{\mathrm{NN}}_{\mu \nu}
     Tr \mathbf{P}
     \exp
          \left(
           i \int d \tau k \cdot A
          \right)
       k_\rho k_\lambda
       \left[ \epsilon_a (\gamma^{\lambda \nu \tau})^{a d} \epsilon_d \right]
       \left[ \epsilon_b (\gamma^{\rho \mu \sigma})^{b c} \epsilon_c \right] \n
   & \times \int d x_2 \int d x_3 \int d x_4
     \frac{z-\bar{z}}{(z-x_2)(\bar{z}-x_2)} \cdot
     \frac{z-x_1}{(z-x_3)(x_1-x_3)} \cdot
     \frac{\bar{z}-x_2}{(x_2-x_4)(\bar{z}-x_4)} .
\end{align}
Taking the limit $x_1 \rightarrow \bar{z}$ and $x_2 \rightarrow z$,
it becomes as:
\begin{align}
  & \frac{1}{4} g_{\tau \sigma}
       \zeta^{\mathrm{NN}}_{\mu \nu}
     STr
     \exp
          \left(
           i k \cdot A
          \right)
       k_\rho k_\lambda
       \left[ \epsilon_a (\gamma^{\lambda \nu \tau})^{a d} \epsilon_d \right]
       \left[ \epsilon_c (\gamma^{\rho \mu \sigma})^{b c} \epsilon_c \right]
    (2 \pi i)^3 \n
= & - 2 \pi^3 i g_{\tau \sigma}
       \zeta^{\mathrm{NN}}_{\mu \nu}
     STr
     \exp
          \left(
           i k \cdot A
          \right)
       k_\rho k_\lambda
       \left[ \epsilon_a (\gamma^{\lambda \nu \tau})^{a d} \epsilon_d \right]
       \left[ \epsilon_b (\gamma^{\rho \mu \sigma})^{b c} \epsilon_c \right] .
\label{AS43}
\end{align}

From (\ref{AS41}), (\ref{AS42}), and (\ref{AS43}), it can be said that
all of these results give the same result as (\ref{Gravitonf4:result});
that is :
\begin{equation}
       \zeta^{\mathrm{NN}}_{\mu \nu}
     STr
     \exp
          \left(
           i k \cdot A
          \right)
       k_\rho k_\lambda
       \left[ \epsilon_a (\gamma^{\rho \mu \tau})^{a b} \epsilon_b \right]
       \left[ \epsilon_c (\gamma^{\lambda \nu \sigma})^{c d} \epsilon_d
      \right] .
\end{equation}

\subsection{Symmetrized trace}
Symmetrized trace is given in \cite{OkawaOoguri3} by
\begin{align}
   &STr [e^{ikx} \mathcal{O}_1 \mathcal{O}_2 \dots \mathcal{O}_n]  \\
 = & \int_0^1 d \tau_1 \int^1_{\tau_1} d \tau_2 \dots \int^1_{\tau_{n-1}} d \tau_n \n
   & \qquad \times Tr [\mathcal{O}_1 e^{i \tau_1 k X} \mathcal{O}_2 e^{i (\tau_2 - \tau_1) k X}
      \dots \mathcal{O}_{n-1} e^{i (\tau_{n-1} - \tau_{n-2}) k X}
      \mathcal{O}_n e^{i (1 - \tau_{n-1}) k X} ] \n
   & + (((m-1)! - 1 ) \text{ more terms to symmetrize}) .
 \end{align} 
The above expression is equal to the following path ordered product:
 \begin{align}
   & Tr \mathbf{P} [ \exp(i \int_0^1 d \tau k X)
          \int_0^1 d \tau_1 \mathcal{O}_1 (\tau_1)
          \int_0^1 d \tau_2  \mathcal{O}_2 (\tau_2)
          \dots
          \int_0^1 d \tau_n \mathcal{O}_n (\tau_n)]\\
 = & \int_0^1 d \tau_1 \int^1_{\tau_1} d \tau_2 \dots \int^1_{\tau_{n-1}} d \tau_n \n
   & \qquad \times Tr [\mathcal{O}_1 (\tau_1)e^{i \tau_1 k X} \mathcal{O}_2 (\tau_2)e^{i (\tau_2 - \tau_1) k X}
      \dots \mathcal{O}_{n-1} (\tau_{n-1})e^{i (\tau_{n-1} - \tau_{n-2}) k X}
      \mathcal{O}_n (\tau_n)e^{i (1 - \tau_{n-1}) k X} ] \n
   & + (((m-1)! - 1 ) \text{ more terms to symmetrize}) ,
\end{align}
when $ \mathcal{O}_i (\tau_i)$ are constant matrices.

\subsection{10D Vertex operators }
Here we list the IIB matrix model vertex operators up to the
4-th rank antisymmetric tensor \cite{KMS}.
(The higher terms are given in \cite{KMS}.)
Here the symbol $\cdot$ in symmetrized trace distinguishes operators to be symmetrized (or
antisymmetrized) in $STr$.
\begin{equation}
 V^{\mathbf{\Phi}}(A,\epsilon) = STr e^{i k \cdot A} .
\end{equation}
\begin{equation}
 V^\Lambda(A,\epsilon) = STr e^{i k \cdot A} \bar{\epsilon}.
\end{equation}
\begin{equation}
   V_{\mu \nu}^B (A,\epsilon)
 = STr e^{i k \cdot A}
   \left( \frac{1}{16}
         k^\rho
         ( \bar{\epsilon} \cdot \gamma_{\mu \nu \rho} \epsilon)
          - \frac{i}{2}
          [A_\mu, A_\nu]
   \right) .
\end{equation}
\begin{equation}
   V_{\mu}^{\mathbf{\Psi}}(A,\epsilon)
 = STr e^{i k \cdot A}
   \left(
         - \frac{i}{12} k^\rho
         (\bar{\epsilon} \cdot \gamma_{\mu \nu \rho} \epsilon)
          - 2 [A_\mu,A_\nu]
   \right) .
\end{equation}
\begin{align}
     V^{h}_{\mu\nu}(A,\epsilon)
 =    & STr e^{ik\cdot A}
       \bigg( -\frac{1}{96} k^{\rho} k^{\sigma}
              \left(\bar{\epsilon} \cdot
              {\gamma_{\mu\rho}}^{\beta} \epsilon \right)
              \cdot \left( \bar{\epsilon} \cdot \gamma_{\nu \sigma \beta} \epsilon \right) \n
      & -\frac{i}{4} k^{\rho}
        \bar{\epsilon} \cdot \gamma_{\rho \beta (\mu} \epsilon \cdot {F_{\nu )}}^{\beta}
        + \frac{1}{2}\bar{\epsilon} \cdot
          \gamma_{(\mu}[A_{\nu )},\epsilon] + 2 {F_{\mu}}^{\rho} \cdot
          F_{\nu\rho} \bigg) .
\end{align}
\begin{align}
      V^{A}_{\mu\nu\rho\sigma}(A,\epsilon)
  =    & STr e^{ik \cdot A}
         \bigg( \frac{i}{8 \cdot 4!}
         k_{\alpha} k_{\beta}
         (\bar{\epsilon} \cdot {\gamma_{[\mu\nu}}^{\alpha} \epsilon)
         \cdot
         (\bar{\epsilon} \cdot {\gamma_{\rho\sigma]}}^{\beta}\epsilon)
        + \frac{i}{3}\bar{\epsilon}
          \cdot
          \gamma_{[\nu\rho\sigma}[\epsilon,A_{\mu]}]  \n
      & + \frac{1}{4}
          F_{[\mu\nu} \cdot (\bar{\epsilon} \cdot
          {\gamma_{\rho\sigma]}}^{\beta}\epsilon)k_{\beta}
          -iF_{[\mu\nu}\cdot F_{\rho \sigma]} \bigg) .
\end{align}

\end{document}